\RequirePackage[T1]{fontenc}
\documentclass[11pt, urlcolor=blue, linkcolor=blue]{article} 
\usepackage{cite}
\usepackage{float}
\usepackage[utf8]{inputenc}

\usepackage{multicol}
\usepackage{multirow}
\usepackage{tablefootnote}
\usepackage{amsmath, amsthm, amssymb,slashed,mathtools,tabu}
\usepackage{ifpdf}
\ifpdf
  \usepackage[pdftex]{graphicx}
  \usepackage{epstopdf}
\else
  \usepackage[dvips]{graphicx}
\fi
\parskip=\baselineskip

\usepackage[usenames, dvipsnames]{color}
\usepackage[svgnames]{xcolor}
\usepackage[colorlinks,citecolor=RoyalBlue, urlcolor=RoyalBlue, linkcolor=RoyalBlue ]{hyperref} 

\allowdisplaybreaks[1]

\numberwithin{equation}{section}

\usepackage{booktabs}

\theoremstyle{definition}

\newcommand{\cblue}[1]{\textcolor{blue}{#1}}

\newcommand{\cblack}[1]{\textcolor{black}{#1}}

\definecolor{mygray}{gray}{0.6}

\usepackage{upgreek}
\usepackage{bbm}

\newcommand{\GSD}{\text{GSD}}

\newenvironment{myfont}[2][]{\csname#2\endcsname[#1]}{}

\usepackage{slashed}
\usepackage[makeroom]{cancel}
\usepackage[normalem]{ulem}
\usepackage{soul}
\newcommand{\stkout}[1]{\ifmmode\text{\sout{\ensuremath{#1}}}\else\sout{#1}\fi}

\usepackage{sseq}
\usepackage[all,cmtip]{xy}
\usepackage{tikz-cd}
\usepackage{tikz}
\usetikzlibrary{matrix}

\newcommand{\bea}{\begin{eqnarray}}
\newcommand{\eea}{\end{eqnarray}}
\def\be{\begin{equation}}
\def\ee{\end{equation}}

\newcommand{\e}{\hspace{1pt}\mathrm{e}}

\newcommand{\ii}{\hspace{1pt}\mathrm{i}\hspace{1pt}}

\def\RP{{\mathbb{RP}}}

\newcommand{\nn}{\nonumber}

\definecolor{red}{rgb}{1,0,0}
\definecolor{blue}{rgb}{0,0,1}
\definecolor{dblue}{rgb}{0,0,0.4}
\definecolor{green}{rgb}{0,1,0}
\definecolor{black}{rgb}{0,0,0}
\definecolor{white}{rgb}{1,1,1}

\definecolor{brn}{rgb}{.8,.4,.0}
\definecolor{redo}{rgb}{1,.5,.0}
\definecolor{ddgrn}{rgb}{0,0.4,0}
\definecolor{dgrn}{rgb}{0,0.55,0}
\definecolor{dbl}{rgb}{0,0,0.5}

\usepackage[bbgreekl]{mathbbol}
\usepackage{amscd}

\newcommand{\Z}{\mathbb{Z}}
\newcommand{\C}{\mathbb{C}}

\newcommand{\dd}{\hspace{1pt}\mathrm{d}}

\newcommand{\Ref}[1]{Ref.~\cite{#1}}
\newcommand{\Eq}[1]{(\ref{#1})}

\newcommand{\Eqn}[1]{Eqn.~(\ref{#1})} 

\newcommand{\Tr}{{\rm Tr}}

\newcommand{\bpm}{\begin{pmatrix}}
\newcommand{\epm}{\end{pmatrix}}
\newcommand{\bmm}{\begin{matrix}}
\newcommand{\emm}{\end{matrix}}

\newcommand{\al}{\alpha} 
\newcommand{\bt}{\beta} 
\newcommand{\del}{\delta}

\newcommand{\ga}{\gamma}

\newcommand{\si}{\sigma}

\def\CA{{\cal A}}

\def\CH{{\cal H}}

\def\Z{{\mathbb{Z}}}

\def\C{{\mathbb{C}}}

\def\Tr{{\mathrm{Tr}}}

\def \H{\operatorname{H}}

\def \Z{\mathbb{Z}}
\def \Pin{\mathrm{Pin}}

\def \RP{\mathbb{RP}}

\newcommand{\Sec}[1]{Sec.~\ref{#1}}

\usetikzlibrary{decorations.markings}
\usetikzlibrary{positioning}
\usetikzlibrary{shadings}

\newcommand{\SO}{{\rm SO}}
\newcommand{\Spin}{{\rm Spin}}
\newcommand{\U}{{\rm U}}
\newcommand{\SU}{{\rm SU}}
\newcommand{\Sp}{{\rm Sp}}

\def\TP{\mathrm{TP}}

\usepackage{datetime}
\usepackage{enumitem} 
\usepackage{moreenum}

\newcommand{\sharpfootnote}[1]{%
\let\oldthefootnote=\thefootnote%
\stepcounter{mpfootnote}%
\addtocounter{footnote}{-1}%
\renewcommand{\thefootnote}{{W$^+$}} 
\footnote{#1}%
\let\thefootnote=\oldthefootnote%
}

\newcommand{\naturalfootnote}[1]{%
\let\oldthefootnote=\thefootnote%
\stepcounter{mpfootnote}%
\addtocounter{footnote}{-1}%
\renewcommand{\thefootnote}{{W$^-\natural$}}
\footnote{#1}%
\let\thefootnote=\oldthefootnote%
}

\newcommand{\flatfootnote}[1]{%
\let\oldthefootnote=\thefootnote%
\stepcounter{mpfootnote}%
\addtocounter{footnote}{-1}%
\renewcommand{\thefootnote}{{W$^-\flat$}}
\footnote{#1}%
\let\thefootnote=\oldthefootnote%
}

\DeclareRobustCommand\sWang
{\cblue{\includegraphics[height=3.2ex]{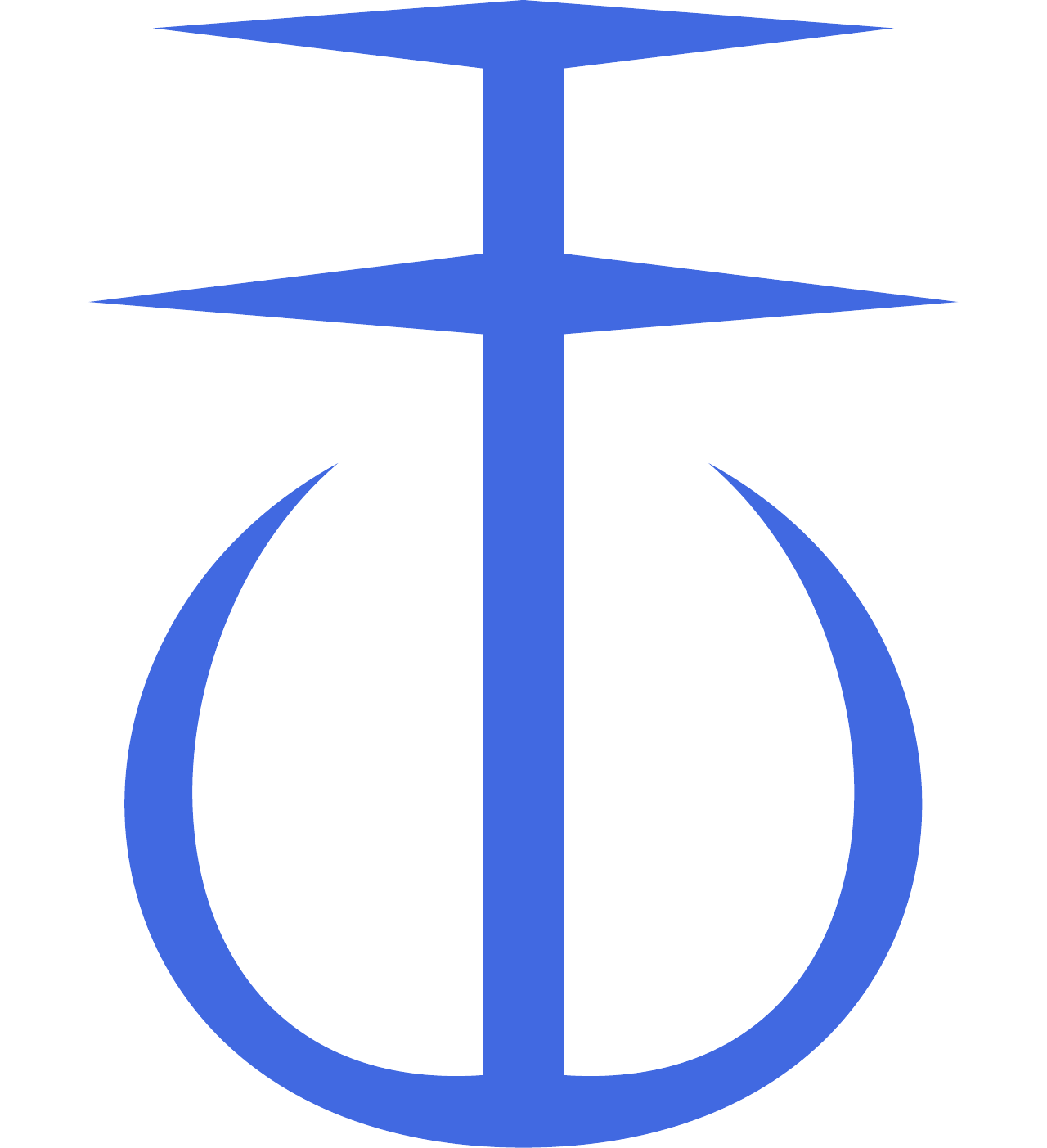}}}

\newcommand{\Wangfootnote}[1]{%
\let\oldthefootnote=\thefootnote%
\stepcounter{mpfootnote}%
\addtocounter{footnote}{-1}%
\renewcommand{\thefootnote}{\sWang}
\footnote{#1}%
\let\thefootnote=\oldthefootnote%
}

\usepackage{comment}

\usepackage{upgreek}
\newcommand{\Fig}[1]{Fig.~\ref{#1}}

\usepackage{mathrsfs}

\newcommand{\Table}[1]{Table \ref{#1}}

\begin{document}
\begin{titlepage}
\vskip1.25in
\begin{center}


{\bf\LARGE{ 
Anomaly and Cobordism Constraints
\\[8mm]
Beyond Grand Unification: Energy Hierarchy 
}}

\vskip0.5cm 
\Large{\quad\quad Juven Wang${\;}^{1}$
\Wangfootnote{
{\tt jw@cmsa.fas.harvard.edu} \quad\; 
\includegraphics[width=3.0in]{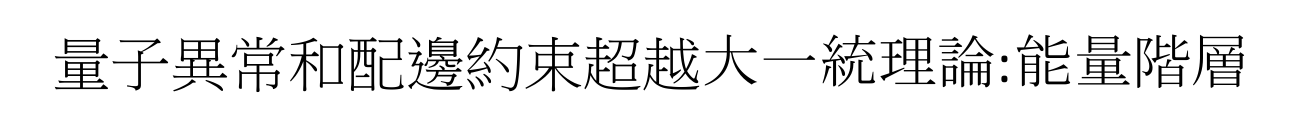}
\quad\quad \hfill \href{http://sns.ias.edu/~juven/}{\bf http://sns.ias.edu/$\sim$juven/} \\[2mm]
All prior supports from
\href{https://www.ntu.edu.tw/index.html}{\cblack{National Taiwan University (Taipei)}},
Massachusetts Institute of Technology, Perimeter Institute for Theoretical Physics, 
Tsinghua University (Beijing),
and Institute for Advanced Study (Princeton), are greatly appreciated. \\[2mm]
} 
\\[2.75mm]  
} 
\vskip.5cm
{ {\small{\textit{$^1${Center of Mathematical Sciences and Applications, Harvard University,  Cambridge, MA 02138, USA}}}}
}

\end{center}

\vskip 0.5cm
\baselineskip 16pt
\begin{abstract}

 A recent work \cite{JW2006.16996} suggests that a 
 4d nonperturbative global anomaly of mod 16 class
 hinting a possible new hidden gapped topological sector beyond the Standard Model (SM) and Georgi-Glashow $su(5)$ Grand
 Unified Theory (GUT) with 15n chiral Weyl fermions
 and a discrete $\mathbb{Z}_{4,X}$ symmetry of $X=5({\bf B -  L})-4Y$.  
This $\mathbb{Z}_{16}$ class global anomaly is a mixed gauge-gravitational anomaly 
between the discrete $X$ and spacetime backgrounds.
The new topological sector has a GUT scale high energy gap, below its low energy
encodes either a 4d noninvertible topological quantum field theory (TQFT), 
or a 5d short-range entangled invertible TQFT, or their combinations. 
This hidden topological sector provides the 't Hooft anomaly matching of 
the missing sterile right-handed neutrinos (3 generations of 16th Weyl fermions),
and possibly also accounts for the Dark Matter sector.
In the SM and $su(5)$ GUT, the discrete $X$ can be either a global symmetry or gauged.
In the $so(10)$ GUT, the $X$ must become gauged,
the 5d TQFT  becomes noninvertible and long-range entangled (which can couple to dynamical gravity). 
In this work, we further examine the anomaly and cobordism constraints 
at higher energy scales above the $su(5)$ GUT to $so(10)$ GUT and $so(18)$ GUT (with Spin(10) and Spin(18) gauge groups precisely).
We also find the  \cite{JW2006.16996}'s proposal on new hidden gapped topological sectors can be consistent with 
anomaly matching under the energy/mass hierarchy. Novel ingredients along tuning the energy
include various energy scales of anomaly-free symmetric mass generation (i.e., Kitaev-Wen mechanism),
 the Topological Mass/Energy Gap from anomalous symmetric topological order (attachable to a 5d $\mathbb{Z}_{4,X}$-symmetric topological superconductor),
 possible {topological quantum phase transitions}, and Ultra Unification that includes GUT with new topological sectors. \\[4mm]

\flushright
July 2020



\end{abstract}

\end{titlepage}

  \pagenumbering{arabic}
    \setcounter{page}{2}
    

\tableofcontents



\hfill ``Lieber Goldberg, spiele mir doch eine meiner Variationen.'' \\[2mm] 
\hspace*{\fill}``Dear Goldberg, play one of my variations.'' \\[2mm]
\hspace*{\fill}Goldberg Variations, BWV 988\\[2mm] 
\hspace*{\fill} Johann Sebastian Bach in 1826

\section{Introduction}

Based on a recent work  \cite{JW2006.16996}, the author examined the 
anomaly and cobordism constraints on 
Glashow-Salam-Weinberg 
Standard Models (SM) with a local Lie algebra $su(3) \times su(2) \times u(1)$
\cite{Glashow1961trPartialSymmetriesofWeakInteractions, Salam1964ryElectromagneticWeakInteractions, Weinberg1967tqSMAModelofLeptons}
of four versions of gauge groups\footnote{The ${\Z_q} \equiv \Z/(q \Z) \equiv (\Z \mod q)$ denotes the finite abelian cyclic group of order $q$.
We also denote 
${\Z_q^p} \equiv {(\Z_q)^p}$ as the $p$th power of $\Z_q$.}  
\bea \label{eq:GSM}
G_{\text{SM}_q} \equiv \frac{\SU(3)\times \SU(2)\times \U(1)}{\Z_q}, \quad q=1,2,3,6,
\eea
and Georgi-Glashow (GG) $su(5)$ Grand Unification \cite{Georgi1974syUnityofAllElementaryParticleForces}, or $su(5)$ Grand Unified Theory (GUT),\footnote{Follow the mathematical convention, we have used the lowercase letter to represent the local Lie algebra. We use the capital uppercase letter to
represent the global Lie group. For example, the $su(5)$ Lie algebra can have a global Lie group SU(5) and others,
the $so(10)$ Lie algebra can have a global Lie group SO(10) or Spin(10), etc.
The reason is that the pioneer work on the Grand Unification 
mostly focus on the local Lie algebra structure\cite{georgi2018liebook}. For example, the $so(10)$ GUT 
of Georgi or Fritzsch-Minkowski GUT  \cite{Fritzsch1974nnMinkowskiUnifiedInteractionsofLeptonsandHadrons} 
does not have the SO(10) Lie group  but instead requires a double covered Spin(10) Lie group.
The $so(18)$ GUT \cite{WilczekZee1981iz1982Spinors,Fujimoto1981SO18Unification}
does not have the SO(18) Lie group  but instead requires a double covered Spin(18) Lie group. Therefore, we decide to stick
to this math convention to avoid the
confusion between Lie algebra (the lowercase letter) and Lie group (the uppercase letter).
}
with additional symmetry such as the baryon ($\mathbf{B}$) minus lepton ($\mathbf{L}$) number.
The constraints from cobordism include all invertible quantum anomalies involving the given internal gauge groups:\footnote{The closely related cobordism classifications for SM and GUT are also pursued by recent pioneer works
including \Ref{GarciaEtxebarriaMontero2018ajm1808.00009, 2019arXiv191011277D} based on Atiyah-Hirzebruch spectral sequence (AHSS), and \Ref{WangWen2018cai1809.11171,
WanWang2018bns1812.11967, WW2019fxh1910.14668, WanWangv2} based on Adams spectral sequence (ASS).
In particular, \Ref{GarciaEtxebarriaMontero2018ajm1808.00009} examined no global anomaly for $su(5)$ GUT alone (without extra global symmetries).
\Ref{WW2019fxh1910.14668, WangWen2018cai1809.11171} also checked and found at most a perturbative $\Z$ class local anomaly for SU(5) chiral fermion theory
and at most a nonperturbative $\Z_2$ class global anomaly for Spin(10) chiral fermion theory (this anomaly is similar to the new SU(2) global anomaly \cite{WangWenWitten2018qoy1810.00844}).
\Ref{WangWen2018cai1809.11171} finds that these anomalies are absence in $su(5)$ GUT and $so(10)$ GUT.
\Ref{2019arXiv191011277D} and  \cite{WW2019fxh1910.14668} checked no global anomaly for four versions of SM models given by the internal gauge group 
\Eq{eq:GSM} (for the cases without extra global symmetries).}
including all \\[-10mm]
\begin{itemize}
\item\emph{perturbative local anomalies}, classified by $\Z$ classes (known as free classes), and  
\item \emph{nonperturbative global anomalies},  classified by $\Z_n$ classes (known as torsion classes). \\[-10mm]
\end{itemize}
The computations of cobordism classifications used in the \cite{JW2006.16996}  are mostly done in \cite{WW2019fxh1910.14668}, based on
Thom-Madsen-Tillmann spectra \cite{thom1954quelques,MadsenTillmann4}, Adams spectral sequence \cite{Adams1958}, 
and Freed-Hopkins theorem \cite{Freed2016},
and the author's prior work jointly with Wan \cite{WanWang2018bns1812.11967, WW2019fxh1910.14668, WanWangv2}.

However, \Ref{GarciaEtxebarriaMontero2018ajm1808.00009} and \cite{WW2019fxh1910.14668} suggested a $\Z_{16}$ (a mod 16 class) 
mixed gauge-gravitational nonperturbative global anomaly,
when there is a discrete $\Z_{4,X}$ symmetry together with a spacetime geometry background probe. 
The $X$ can represent a baryon ($\mathbf{B}$) minus lepton ($\mathbf{L}$) number in the SM case \Eq{eq:GSM},
but the $X$ can also represent a modified version of $({ \mathbf{B}-  \mathbf{L}})$ number up to some electroweak hypercharge $Y$ \cite{Wilczek1979hcZee} in the $su(5)$ GUT:\footnote{Note we choose the convention 
that the $\U(1)_{\rm{EM}}$ electromagnetic charge is
$Q_{\rm{EM}}=T_3 +Y$. 
The $\U(1)_{\rm{EM}}$ is the unbroken (not Higgsed) electromagnetic gauge symmetry
and $T_3= \frac{1}{2}
\begin{pmatrix}
1 & 0\\
0 & -1
\end{pmatrix}$ is a generator of SU(2)$_{\text{weak}}$, 
and other conventions of hypercharges can be related by $\tilde Y = 3 Y_W = 6 Y$
 \cite{WW2019fxh1910.14668}.
}
\bea
X &\equiv&
5({ \mathbf{B}-  \mathbf{L}})-4Y.
\eea
In the SM and $su(5)$ GUT, one can consider such an $X$ charge corresponds to the U(1)$_X$ symmetry.
At different energy scales, the U(1)$_X$ symmetry may:
(i) remain a global symmetry, 
(ii) gauged or
(iii) broken spontaneously or explicitly.
But in the Georgi or Fritzsch-Minkowski $so(10)$ GUT  \cite{Fritzsch1974nnMinkowskiUnifiedInteractionsofLeptonsandHadrons}, it is more natural to keep only 
a discrete order-4 subgroup out of the continuous U(1)$_X$:
$$
\Z_{4,{X}} \subset \U(1)_X
$$ sitting precisely and naturally at the center of Spin(10) gauge group:
\bea
\Z_{4,{X}} = Z(\Spin(10))  \subset \Spin(10).
\eea
The $Z(G)$ denotes the center of $G$.
Thus the $\Z_{4,{X}}$ is at least dynamically gauged in the Spin(10) gauge group for the $so(10)$ GUT.
This mixed gauge-gravitational nonperturbative global anomaly of $\Z_{16}$ classes is characterized by
a 5d cobordism invariant $\eta\big(\text{PD}(\CA_{{\Z_2}})  \big)$\footnote{{The 
$\eta$ is a 4d cobordism invariant of $\Z_{16}$ classe with a Pin$^+$ structure \cite{KT1990}. The PD($\CA_{{\Z_2}}$) defines the Poincar\'e dual (PD) of $\CA_{{\Z_2}}$.
The ${\CA_{{\Z_2}} \in \H^1(M, \Z_{4,X}/\Z_2^F)}$ is locally a $\Z_2=\Z_{4,X}/\Z_2^F$ gauge field. The $\Z_2^F$ is the fermion parity. See
more explanations later.}}
 also called a 5d invertible TQFT (iTQFT or invertible topological order\footnote{In principle, the
iTQFT is the low energy theory description of some gapped phases of invertible topological order. The intrinsic topological order can be long-range entangled,
so some of invertible topological orders are also long-range entangled.
But a subclass of invertible topological orders is in fact short-range entangled known as symmetry-protected topological state (SPTs).
The definition of long-range entangled vs short-range entangled states are based on the modern definition of gapped quantum matter by Wen \cite{Wen2016ddy1610.03911}.
See an overview on the quantum matter terminology \cite{Senthil1405.4015, Wen2016ddy1610.03911}. })
studied in \cite{2018arXiv180502772T, GarciaEtxebarriaMontero2018ajm1808.00009,
Hsieh2018ifc1808.02881, GuoJW1812.11959, WW2019fxh1910.14668,JW2006.16996}.
Its precise 5d partition function (whose boundary has the 4d global anomaly) is explained in  \cite{JW2006.16996}:
\bea \label{eq:Z16anomaly}
{\bf Z}_{\text{5d-iTQFT}}=\exp\bigg(\frac{2\pi \ii}{16} \cdot(-N_{\text{generation}}) \cdot  \eta\big(\text{PD}(\CA_{{\Z_2}})  \big) \bigg\rvert_{M^5}\bigg).
\eea
Here $N_{\text{generation}}$ is the number of generations, which is $N_{\text{generation}}=3$ in the SM.
 The ${\CA_{{\Z_2}} \in \H^1(M, \Z_{4,X}/\Z_2^F)}$ is the first cohomology class of $\Z_2 =  \Z_{4,X}/\Z_2^F$
 (locally and loosely speaking,  ${\CA_{{\Z_2}}}$ is analogous to a 1-form $\Z_2$ gauge field) for ${\Spin \times_{\Z_2} \Z_4}= {\Spin \times_{\Z_2^F} \Z_{4,X}}$ 
 structure.\footnote{Here the spacetime symmetry is commonly denoted as a Spin group omitting the input of the spacetime dimensions $(d+1)$, 
 either for the Lorentz signature Spin($d,1$) or the Euclidean signature Spin($d+1$). See \Sec{sec:Representationofspacetimesymmetrygroup} for more details.}
 The ${\Z_2^F}$ is the fermion parity symmetry shared by the center of the spacetime $\Spin$ group and the normal subgroup of $\Z_{4,X}$.
 The $\eta$ is a $\Z_{16}$ class of  4d cobordism invariant of
 $\Pin^+$ structure \cite{KT1990}.
The \Eq{eq:Z16anomaly} can be detected on a 5-dimensional real projective space $\RP^5$ by computing the partition function
${\bf Z}_{\text{5d-iTQFT}}[M^5=\RP^5]$ \cite{2018arXiv180502772T, JW2006.16996, Kapustin1406.7329, Hason2019akwRyanThorngren1910.14039}.

We can read the classifications of $(d-1)$-dimensional anomalies from the $d$-th mathematical bordism group denoted as
\bea
\Omega_{d}^{G},
\eea
and a specific version of cobordism group (firstly defined to classify Topological Phases [TP] in \cite{Freed2016})
\bea\label{eq:TPG}
\Omega^{d}_{G} &\equiv&
\Omega^{d}_{({\frac{{G_{\text{spacetime} }} \ltimes  {{G}_{\text{internal}} }}{{N_{\text{shared}}}}})} 
\equiv
\TP_d(G).
\eea
The $\ltimes$ is a twisted product known as a semi-direct product.
For example, we can read \Eqn{eq:Z16anomaly} as the 5d cobordism invariant listed inTable 4 of 
\Ref{JW2006.16996} with {$G=\Spin\times_{\Z_2} \Z_4 \times G_{\text{SM}_q} $} and {$G=\Spin\times_{\Z_2} \Z_4 \times \SU(5)$}.

In order to match the non-vanishing anomaly \Eq{eq:Z16anomaly}, it is commonly and n\"aively believed that either of the following scenarios must hold (see the summary in
\Ref{JW2006.16996}'s Sec.~4.3 and Sec.~5):
\begin{enumerate}[leftmargin=4.mm, label=\textcolor{blue}{(\roman*)}., ref={(\roman*)}]

\item The $\Z_{4,{X}}$ symmetry is broken. For example, by the spontaneous or explicit breaking (or by the Dirac or Majorana masses as in the scenario \ref{iii}). 

\item There is the 16th Weyl spinor as the sterile right-handed neutrino being gapless (a free theory or a free conformal field theory (CFT)).
So the $\Z_{4,{X}}$ symmetry can be preserved. The total number of Weyl spacetime spinors are 16n where n is an integer n $\in \Z$.

\item \label{iii}
There is the 16th Weyl spinor as the sterile right-handed neutrino, but it is gapped by Dirac or Majorana masses, such as in the seesaw mechanism \cite{Minkowski1977seesaw, GellMann1979seesaw}.
Thus the $\Z_{4,{X}}$ symmetry is broken.\\[-10mm]
\end{enumerate}
 
However, the novelty of  \cite{JW2006.16996} is suggesting that \emph{none} of the above needs to be obeyed 
(the $\Z_{4,{X}}$ needs not to be broken, \emph{nor} do we need  the 16th Weyl spinor as the sterile right-handed neutrino
 being gapless or having Dirac/Majorana masses). \Ref{JW2006.16996} suggests a new scenario: \\[-10mm]
\begin{enumerate}[leftmargin=.mm, label=\textcolor{blue}{(\roman*)}., ref={(\roman*)}]
\setcounter{enumi}{3}

\item \label{iv} 
The $\Z_{4,{X}}$ symmetry can be preserved but the $\Z_{16}$ class anomaly \Eq{eq:Z16anomaly}
needs to be matched by
 a new gapped topological sector.
The new gapped (previously missing) sector can be either a 4d long-range entangled 
noninvertible topological quantum field theory (TQFT),\footnote{{A TQFT is known as the low energy theory of topological order.
The topological order in condensed matter requires an ultraviolet (UV) lattice completion.}
This phenomenon of symmetric gapped TQFT
 with 't Hooft anomaly is noticed first in a lower dimension 2+1d boundary of 3+1d bulk in condensed matter \Ref{VishwanathSenthil1209.3058}, see an overview 
 \cite{Senthil1405.4015, Wang2017locWWW1705.06728} and a symmetry-extension 
 approach of general construction \cite{Wang2017locWWW1705.06728, Prakash2018ugo1804.11236, Wan2018djlW2.1812.11955} and counter examples \cite{Wan2018djlW2.1812.11955, Cordova2019bsd1910.04962, CordovaOhmori1912.13069}.} 
 or a 5d short-range entangled invertible TQFT, or their combinations. 
This hidden topological sector provides the 't Hooft anomaly matching of 
the missing sterile right-handed neutrinos (with 3 generations),
and possibly also accounts for the Dark Matter sector.\end{enumerate}

Previous work \cite{JW2006.16996} checks explicitly that the anomaly and cobordism constraints 
\emph{below and around} SM, electroweak and Higgs energy scale to the $su(5)$ GUT scales.

\begin{figure}[t!] 
     \includegraphics[width=6.in]{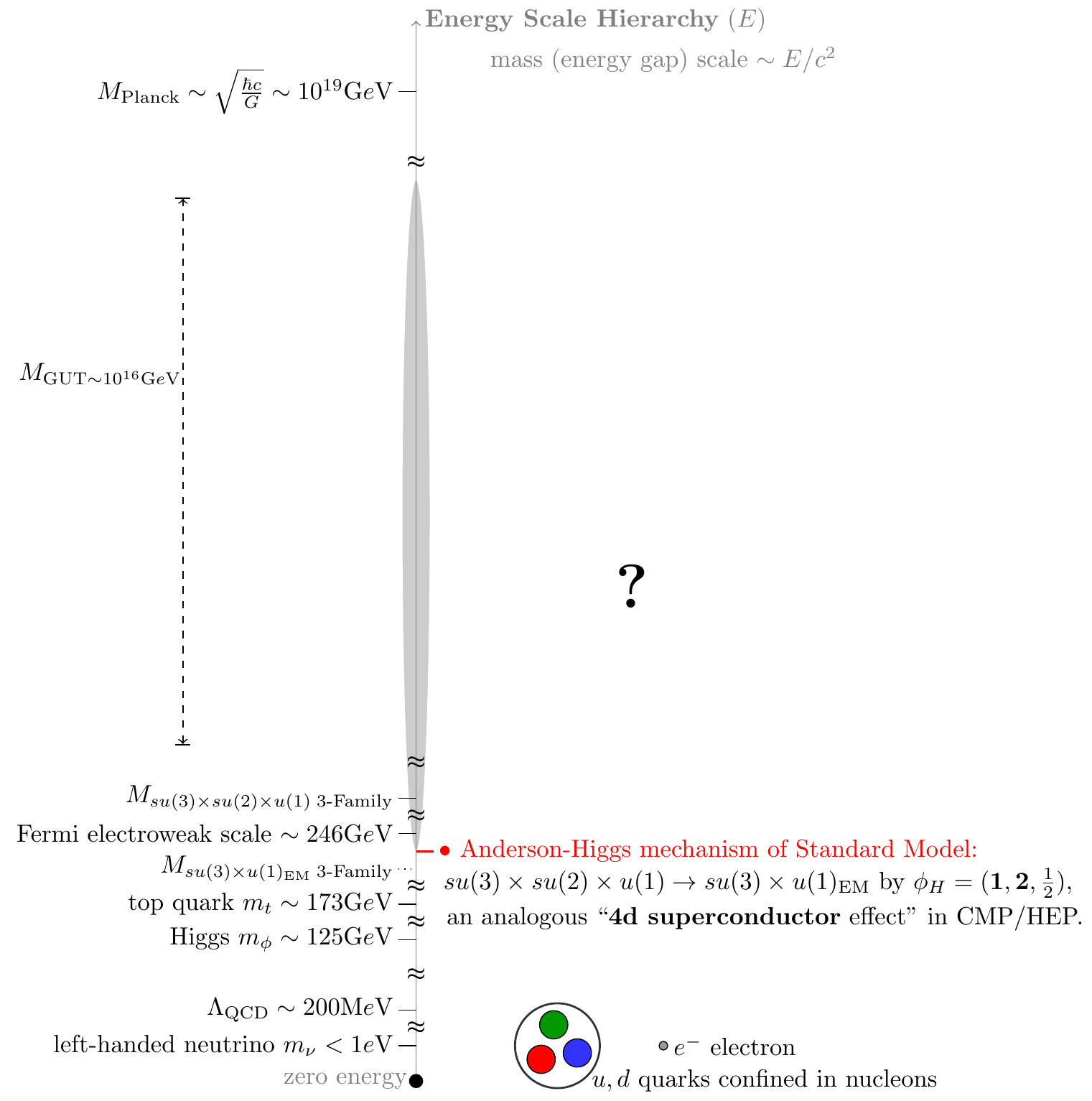}
  \caption{Energy and Mass Hierarchy contemporarily confirmed in the Standard Model:
  In the figure, the breaking structure and hierarchy structure always concern 
  the global Lie group: $\frac{\SU(3)\times \SU(2)\times \U(1)}{\Z_q}, etc$.
  However, we simply denote the local Lie algebra such as $su(3)  \times su(2) \times u(1), etc.$, 
  only for the abbreviation brevity and only to be consistent with the notations of early physics literature.
  The present work address possible the energy hierarchy in the gray region (with a question mark {\bf ?}) around the GUT scale
  above the SM scale and below the Planck scale. See the new proposal in \Fig{fig:energy-hierarchy-lie-algebra-1} and 
  \Fig{fig:energy-hierarchy-lie-algebra-2}.
  }
  \label{fig:energy-hierarchy-lie-algebra-world}
\end{figure}

The purpose of this present work is to check explicitly that the anomaly and cobordism constraints 
\emph{above} the $su(5)$ GUT scale to the higher energy $so(10)$ GUT scale and a further higher energy 
$so(18)$ GUT \cite{WilczekZee1981iz1982Spinors,Fujimoto1981SO18Unification} scale.\footnote{The
$so(10)$ GUT scale and the $so(18)$ GUT have the $so(10)$ and $so(18)$ gauge Lie algebras, 
but precisely we need Spin(10) and Spin(18) gauge Lie groups. The reason is that the fermion matter fields
are not only the spacetime spinor of the Lorentz group (or Spin group) but also in the spinor representation of the internal symmetry.
The $so(10)$ GUT requires the irreducible ${\bf 16}^+$ spinor representation thus which must be in Spin(10).
The $so(18)$ GUT requires the irreducible ${\bf 256}^+$ spinor representation thus which must be in Spin(18).}
In addition, we also aim to understand whether the new proposal in \cite{JW2006.16996}
is still consistent with the additional constraints in the higher energy scales.\footnote{In the present work,
we focus on the anomaly involving the internal symmetry group ${{G}_{\text{internal}} }$ under the spacetime ${G_{\text{spacetime} }}$ background probes. 
After gauging the internal symmetry group ${{G}_{\text{internal}}}$, there could give rise to
higher generalized global $n$-symmetry \cite{Gaiotto2014kfa1412.5148} whose charged objects are $n$-dimensional (in the spacetime picture).
There could be additional new higher 't Hooft anomalies involving higher $n$-symmetries after gauging ${{G}_{\text{internal}}}$.
For example, a pure 4d SU(2) Yang-Mills gauge theory at the topological term $\theta$-term (as the second Chern-class $c_2$ of the SU(2) gauge bundle), with or without Lorentz symmetry enrichment, 
can have higher 't Hooft anomaly mixing between 1-form $\Z_2$ electric symmetry and the time-reversal (or $CP$) symmetry 
\cite{Gaiotto2017yupZoharTTT, Wan2018zqlWWZ1812.11968, Wan2019oyr1904.00994};
a similar phenomenon happens for a 4d U(1) gauge theory \cite{Hsin1904.11550, WangYouZheng1910.14664}.

The additional higher 't Hooft anomalies for SM or GUT, if any, would \emph{not} affect the consistency conditions based on the dynamical gauge anomaly cancellations
that we established (in \cite{JW2006.16996} and the present work).
The additional higher 't Hooft anomalies, if any,  \emph{only} implies that the
higher $n$-symmetries may be emergent and not strictly regularized on the $n$-simplices.
The additional higher 't Hooft anomalies for SM or GUT, if any, can be used to constrain the quantum gauge dynamics, see \cite{MonteroJW}.
} 
{In particular, given the 
present energy hierarchy phenomenological input shown in
\Fig{fig:energy-hierarchy-lie-algebra-world},
``are we able to provide some nonperturbative proposals on the higher energy scales 
(the gray region with a question mark {\bf ?} around the GUT scale in \Fig{fig:energy-hierarchy-lie-algebra-world}), 
 given our knowledge of low energy SM physics,
based on the more complete list of anomaly matching and cobordism constraints?''}
To address this question,
we first need to build up tools of the {spacetime and internal symmetry group embedding} and the representation hierarchy in a systematic careful way in
 \Sec{sec:GaugeGroupHierarchy};
 then we analyze possible scenarios of {energy and mass hierarchy from local and global anomaly constraints}
in \Sec{sec:EnergyandMassHierarchy}.
With some phenomenological and mathematical input, we will be able to come back to address this apostrophe-quoted question in \Sec{sec:Conclusion} in Conclusion.

\section{Gauge Group Embedding and Representation Hierarchy}
\label{sec:GaugeGroupHierarchy}

\subsection{Spacetime and internal symmetry group embedding}

\begin{figure*}[!h] 
\hspace{4mm}
  \includegraphics[width=6.6in]{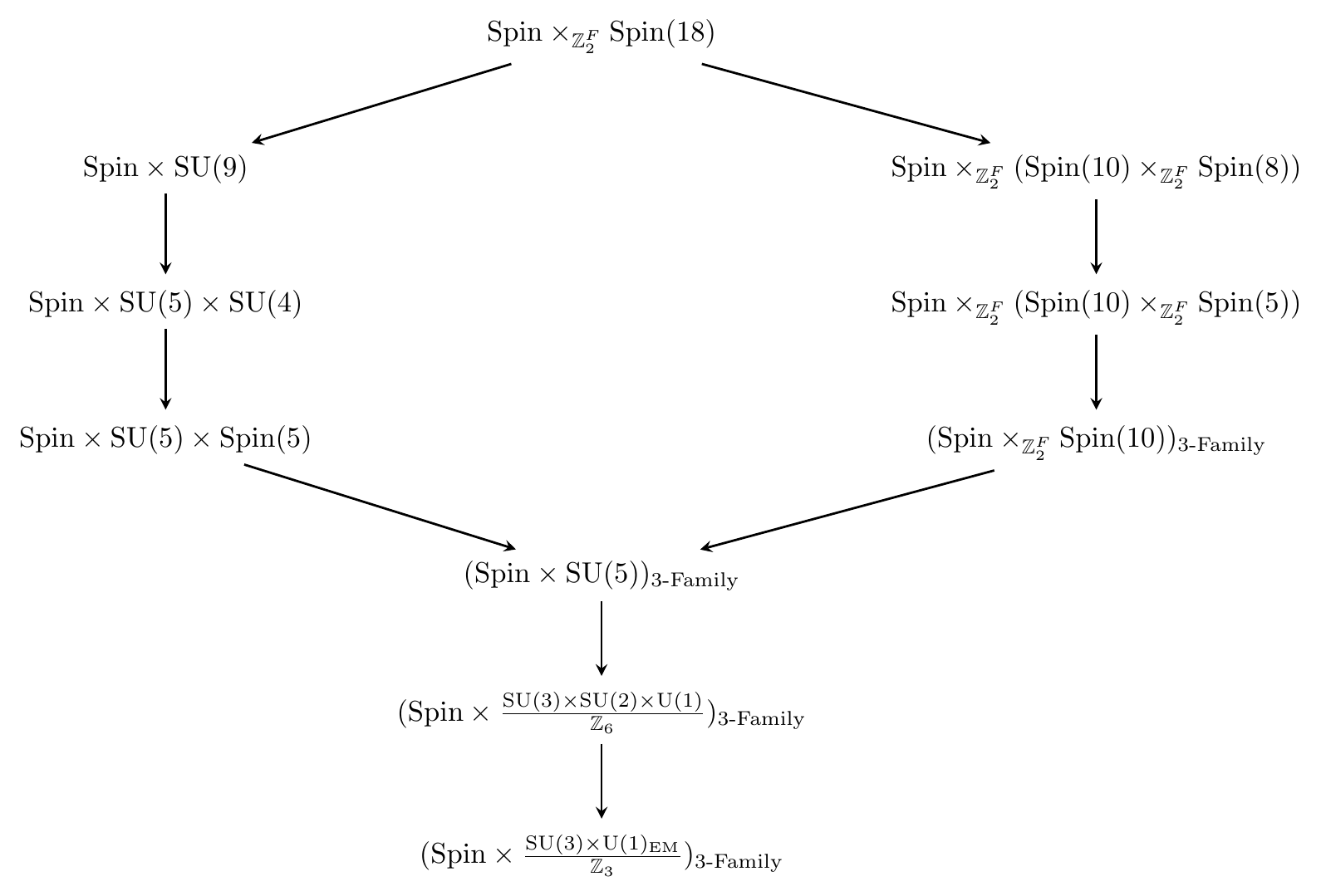}
\caption{The full spacetime-internal symmetry
$G={\frac{{G_{\text{spacetime} }} \ltimes  {{G}_{\text{internal}} }}{{N_{\text{shared}}}}}$ (the precise global symmetry before gauging the ${{G}_{\text{internal}} }$)
for the hierarchy starting from the $so(18)$ GUT with ${{\Spin \times_{\Z_2^F} \Spin(18)}}$, which can be placed on \emph{non-spin manifolds}.
Note that
${\Spin(6)=\SU(4)}$ $\supset$ ${\Spin(5)=\Sp(2)=\U\Sp(4)}$ and
recall 
$G_{\text{SM}_q}$ $\equiv$ $\frac{\SU(3)_{\text{strong}}\times \SU(2)_{\text{weak}}\times \U(1)_Y}{\Z_q}$.
The subscript ``{\text{3-Family}}'' means there are 3 families (or 3 generations) of matter fields, e.g., quarks and leptons. 
Here the arrow from $G_1 \to G_2$ means particularly that $G_1 \supseteq G_2$ contains the later as a subgroup.
This shows the web of full symmetry group embedding, similar to Table 4 of \cite{1711.11587GPW}.
 We have computed the cobordism group 
 $\TP_d(G)$
 of these spacetime-internal symmetry group $G$ in \Ref{WanWangv2}. 
}
 \label{table:sym-web-1}
\end{figure*}

\begin{figure*}[!h] 
  \includegraphics[width=6.4in]{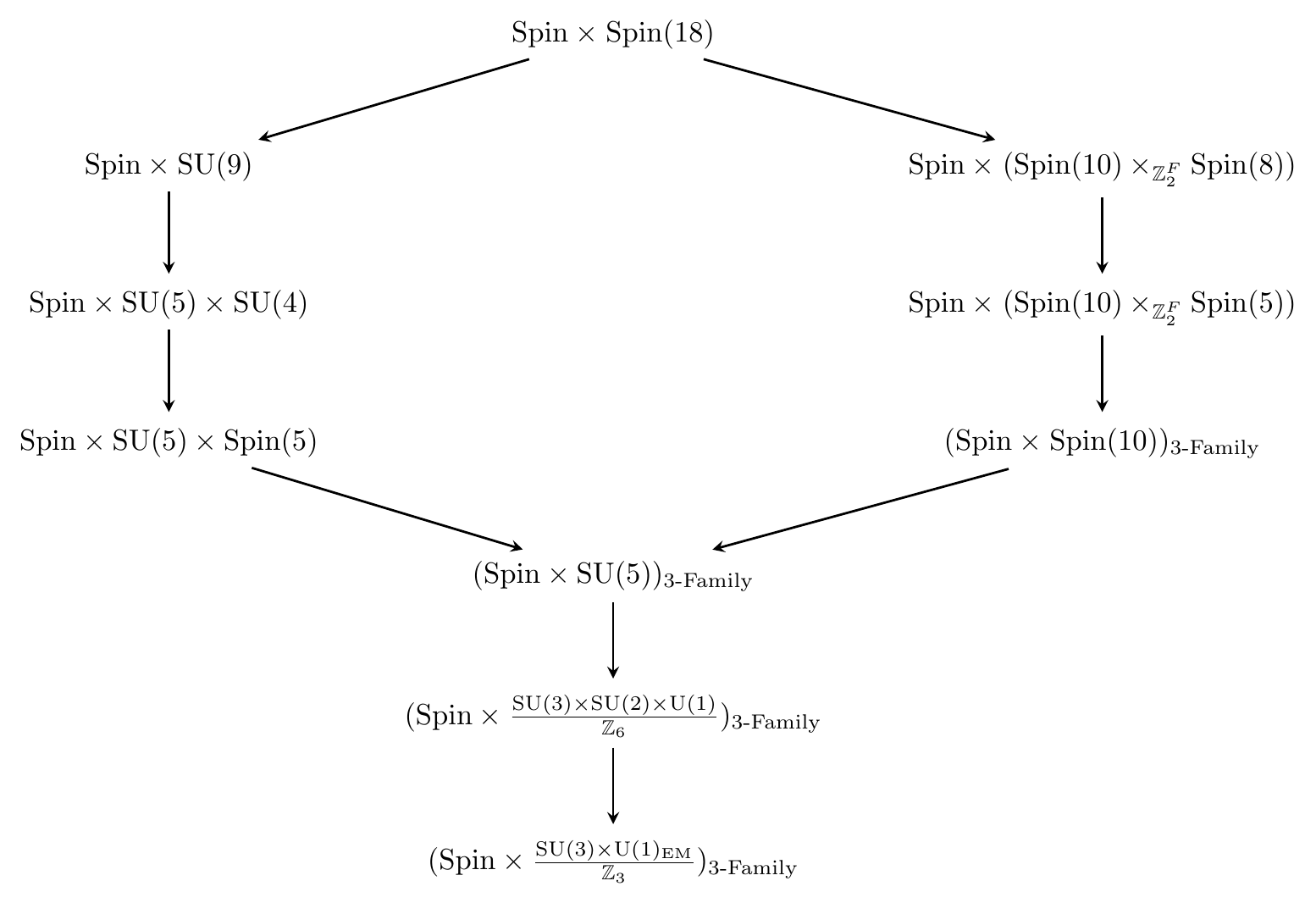}
\caption{The full spacetime-internal symmetry
$G={\frac{{G_{\text{spacetime} }} \ltimes  {{G}_{\text{internal}} }}{{N_{\text{shared}}}}}$ (the precise global symmetry before gauging the ${{G}_{\text{internal}} }$)
for the hierarchy starting from the $so(18)$ GUT with ${{\Spin \times_{} \Spin(18)}}$, which can be placed on \emph{spin manifolds}.
Also we follow the notations/explanations of \Fig{table:sym-web-1}'s caption.
 We have computed the cobordism group 
 $\TP_d(G)$
 of these spacetime-internal symmetry group $G$ in \Ref{WanWangv2}. 
}
 \label{table:sym-web-2}
\end{figure*}

We first write down the precise symmetry group including the 
{Euclidean/Lorentz spacetime symmetry group ${G_{\text{spacetime} }}$ and the internal symmetry group ${{G}_{\text{internal}} }$} in a unified setting as:
\bea
G={\frac{{G_{\text{spacetime} }} \ltimes  {{G}_{\text{internal}} }}{{N_{\text{shared}}}}}.
\eea
Then we discuss the $G$ symmetry embedding in the web in 
\Fig{table:sym-web-1} and \Fig{table:sym-web-2}.
%

In 
\Fig{table:sym-web-1} and \Fig{table:sym-web-2},
we start from the $so(18)$ GUT (with the Spin(18) internal symmetry group or gauge group).
These are two versions of $so(18)$ GUT: One can be placed on manifolds without spin structures (\emph{non-spin manifolds}, where the second Stiefel-Whitney of spacetime
tangent bundle $TM$ to be $w_2(TM) \neq 0$ is nontrivial),
and the other can be placed on manifolds with spin structures (\emph{spin manifolds}, where the second Stiefel-Whitney $w_2(TM) = 0$ is trivial).

In \Fig{table:sym-web-1}, 
the ${{\Spin \times_{\Z_2^F} \Spin(18)}}$ 
 implies\footnote{Here
${\frac{{\Spin} \ltimes  {{G}_{\text{internal}} }}{N_{\text{shared}}}}$
means ${G_{\text{spacetime} }}=\Spin \equiv \Spin(D)$ where $D$ is the relevant spacetime dimensions of manifold $M^D$.}  
 that this $so(18)$ GUT can be placed on \emph{non-spin manifolds} (which also includes \emph{spin manifolds}),
by setting the second Stiefel-Whitney of spacetime
tangent bundle $w_2(TM)=w_2(V_{\SO(18)})$ to be the same as the gauge bundle
from the associated vector bundle 
${\SO(18)}=\frac{\Spin(18)}{\Z_2^F}$.\footnote{More generally,
for ${{\Spin \times_{\Z_2^F} \Spin(N)}}$ structure, say for $N\geq 3$,
the restriction $w_2(TM)=w_2(V_{\SO(N)})$ 
also means that all the odd power of the spinor representation matter field of the internal symmetry $\Spin(N)$
must be associated with the Lorentz/Euclidean spacetime spinor (spacetime fermions) 
as it has a nontrivial $w_2(TM)$ indicating the spacetime $2\pi$-rotation of the matter (fermion) gains a $(-1)$-sign
on its state vector (known as  the fermion self or spin statistics) which must be cancelled by its nontrivial $w_2(V_{\SO(N)})$. 
}
When $w_2(TM)=w_2(V_{\SO(18)})=0$, the theory is on \emph{spin manifolds}.
Similarly, the  ${{\Spin \times_{\Z_2^F} \Spin(10)}}$ implies that this $so(10)$ GUT can be placed on \emph{non-spin manifolds} ---
The $so(10)$ GUT without spin structure is studied in \cite{WangWen2018cai1809.11171,  WangWenWitten2018qoy1810.00844}.

In \Fig{table:sym-web-2}, 
the ${{\Spin \times_{} \Spin(18)}}$ implies that this $so(18)$ GUT can be placed only on \emph{spin manifolds},
limiting to those manifolds with the second Stiefel-Whitney of spacetime
tangent bundle $w_2(TM)=0$ to be zero.

In the present work, we mostly focus on \Fig{table:sym-web-1} starting from ${{\Spin \times_{\Z_2^F} \Spin(18)}}$, since
${{\Spin \times_{\Z_2^F} \Spin(18)}}$ is more general in the following aspects:\footnote{However, the ${{\Spin \times_{} \Spin(18)}}$ 
and its embedding hierarchy in \Fig{table:sym-web-2} is also interesting by its own.
We leave the embedding and breaking pattern on 
\Fig{table:sym-web-2} in a companion work  \cite{WanWangv2}.
}
\begin{enumerate}[leftmargin=4.mm, label=\textcolor{blue}{(\arabic*)}., ref={(\arabic*)}]
\setcounter{enumi}{0}

\item \label{2.1} 
The ${{\Spin \times_{\Z_2^F} \Spin(18)}}$ structure contains \emph{non-spin manifolds} which can be more general. 
Its cobordism theory may detect more exotic anomalies and constraints. 
For example, 
\begin{itemize}
\item The ${{\Spin \times_{} \Spin(3)}}={{\Spin \times_{} \SU(2)}}$ detects only the $\Z_2$ class of
4d familiar SU(2) Witten anomaly \cite{Witten1982fp} by a 5d cobordism invariant on spin manifolds (see also Appendix \ref{subsec:SU(2)anomaly}).
 The co/bordism groups are  \cite{WanWang2018bns1812.11967}
\bea
\Omega_5^{{\Spin \times_{} \SU(2)}} = \Z_2, \quad  \TP_5({{\Spin \times_{} \SU(2)}})= \Z_2
\eea
and its 5d cobordism invariant is \cite{WanWang2018bns1812.11967, WW2019fxh1910.14668}:
\bea
\exp( \ii \pi \int  c_2(V_{\SU(2)}) \tilde \eta)
\eea
where the $c_2(V_{\SU(2)})$ is the second Chern class of SU(2) gauge bundle and
$\tilde \eta$ is the 1d eta invariant or a mod 2 index of 1d Dirac operator,
as the generator of 1d spin bordism group $\Omega_1^\Spin=\Z_2$.

\item The ${{\Spin \times_{\Z_2^F} \Spin(3)}}={{\Spin \times_{\Z_2^F} \SU(2)}}$
detects \emph{not merely} a $\Z_2$ class of
 the familiar 4d SU(2) Witten anomaly \cite{Witten1982fp}, \emph{but also} another new $\Z_2$ class of
the 4d new SU(2) anomaly  \cite{WangWenWitten2018qoy1810.00844} 
 captured 
 by
 the co/bordism group (details in Appendix \ref{subsec:SU(2)anomaly}):
\bea \label{eq:cobordismSpinZ2SU2}
\Omega_5^{{\Spin \times_{\Z_2^F} \SU(2)}} = \Z_2^2, \quad \TP_5({{\Spin \times_{\Z_2^F} \SU(2)}})= \Z_2^2,
\eea
 on non-spin manifolds $M^5$ via 
 another mod 2 class 5d cobordism invariant\footnote{{\bf Notations}: We
denote
the Stiefel-Whitney class of the spacetime tangent bundle $TM$ of spacetime manifold $M$
as $w_j \equiv w_j(TM)$; 
if we do not specify $w_j$ with which bundle, then we implicitly mean $TM$.
We
denote $w_j(V_{\SO(n)}) \equiv w_j({\SO(n)}) $ is the $j$-{th}-Stiefel-Whitney class for the associated vector bundle of an ${\SO(n)}$
gauge bundle.

Throughout the article, {we use the standard notation for characteristic classes \cite{milnor1974characteristic}: $w_i$ for the Stiefel-Whitney class, $c_i$ for the Chern class, $p_i$ for the Pontryagin class, and 
$e_n$ for the Euler class. Note that the Euler class only appears in the total dimension of the vector bundle. 
We may also use the notation $w_i(G)$, $c_i(G)$, $p_i(G)$, and $e_n(G)$ to denote the characteristic classes of the associated vector bundle of the principal $G$ bundle 
(usually denoted as $w_i(V_{G})$, $c_i(V_{G})$, $p_i(V_{G})$, and $e_n(V_{G})$).
\label{footnote:notations}
}} 
 \bea
\exp( \ii \pi \int  w_2(TM)w_3(TM))= \exp( \ii \pi \int  w_2(V_{\SO(3)})w_3(V_{\SO(3)})).
 \eea

\end{itemize}

\item \label{2.2} 
The ${{\Spin \times_{\Z_2^F} \Spin(18)}}$ and ${{\Spin \times_{\Z_2^F} \Spin(10)}}$ structures
may be useful for the lattice regularization from the high-energy ultraviolet (UV) based on local bosons \cite{WangWen2018cai1809.11171} (without the requirement of any local fermions).  Moreover, upon (global symmetry or gauge) group breaking,
when the $2 \pi$ rotation sits at the $\Z_2$ normal subgroup of the internal symmetry Spin group is absent,
we can generate $$
\text{\emph{dynamical spin structures}  \cite{WangWenWitten2018qoy1810.00844} with \emph{emergent fermions}  \cite{WangWen2018cai1809.11171,WangWenWitten2018qoy1810.00844}}.
$$
For example,
${{\Spin \times_{\Z_2^F} \Spin(18)}} \to \Spin\times \SU(9)$ 
and 
${{\Spin \times_{\Z_2^F} \Spin(10)}} \to \Spin\times \SU(5)$ 
generate {dynamical spin structures} \cite{WangWenWitten2018qoy1810.00844}.

\end{enumerate}

\subsection{Decomposition: Lie algebras to Lie groups, and representation theory}
For the convenience of checking the anomaly matching from the cobordism theory,
let us set up some representation (abbreviated as Rep) theory notations for GUT and SM.

\subsubsection{Representation of spacetime symmetry groups}
\label{sec:Representationofspacetimesymmetrygroup}
{\bf Fermions as Lorentz or Euclidean spinors in the spacetime:} 
Fermions are the spinor fields,
as the sections of the spinor bundles of the spacetime manifold $M$. 
The left-handed (chiral) Weyl spinor $\Psi_L$
is a doublet ${\bf 2}$ or the so-called spin-1/2 representation (Rep.) of spacetime symmetry group $G_{\text{spacetime}}$ 
(Minkowski/Lorentz $\Spin(3,1)$ in 3+1d or Euclidean $\Spin(4)$ in 4d), denoted as
\bea
(3,1)\rm{d} \quad \Psi_L &\sim&  {\bf 2}_L \text{ of } {\Spin(3,1)}=\rm{SL}(2,\C),  \text{ complex Rep}. \label{eq:3+1d}\\
{(4,0)\rm{d} \quad  \Psi_L}&\sim& { {\bf 2}_L \text{ of } \Spin(4)=\SU(2)_L \times \SU(2)_R, \text{ pseudoreal Rep}}.  \label{eq:4d}
\eea
The Spin 
group, $\Spin(3,1)$ or $\Spin(4)$, is a
double-cover or universal-cover of the Lorentz group $\SO(3,1)_+$ or Euclidean rotation $\SO(4)$, extended by the fermion parity $\Z_2^F$ which acts on fermion
as $(-1)^F: \Psi \to -\Psi$. 

We will also consider the 5d co/bordism invariants as 5d invertible TQFTs, which can be obtained from integrating out some massive fermions \cite{Witten2015aba1508.04715} 
in  4+1d Lorentz
or in 5d Euclidean spacetime:
\bea
{(4,1)\rm{d} \quad  \Psi} \; \; &\sim& { {\bf 4}  \; \,  \text{ of } \Spin(4,1)=\Sp(1,1), \text{ pseudoreal Rep}}.   \label{eq:4+1d}\\
{(5,0)\rm{d}} \quad  \Psi \; \; &\sim& {\bf 4} \; \,   \text{ of } {\Spin(5)}=\U\Sp(4)=\Sp(2), \text{ pseudoreal Rep}. \label{eq:5d}  
\eea
In the following of this article, we shall use 
{the 5d Euclidean signature's invertible TQFTs 
to capture the anomalies of 3+1d Lorentz signature's quantum field theory (QFT).}
We use the fact \cite{Freed2016} that 
the unitarity of Lorentz QFT is analogous to the reflection positivity of Euclidean QFT.
Therefore, we see a relation that:\\[-8mm]
\bea \label{eq:EuclideanLorentz}
&&\text{the $d$d invertible TQFT  in Euclidean signature with the reflection positivity}\cr
&&\text{$\Rightarrow$  captures the {anomaly of $(d-1)$d Euclidean QFT with the reflection positivity}}\cr
&&\text{$\Rightarrow$ captures the {anomaly of $(d-2,1)$d Lorentz QFT} with the unitary.} 
\eea
If we take the $d=5$, we obtain the relation used in this work:
{the 5d invertible TQFT (from 5d co/bordism invariants) in Euclidean signature with the reflection positivity}
classifies the {(invertible) anomaly of $(3,1)$d Lorentz QFT} with the unitary.
Throughout this article, we may simply denote
5d for Euclidean signature, and 4d = 3+1d for Lorentz signature.
When we refer to a spacetime spinor in 4d = 3+1d, in general we mean the
spinor in \Eq{eq:3+1d} for Lorentz signature;
when we refer to a spacetime spinor in 5d = 4+1d, in general we take the
spinor in \Eq{eq:5d} for Euclidean signature, because we intend to use the relation \Eq{eq:EuclideanLorentz}.

Here the spacetime symmetry is commonly denoted as a Spin group omitting the spacetime dimensions, either for the Lorentz signature Spin($d,1$)
or the Euclidean signature Spin($d+1$).
The readers should still recall that we have implicitly made the representation of fermions as spacetime spinors in the spacetime symmetry group as 
what we have already done in \Sec{sec:Representationofspacetimesymmetrygroup}.
In the following, when we discuss fermions, 
we mainly focus on their {representation of internal/gauge symmetry group}.

\subsubsection{Representation of internal/gauge symmetry groups}

\begin{enumerate}[leftmargin=-4.mm, label=\textcolor{blue}{[\Roman*]}., ref={[\Roman*]}]

\item {\bf Standard Model SM$_q$ with $q=1,2,3,6$:} 
The local gauge structure of Standard Model is  the Lie algebra $su(3)  \times su(2) \times u(1)$.
This means that the Lie algebra valued 1-form gauge fields take values in the Lie algebra generators of $su(3)  \times su(2) \times u(1)$.
There are $8 + 3 + 1 = 12$ Lie algebra generators.
The 1-form gauge fields are the 1-connections of the principals $G_{\text{internal}}$-bundles.

\item {\bf Standard Model SM$_q$ with fermions:} 
In the first generation of SM, the matter fields as left-handed ($L$) or right-handed ($R$) Weyl spinors contain:
\begin{itemize}
\item
The left-handed up and down quarks ($u$ and $d$) form a doublet $\begin{pmatrix}
u\\
d
\end{pmatrix}_L$ in ${\bf 2}$ for the SU(2)$_{\text{weak}}$, and they are in ${\bf 3}$ for the SU(3)$_{\text{strong}}$.
\item
The right-handed up and down quarks, each forms a singlet, $u_R$ and $d_R$, in ${\bf 1}$ for the SU(2)$_{\text{weak}}$. They are in ${\bf 3}$ for the SU(3)$_{\text{strong}}$.
\item
The left-handed electron and neutrino form a doublet $\begin{pmatrix}
\nu_e\\
e
\end{pmatrix}_L$ in ${\bf 2}$ for the SU(2)$_{\text{weak}}$, and they are in ${\bf 1}$ for the SU(3)$_{\text{strong}}$.
\item
The right-handed electron forms a singlet $e_R$ in ${\bf 1}$ for the SU(2)$_{\text{weak}}$, 
and it is in ${\bf 1}$ for the SU(3)$_{\text{strong}}$.
\end{itemize}
There are two more generations of quarks: 
charm and strange quarks ($c$ and $s$),
and top and bottom quarks ($t$ and $b$).
There are also two more generations of leptons:
muon and its neutrino ($\mu$ and $\nu_\mu$), 
and 
tauon and its neutrino ($\tau$ and $\nu_\tau$). 
So there are three generations (i.e., families) of quarks and leptons:
\bea
\Bigg(
\begin{pmatrix}
u\\
d
\end{pmatrix}_L \times {\bf 3}_{\text{color}}, \quad\quad \quad u_R \times {\bf 3}_{\text{color}}, \quad\quad\quad\quad d_R \times {\bf 3}_{\text{color}}, 
\quad\quad\quad
\begin{pmatrix}
\nu_e\\
e
\end{pmatrix}_L,\quad\quad  \quad e_R 
\quad  \Bigg),\nn\\
\Bigg(
\begin{pmatrix}
c\\
s
\end{pmatrix}_L \times {\bf 3}_{\text{color}}, \quad\quad \quad c_R \times {\bf 3}_{\text{color}}, \quad\quad\quad\quad s_R \times {\bf 3}_{\text{color}}, 
\quad\quad\quad
\begin{pmatrix}
\nu_\mu\\
\mu
\end{pmatrix}_L,\quad\quad  \quad \mu_R 
\quad  \Bigg),\nn\\
\Bigg(
\begin{pmatrix}
t\\
b
\end{pmatrix}_L \times {\bf 3}_{\text{color}}, \quad\quad \quad t_R \times {\bf 3}_{\text{color}}, \quad\quad\quad\quad b_R \times {\bf 3}_{\text{color}}, 
\quad\quad\quad
\begin{pmatrix}
\nu_\tau\\
\tau
\end{pmatrix}_L,\quad\quad  \quad \tau_R 
\quad  \Bigg).
\eea
In short, for all of them as three generations, we can denote them as:
\bea\label{eq:3-gene-rep-1}
\Bigg(
\begin{pmatrix}
u\\
d
\end{pmatrix}_L \times {\bf 3}_{\text{color}}, \quad\quad \quad u_R \times {\bf 3}_{\text{color}}, \quad\quad\quad\quad d_R \times {\bf 3}_{\text{color}}, 
\quad\quad\quad
\begin{pmatrix}
\nu_e\\
e
\end{pmatrix}_L,\quad\quad  \quad e_R 
\quad  \Bigg)\times \text{3 generations}. \quad
\eea
We can also denote this \Eq{eq:3-gene-rep-1} as the left-handed spacetime Weyl spinor representation as:
\bea\label{eq:3-gene-rep-2}
\Bigg(
q_L \times {\bf 3}_{\text{color}}, \quad\quad \quad \bar{u}_R \times {\bf 3}_{\text{color}}, \quad\quad\quad\quad \bar{d}_R \times {\bf 3}_{\text{color}}, 
\quad\quad\quad
l_L,\quad\quad  \quad \bar{e}_R 
\quad  \Bigg)\times \text{3 generations} \quad\\
\label{eq:3-gene-rep-3}
\equiv \Bigg(
q_L \times {\bf 3}_{\text{color}}, \quad\quad \quad u^c \times {\bf 3}_{\text{color}}, \quad\quad\quad\quad d^c \times {\bf 3}_{\text{color}}, 
\quad\quad\quad
l_L,\quad\quad  \quad e^c 
\quad  \Bigg)\times \text{3 generations}. 
\eea
In fact, all the following \emph{four} kinds of
$
G_{\text{SM}_q}=\frac{\SU(3)\times \SU(2)\times \U(1)}{\Z_q}
$
with $q=1,2,3,6$ are compatible with the above representations of fermion fields.
(See an excellent exposition in a recent work by Tong \cite{Tong2017oea1705.01853}.)
These $15  \times 3$ Weyl spinors can be written in the following more succinct forms of representations for any of the internal symmetry group $G_{\text{internal}}$ with $q=1,2,3,6$:
\bea \label{eq:rep-3generations}
\Bigg( ({\bf 3},{\bf 2}, 1/6)_L,({\bf 3},{\bf 1}, 2/3)_R,({\bf 3},{\bf 1},-1/3)_R,({\bf 1},{\bf 2},-1/2)_L,({\bf 1},{\bf 1},-1)_R  \Bigg)\times \text{3 generations} \nn\\
\Rightarrow
\Bigg( ({\bf 3},{\bf 2}, 1/6)_L,(\overline{\bf 3},{\bf 1}, -2/3)_L,(\overline{\bf 3},{\bf 1},1/3)_L, ({\bf 1},{\bf 2},-1/2)_L,({\bf 1},{\bf 1},1)_L  \Bigg)\times \text{3 generations}.
\eea
Each of the triplet given above is listed by their  representations:
\bea \label{eq:rep}
\text{(SU(3) representation, SU(2) representation, hypercharge $Y$)}.
\eea

\item {\bf $su(5)$ GUT with $su(5)$ Lie algebra and SU(5) Lie group:}  If we include the 
$3 \times 2 + 3 + 3 + 2 +1 =15$ left-handed Weyl spinors from one single generation in \Eq{eq:rep-3generations}, 
we can combine them as a multiplet of $\overline{\bf 5}$ and {\bf 10} left-handed Weyl spinors of SU(5).
Recall that the SM matter field contents can be embedded into SU(5) as follows, in terms of their representations:
\bea
\begin{array}{rcccccc}
(\overline{\bf 3},{\bf 1},1/3)_L \oplus ({\bf 1},{\bf 2},-1/2)_L &\sim & \bar{d}_R \oplus l_L &\sim&  {d}^c \oplus l &\sim& \overline{\bf 5} \text{ of } \SU(5).\\
  ({\bf 3},{\bf 2}, 1/6)_L \oplus (\overline{\bf 3},{\bf 1}, -2/3)_L \oplus ({\bf 1},{\bf 1},1)_L  &\sim & q_L \oplus \bar{u}_R \oplus \bar{e}_R &\sim&
    q\oplus {u}^c \oplus {e}^c
   &\sim& {\bf 10} \text{ of } \SU(5).
\end{array}
\eea
More explicitly, in terms of the anti-fundamental ${\bf 5}$ and the anti-symmetric matrix representation ${\bf 10}$:
\bea
 \overline{\bf 5} \text{ of } \SU(5)
  &=&\begin{pmatrix}
\psi_{\al}\\
\psi_i
\end{pmatrix} 
 =\begin{pmatrix}
\bar{d}_{R,\al}\\
\begin{pmatrix}
\nu_e\\
e
\end{pmatrix}_L
\end{pmatrix} 
=
\begin{pmatrix}
\bar{d}_{R,r}\\
\bar{d}_{R,g}\\
\bar{d}_{R,b}\\
\nu_{e,L}\\
e_L
\end{pmatrix}. \\
 {\bf 10} \text{ of } \SU(5)
  &=&
\begin{pmatrix}
0 & \psi^{\al \bt} & - \psi^{\al \bt} &  \psi^{\al i} &  \psi^{\al i} \\
- \psi^{\al \bt} & 0 &  \psi^{\al \bt} &  \psi^{\al i} & \psi^{\al i}  \\
 \psi^{\al \bt} & -  \psi^{\al \bt}& 0 & \psi^{\al i}  & \psi^{\al i} \\
-  \psi^{\al i}  & -  \psi^{\al i}  & - \psi^{\al i}   &0 &  \psi^{ ij }\\
-  \psi^{\al i}  & -  \psi^{\al i}  & - \psi^{\al i}  &   \psi^{ ij }& 0
\end{pmatrix} 
=
\begin{pmatrix}
0 & \bar{u} & - \bar{u} &   d&  u \\
- \bar{u} & 0 & \bar{u} &  d & u  \\
 \bar{u} & -\bar{u}& 0 & d  & u \\
-  d & -  d  & - d   &0 & \bar{e}\\
-  u  & - u  & - u  &   - \bar{e}& 0
\end{pmatrix}. 
\eea
Hence these are matter field representations of the $su(5)$ GUT with an SU(5) gauge group.

\item {\bf Break SU(5) to $\frac{\SU(3) \times   \SU(2) \times \U(1)}{\Z_6}$ and to $\frac{\SU(3) \times \U(1)_{\text{EM}}}{\Z_3}$:}
\label{BreakSU(5)}
Other than the electroweak Higgs $\phi_H$, we also need to introduce a different GUT Higgs field $\phi_{GG}$ 
to break down SU(5) to $\frac{\SU(3) \times   \SU(2) \times \U(1)}{\Z_6}$.
The $\phi_{GG}$ can be 
$\phi_{su(5)_{\bf 24}}={\bf 24}$ in the adjoint representation of SU(5) as 
\begin{multline}
\phi_{su(5)_{\bf 24}} \sim {\bf 24}  \text{ of } \SU(5) \\
=({\bf 8},{\bf 1}, {Y}=0) \oplus ({\bf 1},{\bf 3}, {Y}=0) \oplus  ({\bf 1},{\bf 1}, {Y}=0) \oplus({\bf 3},{\bf 2}, {Y}=-\frac{5}{6})
 \oplus(\bar{\bf 3},{\bf 2}, {Y}=\frac{5}{6}) \text{ of } \frac{\SU(3) \times   \SU(2) \times \U(1)}{\Z_6} .
 \end{multline}
Moreover, to give fermion mass by Yukawa-Higgs Dirac term via a Higgs mechanism,
we can introduce additional new Higgs fields for the electroweak Higgs $\phi_H$ which further breaks 
SU(5) to $\frac{\SU(3) \times  \U(1)_{\text{EM}}}{\Z_3}$:
\bea
\phi_{su(5)_{{\bf 5}}} \sim {{\bf 5}} \text{ of } \SU(5) , \quad\quad\quad \text{ also } \quad\quad\quad  \phi_{su(5)_{{\bf 45}}} \sim {{\bf 45}} \text{ of } \SU(5) .
\eea
In short, for $su(5)$ GUT, 
the GUT Higgs field is in the adjoint ${\bf 24}$,
while the electroweak Higgs field is in ${\bf 5}$, and also another ${\bf 45}$.\footnote{To choose 
an electroweak Higgs $\phi_H$,
we check that Yukawa-Higgs-Dirac term
$\phi_H \psi_R^\dagger \psi_L +h.c.$ or 
$\phi_H^* \psi_R^\dagger \psi_L +h.c.$ gives the trivial representation in SU(5).
We know the forms of electron mass term $\bar{e}_R e_L \sim {\bf 10 }\otimes \overline{\bf 5}$,
the up quark mass term $\bar{u}_R u_L \sim {\bf 10 }\otimes {\bf 10}$,
and the down mass term $\bar{d}_R d_L \sim \overline{\bf 5} \otimes {\bf 10 }$. 
Together with the fact ${\bf 10 }\otimes \overline{\bf 5}= {\bf 5} \oplus {\bf 45}$
and ${\bf 10 }\otimes {\bf 10}=  \overline{\bf 5} \oplus  \overline{\bf 45}  \oplus  \overline{\bf 50}$
implies that 
$\phi_H^* \bar{e}_R e_L +h.c.$,
$\phi_H \bar{u}_R u_L +h.c.$ and
$\phi_H^*\bar{d}_R d_L +h.c.$
give the correct Yukawa-Higgs-Dirac terms choosen
from either $\phi_{su(5)_{{\bf 5}}} \sim {{\bf 5}}$
(by using the fact $\overline{\bf 5 }\otimes {\bf 5}={\bf 1} \oplus {\bf 24}$ contains ${\bf 1}$)
or 
$\phi_{su(5)_{{\bf 45}}} \sim {{\bf 45}}$ (by using the fact ${\bf 45 }\otimes \overline{\bf 45}$ contains ${\bf 1}$).
}

We can add a right-handed neutrino ${\nu}_R$ (or ${\nu}^c$ known as the sterile neutrino which does not interact with any SU(5) gauge bosons) into the SU(5) with a trivial representation:
\bea \label{eq:sterileneutrino16}
\begin{array}{rcccccc}
  ({\bf 1},{\bf 1},0)_L  &\sim &  \bar{\nu}_R &\sim &  {\nu}^c  &\sim& {\bf 1} \text{ of } \SU(5),
\end{array}
\eea

\item {\bf $so(10)$ GUT with $so(10)$ Lie algebra and Spin(10) Lie group:} If we include the $3 \times 2 + 3 + 3 + 2 +1 =15$ left-handed Weyl spinors from one single generation, 
and also a right-handed neutrino \Eq{eq:sterileneutrino16}, we can combine them as a multiplet of 16 left-handed Weyl spinors:
\bea
\quad  \Psi_L \sim {\bf 16}^+ \text{ of } \Spin(10),
\eea
which sits at the 16-dimensional irreducible spinor representation of Spin(10).
The two irreducible spinor representations together
\bea
{\bf 16}^+ \oplus {\bf 16}^- ={\bf 32}
\eea
 form a ${\bf 32}$-dimensional reducible spinor representation of Spin(10).
Namely, 
we must regard the $so(10)$ GUT with a Spin(10) gauge group.
Based on the Nielsen-Ninomiya \emph{fermion doubling} of the free fermion theory, 
the ${\bf 16}^+ \oplus {\bf 16}^-$ can be regarded as the realization of
$$\text{ the \emph{chiral matter} ${\bf 16}^+$}
\text{  and } 
\text{ the \emph{mirror matter} ${\bf 16}^-$}$$ 
(anti-chiral with complex conjugated representation).
Based on a generalization of gapping mirror fermion \cite{Eichten1985ftPreskill1986} by suitable nonperturbative interactions 
(see an overview from \cite{Kaplan1992bt9206013, Poppitz2010atShang1003.5896, Wang2013ytaJW1307.7480, Kikukawa2017gvk1710.11101, Wang2018ugfJW1807.05998}),
References \cite{Wen2013ppa1305.1045, BenTov2015graZee1505.04312, Kikukawa2017ngf1710.11618, WangWen2018cai1809.11171}
suggested that the {mirror matter} ${\bf 16}^-$
can be fully gapped without breaking the Spin(10) group.
It is shown in \cite{WangWen2018cai1809.11171} that the gapping ${\bf 16}^-$ without breaking 
 Spin(10) is consistent with the classification of all invertible local and global anomalies from the cobordism classification \cite{WangWen2018cai1809.11171, WanWang2018bns1812.11967}.
 (We will explain more details in \Sec{sec:newSU2anom}.)
It is also shown in \cite{WangWen2018cai1809.11171}  that the gapping ${\bf 16}^-$ without breaking 
 Spin(10) is consistent with the perspective of Seiberg's deformation class of quantum field theories (QFT) \cite{NSeiberg-Strings-2019-talk}.

\item {\bf Break Spin(10) to SU(5):} \label{BreakSpin(10)toSU(5)}
To break the Lie group ${\Spin(10)\to \SU(5)}$ (which is a stronger statement than the breaking of Lie algebra ${so(10)\to su(5)}$),
we can implement 
 the following Higgs fields (see \Fig{fig:energy-hierarchy-lie-algebra-1} and  \Fig{fig:energy-hierarchy-lie-algebra-2}):
\begin{enumerate}[leftmargin=4.mm, label=\textcolor{blue}{(\roman*)}., ref={(\roman*)}]
\item
\bea
\phi_{so(10)_{\bf 16}}\sim {\bf 16} \text{ of } \Spin(10), 
\eea which ${\bf 16}$ is also in ${\bf 16}^+$. Conventionally, an old wisdom said that
$\phi_{so(10)_{\bf 16}}$ requires an additional (17th) Weyl fermion \cite{georgi2018liebook} (in a trivial representation ${\bf 1}$ of Spin(10))  to pair with the 
16th Weyl fermion in ${\bf 16}$ and $\phi_{so(10)_{\bf 16}} \sim {\bf 16}$ to give it a mass via Higgs mechanism.\footnote{This is simply based on
the Yukawa-Higgs-Dirac term
$\phi_{so(10)_{\bf 16}}^*  \psi_R^\dagger \psi_L
\sim \overline{\bf 16}\otimes  {\bf 1}  \otimes {\bf 16} $
and the fact 
$\overline{\bf 16} \otimes  {\bf 1} \otimes {\bf 16} =\overline{\bf 16} \otimes {\bf 16}= {\bf 1}\oplus{\bf 45}\oplus{\bf 210}$
contains the trivial ${\bf 1}$ allowed for Yukawa-Higgs-Dirac term.
But the 17th Weyl fermion in ${\bf 1}$ has an disadvantage to mismatch the $\Z_{16}$ anomaly if we wish to maintain the $\Z_4$ discrete subgroup of
$X\equiv
5({ \mathbf{B}-  \mathbf{L}})-4Y$  \cite{JW2006.16996}.
So \Ref{JW2006.16996} proposes a new way out using a new 4d TQFT (without adding any 17th Weyl fermion) which can match the $\Z_{16}$ anomaly,
while the vev of $\phi_{so(10)_{\bf 16}} \sim {\bf 16}$ can still break ${\Spin(10)\to \SU(5)}$.
} However, as we learned, \Ref{JW2006.16996} suggested
the $\Z_{16}$ anomaly  cannot be matched by all 17 Weyl fermions.

A new more promising proposal from \Ref{JW2006.16996}
is that we can still use the vacuum expectation value (vev) of $\phi_{so(10)_{\bf 16}}={\bf 16}$
to break  ${\Spin(10)\to \SU(5)}$, but 
 we introduce a new 4d TQFT sector (but keep only 16 Weyl fermions in ${\bf 16}^+$, without
introducing the 17th Weyl fermion in ${\bf 1}$) with a huge energy gap 
$
\Delta_{\mathrm{TQFT}}
$
of GUT scale shown in \Fig{fig:energy-hierarchy-lie-algebra-1} and \Fig{fig:energy-hierarchy-lie-algebra-2}. 
The $\Delta_{\mathrm{TQFT}}$ means the energy gap between the ground state $| \Psi_{g.s.}\rangle$ of its topological order
and the first excited state(s) ${| \Psi_{\text{1st excited}}\rangle}$ of fractionalized excitations (anyonic strings with fractional braiding statistics of 4d TQFT  \cite{WangLevin1403.7437, Jiang1404.1062, Wang1404.7854, 1602.05951WWY,Putrov2016qdo1612.09298PWY,Wang2019diz1901.11537}).
So that the energy difference between ${| \Psi_{\text{1st excited}}\rangle}$ and $| \Psi_{g.s.}\rangle$ is the TQFT/topological order gap defined as:
\bea\label{eq:DeltaTQFT}
\Delta_{\mathrm{TQFT}} \equiv E_{| \Psi_{\text{1st excited}}\rangle} - E_{| \Psi_{g.s.}\rangle}.
\eea
The $\Z_{16}$ anomaly is compensated by the \emph{4d TQFT} (replacing the 16th Weyl fermion of the sterile right-handed neutrino)
plus the remaining \emph{15 Weyl fermions} of $su(5)$ GUT. 

\item
\bea
\phi_{so(10)_{\bf 126}}\sim {\bf 126} \text{ of } \Spin(10). 
\eea
Alternatively, we can introduce the Majorana mass to the 16th Weyl fermion in ${\bf 16}^+$
by the Higgs field $\phi_{so(10)_{\bf 126}}$ via Yukawa-Higgs mechanism with a Yukawa-Higgs-Majorana mass term.\footnote{Recall a Majorana mass term 
$
\psi \psi \sim {\bf 16} \otimes {\bf 16} 
$
and the fact
${\bf 16} \otimes {\bf 16} =  {\bf 10}\oplus {\bf 120}\oplus \overline{\bf 126}$ where ${\bf 126}$ is complex, self-dual, total anti-symmetric 5-index tensor in Spin(10).
Then a Yukawa-Higgs-Majorana term $\phi_{so(10)_{\bf 126}} \psi \psi \sim  {\bf 126} \otimes {\bf 16} \otimes {\bf 16}$ can contain ${\bf 126} \otimes \overline{\bf 126}$, which can also contain
the desired trivial ${\bf 1}$.
(Note that ${\bf 126} \otimes \overline{\bf 126} = {\bf 1}\oplus {\bf 45} \oplus {\bf 210}\oplus {\bf 770}\oplus {\bf 5940}\oplus {\bf 8910}$.)}

\end{enumerate}

\item {\bf Break Spin(10) to SU(5) then to $\frac{\SU(3) \times   \SU(2) \times \U(1)}{\Z_6}$ and to $\frac{\SU(3) \times   \U(1)_{\text{EM}}}{\Z_3}$:}
We have explained breaking Spin(10) to SU(5) in \ref{BreakSpin(10)toSU(5)}
and breaking SU(5) to $\frac{\SU(3) \times   \SU(2) \times \U(1)}{\Z_6}$ in \ref{BreakSU(5)}.
\begin{enumerate}[leftmargin=4.mm, label=\textcolor{blue}{(\roman*)}., ref={(\roman*)}]
\item
The Spin(10) can be further broken down to $\frac{\SU(3) \times   \SU(2) \times \U(1)}{\Z_6}$ by adding
a new Higgs condensate in addition to the Higgs $\phi_{so(10)_{\bf 16}}$ or $\phi_{so(10)_{\bf 126}}$ of \ref{BreakSU(5)}
which already breaks Spin(10) to SU(5). The new Higgs condensate can be 
\bea
\phi_{so(10)_{\bf 45}}\sim {\bf 45} \text{ of } \Spin(10) = {\bf 1} \oplus {\bf 10} \oplus \overline{\bf 10}\oplus {\bf 24} \text{ of } \SU(5), 
\eea
since the branching rule contains $\phi_{su(5)_{{\bf 24}}} \sim {\bf 24}$ of SU(5) that breaks SU(5) to $\frac{\SU(3) \times   \SU(2) \times \U(1)}{\Z_6}$.
Another option is the $\phi_{so(10)_{\bf 54}}\sim {\bf 54} $ that also contains $\phi_{su(5)_{{\bf 24}}} \sim {\bf 24}$ of SU(5): 
\bea
\phi_{so(10)_{\bf 54}}\sim {\bf 54} \text{ of } \Spin(10) = {\bf 15}   \oplus \overline{\bf 15}\oplus {\bf 24} \text{ of } \SU(5). 
\eea
\item The Spin(10) can be further broken down to $\frac{\SU(3) \times   \U(1)_{\text{EM}}}{\Z_3}$ just as 
SU(5) broken to $\frac{\SU(3) \times   \U(1)_{\text{EM}}}{\Z_3}$ by 
$\phi_{su(5)_{{\bf 5}}} \sim {{\bf 5}}$ and/or $\phi_{su(5)_{{\bf 45}}} \sim {{\bf 45}}$  of SU(5).
To do so, we look at the Yukawa-Higgs Dirac mass term
$\phi^* \psi_R^\dagger \psi_L + h.c.$ and see the fermion bilinear ${\bf 16} \otimes {\bf 16}= {\bf 10}\oplus {\bf 120}\oplus \overline{\bf 126}$
see that 
\bea
\phi_{so(10)_{\bf 10}}&\sim& {\bf 10} \text{ of } \Spin(10) = {\bf 5}  \oplus \overline{\bf 5} \text{ of } \SU(5), \\
\phi_{so(10)_{\bf 120}}&\sim& {\bf 120} \text{ of } \Spin(10) = {\bf 5}  \oplus \overline{\bf 5} \oplus {\bf 10}  \oplus \overline{\bf 10}\oplus {\bf 45}  \oplus \overline{\bf 45}  \text{ of } \SU(5), \\
\phi_{so(10)_{\overline{\bf 126}}}&\sim& {\overline{\bf 126}} \text{ of } \Spin(10) = {\bf 5}  \oplus \overline{\bf 5} \text{ of } \SU(5).
\eea
We see all $\phi_{so(10)_{\bf 10}}, \phi_{so(10)_{\bf 120}}$ and $\phi_{so(10)_{\overline{\bf 126}}}$
contain the Higgs $\phi_{su(5)_{{\bf 5}}} \sim {{\bf 5}}$ and/or $\phi_{su(5)_{{\bf 45}}} \sim {{\bf 45}}$ of SU(5) thus they can suit the job, breaking down to
$\frac{\SU(3) \times   \U(1)_{\text{EM}}}{\Z_3}$.
\end{enumerate}
\item {\bf $so(18)$ GUT with $so(18)$ Lie algebra and Spin(18) Lie group:}
The 256 left-handed Weyl spinors (each as a 2-component spacetime spinor) has 16 copies of ${\bf 16}^+ \text{ of } \Spin(10)$
\bea
\quad  \Psi_L \sim {\bf 256}^+ \text{ of } \Spin(18),
\eea
 sits at the 256-dimensional irreducible spinor representation of Spin(18).
The two irreducible spinor representations together
\bea
{\bf 256}^+ \oplus {\bf 256}^- ={\bf 512}
\eea
 form a ${\bf 512}$-dimensional reducible spinor representation of Spin(18).\\[-10mm]
\end{enumerate}
We also know ${\Spin(6)=\SU(4)}$ $\supset$ ${\Spin(5)=\Sp(2)=\U\Sp(4)}$,
thus $\SU(9) \supset \SU(5) \times \SU(4)  \supset \SU(5) \times \Spin(4)$.
With the above information and 
\cite{WW2019fxh1910.14668, JW2006.16996, WanWangv2} and some basics of GUT \cite{Zee2003book, Slansky1981yr1981} in mind,
we obtain the embedding web in \Fig{table:sym-web-1} and \Fig{table:sym-web-2}. 


\section{Energy and Mass Hierarchy: Local and Global Anomaly Constraints}
\label{sec:EnergyandMassHierarchy}
%
With \Ref{JW2006.16996} and \Sec{sec:GaugeGroupHierarchy} in mind, we suggest that 
the anomaly can be matched at different energy or mass scales in different Scenarios, 
see \Fig{fig:energy-hierarchy-lie-algebra-1}
and \Fig{fig:energy-hierarchy-lie-algebra-2}. 
We enlist several Scenarios \ref{SO(18)GUTgap}, \ref{SO(18)SO(10)SO(8)GUTgap}, 
\ref{SO(10)SO(8)SO(6)GUTgap}, \ref{SO(10)SO(8)SO(5)GUTgap},
at different energy scales into subsections.

\subsection{$so(18)$ GUT without mirror fermion doubling and the energy scale $\Delta_{\mathrm{KW}.so(18)}$}
\label{sec:so(18)GUTnofermion doubling}

\begin{enumerate}[leftmargin=-4.mm, label=\textcolor{blue}{[\Roman*]}., ref={[\Roman*]}]
\item {\bf $so(18)$ GUT without mirror fermion doubling and the energy gap scale $\Delta_{\mathrm{KW}.so(18)}$}:\\
\label{SO(18)GUTgap}
In fact, based on  \cite{WangWen2018cai1809.11171},
for $so(18)$ GUT with fermions in the spinor representation of Spin(18)
we can consider a quantum model with UV completion (but without gravity) of the followings:\\[2mm]
{\bf (1)} We can start from a 4d Universe with fermion doublings  $\Psi_L^{\Spin(18)} \sim {\bf 256}^+$ and
$\Psi_R^{\Spin(18)} \sim {\bf 256}^-$. Then we can gap $\Psi_R^{\Spin(18)} \sim {\bf 256}^-$ by nonperturbative interactions  without breaking ${\Spin(18)}$.
The reason is that for the all $G$-anomaly-free theory (as we can check $so(18)$ GUT 
with Spin(18) chiral fermions is fully anomaly free \cite{WangWen2018cai1809.11171}, see also later in \Sec{sec:newSU2anom}),
we can gap the theory as a $G$-symmetric gapped boundary of a trivial $G$-symmetric gapped bulk.
Here the ``trivial'' means a trivial SPT state thus a trivial cobordism class.
The gapped boundary is in fact an gapped interface between a ``trivial $G$-symmetric gapped bulk'' and a ``trivial $G$-symmetric gapped vacuum.''
Thus, the trivial $G$-symmetric gapped bulk can be smoothly crossed over to the trivial vacuum without breaking $G$ symmetry and without closing the energy gap
at the $G$-symmetric gapped interface.\\[2mm]
{\bf (2)} We can start from a 4d Universe without fermion doublings and with only $\Psi_L^{\Spin(18)} \sim {\bf 256}^+$.
This is true since we can simply choose to 
start with a $G$-symmetric gapped boundary on the mirror sector for the bulk with a trivial cobordism class in $G$ \cite{WangWen2018cai1809.11171} (whose boundary contains
any all-$G$-anomaly-free theory \cite{WangWen2018cai1809.11171}).

To establish this claim, we can check that $so(18)$ GUT is free from the $\Z_2$ class global anomaly \cite{WangWen2018cai1809.11171, WW2019fxh1910.14668}
(which turns out to be the same $\Z_2$ class anomaly of $so(10)$ GUT if we embed $\Spin(10) \subset \Spin(18)$:
\bea
&& 
\Omega_5^{{\Spin \times_{\Z_2} \Spin(18)}} =\Omega_5^{{\Spin \times_{\Z_2} \Spin(10)}} = \Z_2, \quad \\ 
&&
\TP_5({{\Spin \times_{\Z_2} \Spin(18)}})=\TP_5({{\Spin \times_{\Z_2} \Spin(10)}})=  \Z_2.
\eea
This potential $\Z_2$ class 4d global anomaly in Spin(18) and Spin(10) chiral fermion theory
is analogous to the 4d new SU(2) anomaly  \cite{WangWenWitten2018qoy1810.00844} occurred in \Eq{eq:cobordismSpinZ2SU2},
based on $\SU(2) \subset \Spin(3) \subset \Spin(10) \subset \Spin(18)$.
Each is respectively generated by 5d cobordism invariants (see footnote \ref{footnote:notations} for notations): 
 \bea \label{eq:so(10)so(18)anomaly}
\exp( \ii \pi   \int  w_2(V_{\SO(18)})w_3(V_{\SO(18)})) \quad \text{ and } \quad \exp( \ii \pi  \int   w_2(V_{\SO(10)})w_3(V_{\SO(10)}) ).
 \eea
 Follow 
 \cite{WangWen2018cai1809.11171, WangWenWitten2018qoy1810.00844}, we can explicitly check 
these anomalies \Eq{eq:so(10)so(18)anomaly} are absent in the $so(10)$ and $so(18)$ GUT (also shown later in  \Sec{sec:newSU2anom}).
 The new SU(2) anomaly requires an isospin-3/2 fermion of SU(2) = Spin(3) (i.e., in the representation ${\bf 4}$ of SU(2)) to realize the anomaly  \cite{WangWenWitten2018qoy1810.00844}. 
 However, the fermions in $so(10)$ GUT and $so(18)$ GUT
are in the spinor representation ${\bf 16}^+$ of Spin(10) and ${\bf 256}^+$ of Spin (18), which after projection to the Spin(3) can be decomposed 
as the direct sums of 
8 copies of ${\bf 2}$ (an isospin-1/2 fermion) of  SU(2),
and 
128 copies of ${\bf 2}$ (an isospin-1/2 fermion) of SU(2) respectively.
Same for the ${\bf 16}^-$ of Spin(10) and ${\bf 256}^-$ of Spin (18).
Namely, we obtain
\bea
\quad  \Psi_L \sim {\bf 256}^+ \text{ of } \Spin(18) \text{ or }   \Psi_R \sim {\bf 256}^- \text{ of } \Spin(18) &\sim& 128 \cdot {\bf 2} \text{ of } \Spin(3) = \SU(2).\label{eq:256-}\\
\quad  \Psi_L \sim {\bf 16}^+ \text{ of } \Spin(10)  \text{ or }   \Psi_R \sim {\bf 16}^- \text{ of } \Spin(10)  &\sim&  8\cdot {\bf 2}\text{ of }  \Spin(3) = \SU(2).
\eea
So the $so(10)$ GUT and $so(18)$ GUT both have some \emph{even} numbers of isospin-1/2 fermions, which do not have the familiar Witten SU(2) anomaly \cite{Witten1982fp},
nor do they have the new SU(2) anomaly \cite{WangWen2018cai1809.11171, WangWenWitten2018qoy1810.00844} in the absence of {\bf 4} of SU(2). 
So the $so(10)$ GUT and $so(18)$ GUT can have gapped mirror sectors without the unwanted fermion doubling \cite{WangWen2018cai1809.11171}.
%
%
%
\begin{figure}[t!] 
  \hspace{-24mm}
     \includegraphics[width=8.4in]{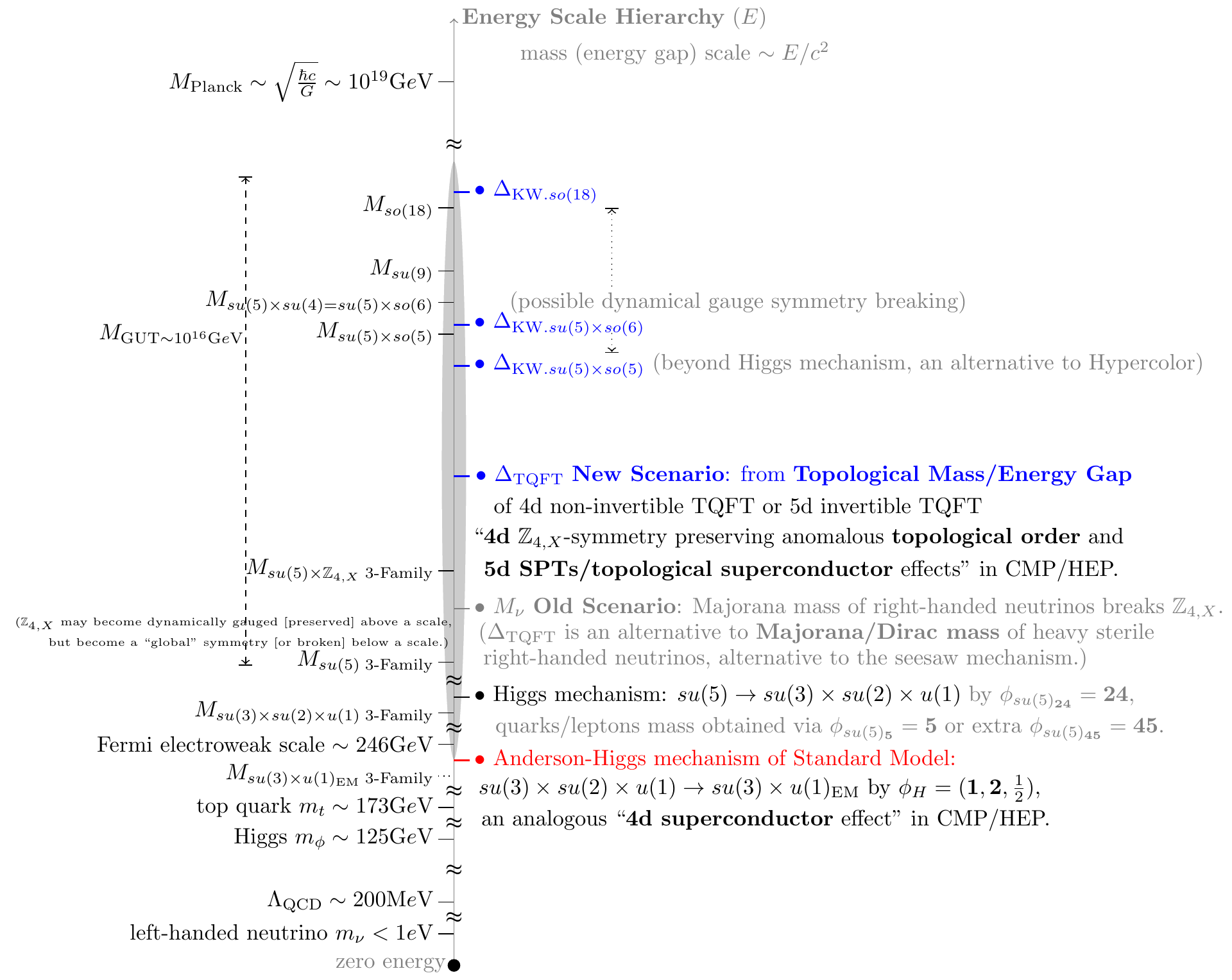}
  \caption{Energy and Mass Hierarchy Proposal: Follow \Fig{fig:energy-hierarchy-lie-algebra-world},
   the breaking structure and hierarchy structure always concern 
  the global Lie group: $\Spin(18), \SU(5), \frac{\SU(3)\times \SU(2)\times \U(1)}{\Z_q}, etc$.
  However, we simply denote the local Lie algebra such as $so(18),  su(5), su(3)  \times su(2) \times u(1), etc.$, only for the abbreviation brevity and only to be consistent with the notations of early physics literature.
  }
  \label{fig:energy-hierarchy-lie-algebra-1}
\end{figure}

\begin{figure}[t!] 
  \hspace{-24mm}
     \includegraphics[width=8.4in]{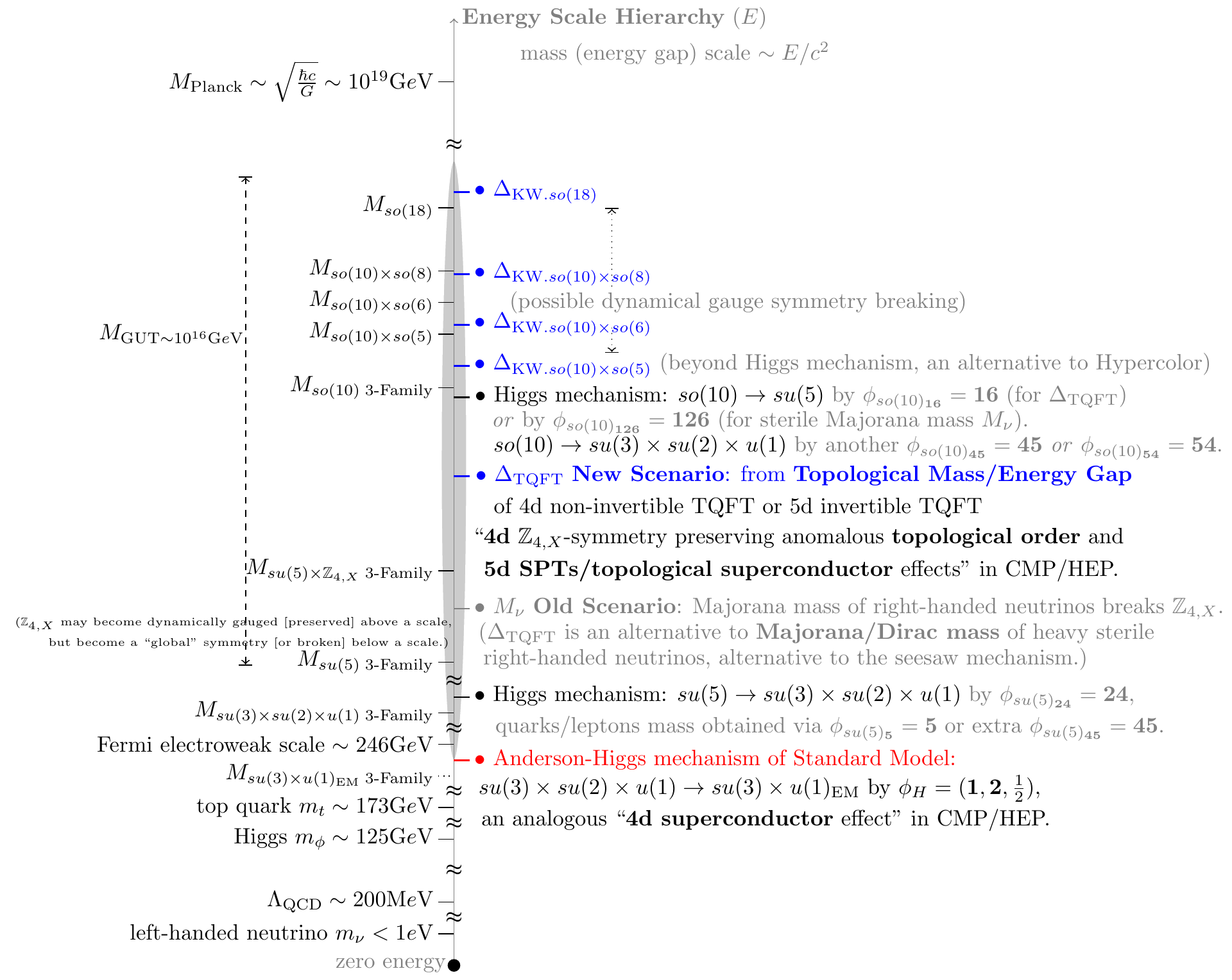}
  \caption{Energy and Mass Hierarchy Proposal:
  Follow \Fig{fig:energy-hierarchy-lie-algebra-world} and \Table{fig:energy-hierarchy-lie-algebra-1}'s 
  notations/explanations,
  what we have in mind about is always the global Lie group: $\frac{\Spin(10)\times \Spin(5)}{\Z_2}, etc$.
  However, we simply denote the local Lie algebra such as $so(10) \times so(5), etc$.
}
  \label{fig:energy-hierarchy-lie-algebra-2}
\end{figure}

In any case, the gapped mirror sector via {\bf (1)} or {\bf (2)} must have a huge energy gap of GUT scale somehow slightly higher than the 
$so(18)$ GUT scale but much lower than {$M_{\mathrm{Planck}}\sim \sqrt{\frac{\hbar c}{G}}\sim 10^{19}{\rm G}e{\rm V}$}, which we call 
\bea
\Delta_{\mathrm{KW}.so(18)}.
\eea
Here KW stands for Kitaev-Wen (KW) mechanism due to the original 
pioneer work of \cite{FidkowskifSPT1, FidkowskifSPT2, Kitaev2015, Wen2013oza1303.1803, Wen2013ppa1305.1045}.
The ${\mathrm{KW}.so(18)}$ means the \emph{first} Kitaev-Wen mechanism that we implement starting around from the scale of $so(18)$ GUT.
This $\Delta_{\mathrm{KW}.so(18)}$ must be in a larger energy gap than the other dynamical gauge symmetry breaking or GUT Higgs scales. See 
our \Fig{fig:energy-hierarchy-lie-algebra-1} and \Fig{fig:energy-hierarchy-lie-algebra-2}.

\subsection{$so(18)$ GUT to $so(10)\times so(8)$ GUT and $\Delta_{\mathrm{KW}.so(10)\times so(8)}$}

\item 
\label{SO(18)SO(10)SO(8)GUTgap}
{\bf Break $so(18)$ GUT to $so(10)\times so(8)$ GUT}:\\
Suppose the breaking
${{\Spin \times_{\Z_2^F} \Spin(18)}} \to {\Spin \times_{\Z_2^F} ({\Spin(10) \times_{\Z_2^F} \Spin(8)})}$ occurs at 
the energy scale $M_{so(10)\times so(8)}$,\footnote{It is possible
to achieve the breaking by \emph{dynamical symmetry breaking} or by \emph{Higgs mechanism}.
We will not pursuit the details of the possibility of {dynamical symmetry breaking} in this article. 
Instead, as the old wisdom goes,
we can simply follow the similar route of the breaking analysis by Higgs field as performed in \cite{Zee2003book, BenTov2015graZee1505.04312}.
We leave the analysis for synthesizing the old idea of 
{dynamical symmetry breaking} and the new idea of nonperturbative anomaly/cobordism constraints, 
together with heavy color, technicolor, or hypercolor types of ideas \cite{Weinberg1975gmDynamical,Susskind1978msDynamical,Farhi1980xsSusskindDynamical1981,Hill2002apDynamical0203079}, in a future work.} 
we have the representation branching rule decomposition for fermions in the spinor representations of internal symmetry groups:
\bea
\label{eq:so10so8L}
\quad  \Psi_L^{\Spin(18)} \sim {\bf 256}^+ \text{ of } \Spin(18) \sim ({\bf 16}^+, {\bf 8}^+) \oplus ({\bf 16}^-, {\bf 8}^-)  \text{ of } {\Spin(10) \times_{\Z_2^F} \Spin(8)},\\
\label{eq:so10so8R}
\quad  \Psi_R^{\Spin(18)} \sim {\bf 256}^- \text{ of } \Spin(18) \sim ({\bf 16}^+, {\bf 8}^-) \oplus ({\bf 16}^-, {\bf 8}^+)  \text{ of } {\Spin(10) \times_{\Z_2^F} \Spin(8)}.
\eea
Again the mirror fermion $\Psi_R^{\Spin(18)} \sim {\bf 256}^-  \sim ({\bf 16}^+, {\bf 8}^-) \oplus ({\bf 16}^-, {\bf 8}^+) $ is already fully gapped out, on and above the scale $\Delta_{\mathrm{KW}.so(18)}$ in Scenario \ref{SO(18)GUTgap}.

Follow the similar idea of \Ref{BenTov2015graZee1505.04312}, 
we could check whether the Kitaev-Wen analogous mechanism can gap $({\bf 16}^-, {\bf 8}^-)$ without breaking $({\Spin(10) \times_{\Z_2^F} \Spin(8)})$
by checking whether all the anomalies vanish
for the $({\bf 16}^-, {\bf 8}^-)$ chiral fermion. First, the 
cobordism classification of the anomalies show \cite{WanWangv2}:
\bea
&& \Omega_5^{\Spin \times_{\Z_2} { \frac{\Spin(10) \times {\Spin(8) }}{\Z_2}}}  = \Z_2^2,\quad\\
 &&\TP_5({\Spin \times_{\Z_2} { \frac{\Spin(10) \times {\Spin(8) }}{\Z_2}}})  = \Z_2^2.
 \eea
 The $\Z_2^2$ are two $\Z_2$ classes of 
4d nonperturbative global anomalies generated by 5d cobordism invariants (see footnote \ref{footnote:notations} for notations)
\bea
 \quad \exp( \ii \pi   \;  \int \Big( n_{10}\;  w_2(V_{\SO(10)})w_3(V_{\SO(10)}) + n_{8} \; w_2(V_{\SO(8)})w_3(V_{\SO(8)}) \Big)  ),
\eea
 where $(n_{10}, n_{8}) \in \Z_2^2$. 
These $\Z_2^2$ class anomalies are similar to the new SU(2) anomaly \cite{WangWenWitten2018qoy1810.00844} thanks to $\Spin(10)\supset\Spin(8)\supset\Spin(3)$. 
Follow the same projection checking in Scenario \ref{SO(18)GUTgap},
 \Ref{BenTov2015graZee1505.04312} wishes to gap 
the fermions in the spinor representation ${\bf 16}^-$ of Spin(10) and ${\bf 8}^-$ of Spin (8), which after projection to the Spin(3) can be decomposed 
as the direct sums of 
8 copies of ${\bf 2}$ (an isospin-1/2 fermion) of  SU(2),
and 
4 copies of ${\bf 2}$ (an isospin-1/2 fermion) of SU(2) respectively.
Namely, we obtain
\bea
\quad   {\bf 16}^- \text{ of } \Spin(10)&\sim&  8\cdot {\bf 2}\text{ of }  \Spin(3) = \SU(2).\\
\quad  {\bf 8}^- \text{ of } \Spin(8) &\sim& 4\cdot {\bf 2} \text{ of } \Spin(3) = \SU(2).
\eea
So the $so(10)$ GUT and $so(18)$ GUT both have some \emph{even} numbers of isospin-1/2 fermions {\bf 2} of SU(2) ({\bf 4}) of SU(2), which do not have the familiar Witten SU(2) anomaly \cite{Witten1982fp},
nor do they have the new SU(2) anomaly \cite{WangWen2018cai1809.11171, WangWenWitten2018qoy1810.00844} in the absence of isospin-3/2 fermions {\bf 4} of SU(2). 
So the $({\bf 16}^-, {\bf 8}^-)$ can be a gapped sector without breaking the symmetry, agreeing with \cite{BenTov2015graZee1505.04312}.
If there is an energy gap scale for this gapping $({\bf 16}^-, {\bf 8}^-)$
scenario, we can name it as:
\bea
\Delta_{\mathrm{KW}.so(10)\times so(8)},
\eea
which is around the breaking scale $M_{so(10)\times so(8)}$.

\subsection{$so(10)\times so(8)$ GUT to $so(10)\times so(6)$ GUT and $\Delta_{\mathrm{KW}.so(10)\times so(6)}$}
\label{sec:SO(10)SO(8)SO(6)GUTgap}

\item 
\label{SO(10)SO(8)SO(6)GUTgap}
Motivated by the dynamical symmetry breaking and/or Higgs mechanism of the $so(18)$ GUT  \cite{WilczekZee1981iz1982Spinors, Fujimoto1981SO18Unification, Zee2003book, BenTov2015graZee1505.04312}, 
we can consider the following breaking pattern:
\begin{multline}
{\Spin \times_{\Z_2^F} ({\Spin(10) \times_{\Z_2^F} \Spin(8)})}\\
\longrightarrow
{\Spin \times_{\Z_2^F} ({\Spin(10) \times_{\Z_2^F} (\Spin(6) \times_{\Z_2^F} \Spin(2))  })}\\
\longrightarrow
{\Spin \times_{\Z_2^F} ({\Spin(10) \times_{\Z_2^F} \Spin(6)})},
\end{multline}
where $\Spin(6)=\SU(4)$ and $\Spin(2)=\U(1)$. The ${\bf 8}^+$ and ${\bf 8}^-$ in \Eq{eq:so10so8L}
are the 8-dimensional representations of Spin(8). There are three
8-dimensional representations: the vector representation ${\bf 8}_v$, the spinor representation ${\bf 8}_s$, and the conjugate of spinor representation ${\bf 8}_c$ related by the Spin(8) triality.
The ${\bf 8}_v, {\bf 8}_s$, and ${\bf 8}_c$ can be transformed to each other via the outer automorphism $S_3$, which is the symmetric group of the 
order $3! = 6$ as the permutation of 3 elements.
In particular, for the convenience of obtaining a desirable breaking pattern,
we choose that
${\bf 8}^+$ is ${\bf 8}_v$ and ${\bf 8}^-$ is ${\bf 8}_c$, also we choose the decomposition branching rules for
$\Spin(8) \longrightarrow (\Spin(6) \times_{\Z_2^F} \Spin(2)) \longrightarrow  \Spin(6)$ as
\bea
\begin{array}{lcl}
{\bf 8}_v  \text{ of } \Spin(8)  &=& ({\bf 1},2) \oplus ({\bf 1},-2)\oplus({\bf 6},0) \text{ of $(\Spin(6)  \times_{\Z_2^F} \Spin(2))$}
={\bf 1}\oplus{\bf 1}\oplus{\bf 6} \text{ of $\Spin(6)$}.\\
{\bf 8}_s \text{ of } \Spin(8)  &=& ({\bf 4},-1)\oplus(\overline{\bf 4},1) 
\text{ of $(\Spin(6)  \times_{\Z_2^F} \Spin(2))$} ={\bf 4}\oplus \overline{\bf 4}  \text{ of $\Spin(6)$}.\\
{\bf 8}_c  \text{ of } \Spin(8)  &=& ({\bf 4},1) \oplus(\overline{\bf 4},-1) \text{ of $(\Spin(6)  \times_{\Z_2^F} \Spin(2))$} ={\bf 4}\oplus \overline{\bf 4}  \text{ of $\Spin(6)$}.
\end{array}
\eea
After $({\bf 16}^-, {\bf 8}^-)$ can be already gapped out by Scenario \ref{SO(18)SO(10)SO(8)GUTgap}, we can check whether any additional sector of  $({\bf 16}^+, {\bf 8}^+) \text{ of } {\Spin(10) \times_{\Z_2^F} \Spin(8)}$ can be gapped out by Kitaev-Wen analogous mechanism.
The ${\bf 16}^+ \text{ of } \Spin(10)\sim  8\cdot {\bf 2}\text{ of }  \Spin(3) = \SU(2)$ is free from the old (familiar Witten) SU(2) anomaly \cite{Witten1982fp} and the new SU(2) anomaly \cite{WangWenWitten2018qoy1810.00844}.
The ${\bf 8}^+ = {\bf 8}_v  \text{ of } \Spin(8)$ can be decomposed as ${\bf 1}\oplus{\bf 1}\oplus{\bf 6}$ of $\Spin(6)$.
Thus we can check whether some of the components in 
 \bea \label{eq:so10so6-16-11}
 ({\bf 16}^+, {\bf 8}^+) \text{ of } {\Spin(10) \times_{\Z_2^F} \Spin(8)} =
  ({\bf 16}^+, {\bf 1}\oplus{\bf 1}\oplus{\bf 6}) \text{ of } {\Spin(10) \times_{\Z_2^F} \Spin(6)} \label{eq:so10so6-16-116}
 \eea
 are free from the anomalies classified by the cobordism group  \cite{WanWangv2}:
\bea 
&& \Omega_5^{\Spin \times_{\Z_2} { \frac{\Spin(10) \times {\Spin(6) }}{\Z_2}}}  = \Z_2^2,\quad\\
 &&\TP_5({\Spin \times_{\Z_2} { \frac{\Spin(10) \times {\Spin(6) }}{\Z_2}}})  = \Z_2^2. \label{eq:TPSO10SO6}
 \eea
 The $\Z_2^2$ are two $\Z_2$ classes of 
4d nonperturbative global anomalies 
generated by 5d cobordism invariants (see footnote \ref{footnote:notations} for notations):
\bea
 \quad \exp( \ii \pi   \;  \int \Big( n_{10}\;  w_2(V_{\SO(10)})w_3(V_{\SO(10)}) + n_{6} \; w_2(V_{\SO(6)})w_3(V_{\SO(6)}) \Big)  ),
\eea
 where $(n_{10}, n_{6}) \in \Z_2^2$. 
These $\Z_2^2$ class anomalies are similar to the new SU(2) anomaly \cite{WangWenWitten2018qoy1810.00844} 
thanks to $\Spin(10)\supset\Spin(6)\supset\Spin(3)$. We can project the Spin(6) to Spin(3) = SU(2) representation
 \bea
  {\bf 1}\oplus{\bf 1}\oplus{\bf 6} \text{ of } {\Spin(6)}  =   {\bf 1}\oplus{\bf 1}\oplus (   {\bf 1}\oplus{\bf 1} \oplus{\bf 2} \oplus{\bf 2}) \text{ of } {\Spin(3) =  \SU(2)}. 
 \eea
 In particular, the ${\bf 6}$ of Spin(6) as $({\bf 1}\oplus{\bf 1} \oplus{\bf 2} \oplus{\bf 2})$ of ${\Spin(3) =  \SU(2)}$,\footnote{Depending on how do we embed
$\Spin(6) \supset \Spin(3)$, it is possible that we can obtain ${\bf 6}$ of Spin(6) as $({\bf 1}\oplus{\bf 1} \oplus{\bf 2} \oplus{\bf 2})$ of ${\Spin(3)}$,
or ${\bf 6}$ of Spin(6) as $({\bf 1}\oplus{\bf 1} \oplus{\bf 1} \oplus{\bf 3})$ of ${\Spin(3)}$. In any case, it still has an even number of ${\bf 2}$ and an even number of ${\bf 4}$ of SU(2), 
which we confirm to be free from the old and the new SU(2) anomalies.} 
 due to an even number of ${\bf 2}$ and no ${\bf 4}$ of SU(2) (thus their mod 2 classes are zeros), 
 is now confirmed to be free from the mod 2 classes of old and the new SU(2) anomalies. Thus the
 $({\bf 16}^+, {\bf 6}) \text{ of } {\Spin(10) \times_{\Z_2^F} \Spin(6)} $ is free from all anomalies of \Eq{eq:TPSO10SO6}.\footnote{Similarly,  
 the $({\bf 16}^+, {\bf 1}\oplus{\bf 1}) \text{ of } {\Spin(10) \times_{\Z_2^F} \Spin(6)}$
 is also free from all anomalies of \Eq{eq:TPSO10SO6}  and thus can be gapped, but we wish to keep the $({\bf 16}^+, {\bf 1}\oplus{\bf 1}\oplus \dots)$ intact
 for the nearly gapless sector for the SM phenomenology.} 
In summary,  we can gap $({\bf 16}^+, {\bf 6})$ by nonperturbative interactions without breaking the ${\Spin(10) \times_{\Z_2^F} \Spin(6)}$ symmetry, 
but we keep the gapless $({\bf 16}^+, {\bf 1}\oplus{\bf 1})$ intact.
{If there is an energy gap scale for this gapping $({\bf 16}^+, {\bf 6}) $ scenario, we can name it as:
\bea
\Delta_{\mathrm{KW}.so(10)\times so(6)},
\eea
which is around the breaking scale $M_{so(10)\times so(6)}$. However, this is not desirable because we are left with the nearly
gapless $({\bf 16}^+, {\bf 1}\oplus{\bf 1})$ with only \emph{two} generations instead of \emph{three} generations of quarks and leptons.}

\subsection{$so(10)\times so(8)$ GUT to $so(10)\times so(5)$ GUT,  $\Delta_{\mathrm{KW}.so(10)\times so(5)}$, and Three Generations}
\label{sec:SO(10)SO(8)SO(5)GUTgap}

\item 
\label{SO(10)SO(8)SO(5)GUTgap}
We can also consider the following breaking pattern:
\begin{multline}
{\Spin \times_{\Z_2^F} ({\Spin(10) \times_{\Z_2^F} \Spin(8)})}\\
\longrightarrow
{\Spin \times_{\Z_2^F} ({\Spin(10) \times_{\Z_2^F} (\Spin(5)  \times_{\Z_2^F} \Spin(3))})}\\
\longrightarrow
{\Spin \times_{\Z_2^F} ({\Spin(10) \times_{\Z_2^F} \Spin(5)})},
\end{multline}
%
where $\Spin(5)=\Sp(2)=\U\Sp(4)$ and $\Spin(3)=\SU(2)$.
Again, for the convenience of obtaining a desirable breaking pattern,
we choose that
${\bf 8}^+$ is ${\bf 8}_v$ and ${\bf 8}^-$ is ${\bf 8}_c$, also we choose the decomposition branching rules for
$\Spin(8) \longrightarrow (\Spin(5) \times_{\Z_2^F} \Spin(3)) \longrightarrow  \Spin(5)$ as
\bea
\begin{array}{lcl}
{\bf 8}_v \text{ of } \Spin(8)  &=& {\bf (1,3)}\oplus  {\bf (5,1)} \text{ of $(\Spin(5)  \times_{\Z_2^F} \Spin(3))$} ={\bf 1}\oplus{\bf 1}\oplus{\bf 1}\oplus{\bf 5} \text{ of $\Spin(5)$}.\\
{\bf 8}_s  \text{ of } \Spin(8)  &=&{\bf (4,2)} \text{ of $(\Spin(5)  \times_{\Z_2^F} \Spin(3))$} ={\bf 4}\oplus{\bf 4} = 2 \cdot {\bf 4} \text{ of $\Spin(5)$}.\\
{\bf 8}_c  \text{ of } \Spin(8)  &=&{\bf (4,2)} \text{ of $(\Spin(5)  \times_{\Z_2^F} \Spin(3))$} ={\bf 4}\oplus{\bf 4} = 2 \cdot {\bf 4} \text{ of $\Spin(5)$}.
\end{array}
\eea
Similar to Scenario \ref{SO(10)SO(8)SO(6)GUTgap}, after $({\bf 16}^-, {\bf 8}^-)$ is already gapped out by Scenario \ref{SO(18)SO(10)SO(8)GUTgap},
we can check whether any additional sector of  $({\bf 16}^+, {\bf 8}^+) \text{ of } {\Spin(10) \times_{\Z_2^F} \Spin(8)}$ can be gapped out by Kitaev-Wen analogous mechanism.
The ${\bf 16}^+ \text{ of } \Spin(10)\sim  8\cdot {\bf 2}\text{ of }  \Spin(3) = \SU(2)$ is free from the old (familiar Witten) SU(2) anomaly \cite{Witten1982fp} and the new SU(2) anomaly \cite{WangWenWitten2018qoy1810.00844}.
The ${\bf 8}^+ = {\bf 8}_v  \text{ of } \Spin(8)$ can be decomposed as ${\bf 1}\oplus{\bf 1}\oplus{\bf 1}\oplus{\bf 5}$ of $\Spin(5)$.
Thus we can check whether some of the components in 
 \bea
 ({\bf 16}^+, {\bf 8}^+) \text{ of } {\Spin(10) \times_{\Z_2^F} \Spin(8)} =
  ({\bf 16}^+, {\bf 1}\oplus{\bf 1}\oplus{\bf 1}\oplus{\bf 5}) \text{ of } {\Spin(10) \times_{\Z_2^F} \Spin(5)} 
  \label{eq:so10so5-16-1115}
 \eea
 are free from the anomalies classified by the cobordism group  \cite{WanWangv2}:
\bea
&& \Omega_5^{\Spin \times_{\Z_2} { \frac{\Spin(10) \times {\Spin(5) }}{\Z_2}}}  = \Z_2^2,\quad\\
 &&\TP_5({\Spin \times_{\Z_2} { \frac{\Spin(10) \times {\Spin(5) }}{\Z_2}}})  = \Z_2^2.\label{eq:TPSO10SO5}
 \eea
 The $\Z_2^2$ are two $\Z_2$ classes of 
4d nonperturbative global anomalies generated by 5d cobordism invariants  (see footnote \ref{footnote:notations} for notations)
\bea
 \quad \exp( \ii \pi   \;  \int \Big( n_{10}\;  w_2(V_{\SO(10)})w_3(V_{\SO(10)}) + n_{5} \; w_2(V_{\SO(5)})w_3(V_{\SO(5)}) \Big)  ),
\eea
 where $(n_{10}, n_{5}) \in \Z_2^2$, similar to the new SU(2) anomaly \cite{WangWenWitten2018qoy1810.00844} 
thanks to $\Spin(10)\supset\Spin(5)\supset\Spin(3)$. We can project the Spin(5) to Spin(3) = SU(2) representation
\bea
  {\bf 1}\oplus{\bf 1}\oplus{\bf 1}\oplus{\bf 5} \text{ of } {\Spin(5)}  =   {\bf 1}\oplus{\bf 1}\oplus{\bf 1}\oplus (   {\bf 1}\oplus{\bf 2} \oplus{\bf 2}) \text{ of } {\Spin(3) =  \SU(2)}. 
 \eea
  In particular, the ${\bf 5}$ of Spin(5) as $({\bf 1}\oplus{\bf 2} \oplus{\bf 2})$ of ${\Spin(3) =  \SU(2)}$,\footnote{Depending on how do we embed
$\Spin(5) \supset \Spin(3)$, it is possible that we can obtain ${\bf 5}$ of Spin(5) as $({\bf 1} \oplus{\bf 2} \oplus{\bf 2})$ of ${\Spin(3)}$;
 it may also be possible to
choose the ${\bf 5} \text{ of } {\Spin(5)}$ as the $(   {\bf 1}\oplus{\bf 1} \oplus{\bf 3})$ of ${\Spin(3) =  \SU(2)}$.
In any case, it still has an even number of ${\bf 2}$ and an even number of ${\bf 4}$ of SU(2), 
which we confirm to be free from the old and the new SU(2) anomalies. \label{ft:Spin5-122-113}
} due to an even number of ${\bf 2}$ and no ${\bf 4}$ of SU(2) (thus their mod 2 classes are zeros), 
 is now confirmed to be free from the old and the new SU(2) anomalies. Thus the
 $({\bf 16}^+, {\bf 5}) \text{ of } {\Spin(10) \times_{\Z_2^F} \Spin(5)} $ is free from all anomalies of \Eq{eq:TPSO10SO5}.\footnote{Similarly,  
 the $({\bf 16}^+, {\bf 1}\oplus{\bf 1}\oplus{\bf 1})$ { of } ${\Spin(10) \times_{\Z_2^F} \Spin(5)}$
 is also free from all anomalies of \Eq{eq:TPSO10SO5} and thus can be gapped, but we wish to keep the $({\bf 16}^+, {\bf 1}\oplus{\bf 1}\oplus {\bf 1})$ intact
 for the nearly gapless sector for the SM phenomenology.} 
 In summary,  we can gap $({\bf 16}^+, {\bf 5})$ by nonperturbative interactions, 
but we keep the gapless 
\bea \label{eq:so10so5-16-111}
({\bf 16}^+, {\bf 1}\oplus{\bf 1}\oplus{\bf 1})
\eea intact, while preserving the ${\Spin(10) \times_{\Z_2^F} \Spin(5)}$ symmetry.
{If there is an energy gap scale for this gapping $({\bf 16}^+, {\bf 5}) $ scenario, we can name it as:
\bea
\Delta_{\mathrm{KW}.so(10)\times so(5)},
\eea
which is around the breaking scale $M_{so(10)\times so(5)}$. This seems to be desirable because we are left with the nearly
gapless $({\bf 16}^+, {\bf 1}\oplus{\bf 1}\oplus{\bf 1})$ with exactly three generations of quarks and leptons.}

\subsection{$so(18)$ GUT to $su(9)$ GUT and $su(5) \times so(6)$ GUT, and $\Delta_{\mathrm{KW}.su(5)\times so(6)}$}
\label{sec:SO(18)SU(9)SU(5)SU(4)GUTgap}

\item {\bf Break $so(18)$ GUT from $\Spin(18)$ to $\SU(9)$, to $\SU(5) \times \SU(4)=\SU(5) \times \Spin(6)$ and their GUT}:\\
Here we attempt to make the historical {$so(18)$ GUT to $su(9)$ GUT and $su(5) \times su(4)$ GUT} (or $su(5) \times so(6)$ GUT, by the Lie algebra  $su(4)=so(6)$)
breaking process in 
\cite{WilczekZee1981iz1982Spinors,Fujimoto1981SO18Unification,BenTov2015graZee1505.04312} more mathematically precise, 
by considering the embedding web \Fig{table:sym-web-1}.  Breaking
${{\Spin \times_{\Z_2^F} \Spin(18)}} \to {\Spin \times \SU(9)}$,
we have the representation branching rule:
\bea
\quad  \Psi_L^{\Spin(18)} \sim {\bf 256}^+ \text{ of } \Spin(18) \sim  
[0] \oplus [2] \oplus [4] \oplus [6] \oplus [8] =
{\bf 1} \oplus  {\bf 36} \oplus  {\bf 126} \oplus \overline  {\bf 84} \oplus \overline  {\bf 9} \text{ of } {\SU(9)}.
\eea
Follow the setup and notation in \cite{BenTov2015graZee1505.04312}, the [N] is an N-index anti-symmetric tensor from the fundamental representation of SU(9).
Let us decompose the above matter field representations from the viewpoint of
${\Spin \times \SU(9)} \to \Spin \times \SU(5) \times \SU(4)= \Spin \times  \SU(5) \times \Spin(6)$.
Below we follow the notations in \cite{BenTov2015graZee1505.04312}, the subscript [N] of the 2-tuple ``(SU(5) representation, Spin(6) representation)$_{\text{[N]}}$''
means where the 2-tuple is from the [N] of SU(9).
We see that part of 120 Weyl fermions in ${\bf 256}^+$ is analogous to matter fields,
\bea
&&\text{Representation of } { \SU(5) \times \SU(4)= \SU(5) \times \Spin(6)}: \nn \\[2mm]
&&( \overline  {\bf 5}, {\bf 1})_{[8]} \oplus ( \overline  {\bf 5} , {\bf 1})_{[4]} \oplus ( \overline  {\bf 5} , {\bf 6})_{[6]} 
\oplus (   {\bf 10} , {\bf 1})_{[2]}  \oplus (   {\bf 10} , {\bf 1})_{[6]} \oplus (   {\bf 10} , {\bf 6})_{[4]} \cr
&&={( \overline  {\bf 5}, {\bf 1}) \oplus ( \overline  {\bf 5} , {\bf 1}) \oplus ( \overline  {\bf 5} , {\bf 6}) 
\oplus (   {\bf 10} , {\bf 1})  \oplus (   {\bf 10} , {\bf 1}) \oplus (   {\bf 10} , {\bf 6})}
=\Big(( \overline  {\bf 5} \oplus {\bf 10} ),\; ( {\bf 1} \oplus {\bf 1} \oplus  {\bf 6}) \Big)  .  \label{eq:SU5SU4-SU5Spin6-gapless-multiplet}
\eea
We also see that additional part of 120 Weyl fermions in ${\bf 256}^+$ is analogous to extra matter fields,
\bea
&&\text{Representation of } { \SU(5) \times \SU(4)= \SU(5) \times \Spin(6)}: \nn \\[2mm]
&&(   {\bf 5}, {\bf 4})_{[4]} \oplus (   {\bf 5} , \overline {\bf 4})_{[2]} \oplus ( \overline  {\bf 10} , {\bf 4})_{[4]} \oplus ( \overline  {\bf 10} , \overline{\bf 4})_{[6]}\cr
&&={(   {\bf 5}, {\bf 4}) \oplus (   {\bf 5} , \overline {\bf 4}) \oplus ( \overline  {\bf 10} , {\bf 4}) \oplus ( \overline  {\bf 10} , \overline{\bf 4})} 
=\Big((   {\bf 5} \oplus \overline {\bf 10} ),\; ( {\bf 4} \oplus  \overline{\bf 4}) \Big) . \label{eq:SU5SU4-SU5Spin6-multiplet}
\eea
There is an additional part of 16 Weyl fermions in ${\bf 256}^+$ analogous to 16 copies of a right-handed ``sterile'' neutrino,
\bea
&&\text{Representation of } { \SU(5) \times \SU(4)= \SU(5) \times \Spin(6)}: \nn \\[2mm]
&&(   {\bf 1}, {\bf 1})_{[0]} \oplus (   {\bf 1}, {\bf 1})_{[4]}
\oplus (   {\bf 1}, {\bf 4})_{[6]}
\oplus (   {\bf 1}, \overline {\bf 4})_{[8]}
\oplus (   {\bf 1}, {\bf 6})_{[2]}\cr
&&(   {\bf 1}, {\bf 1}) \oplus (   {\bf 1}, {\bf 1})
\oplus (   {\bf 1}, {\bf 4})
\oplus (   {\bf 1}, \overline {\bf 4})
\oplus (   {\bf 1}, {\bf 6})
=\Big(  {\bf 1}   ,\; ( {\bf 1} \oplus {\bf 1} \oplus {\bf 4} \oplus  \overline{\bf 4}\oplus  {\bf 6}) \Big).
\label{eq:SU5SU4-SU5Spin6-neutrino-multiplet}
\eea
The $(   {\bf 1}, {\bf 1})_{[0]}$ and $(   {\bf 1}, {\bf 1})_{[4]}$ are indeed sterile and they carry
no gauge charge under $\SU(5) \times \Spin(6)$ (thus not charged under $G_{\text{SM}_6}$ and SM gauge forces).
But the $(   {\bf 1}, {\bf 4})_{[6]}$,
$(   {\bf 1}, \overline {\bf 4})_{[8]}$, and $(   {\bf 1}, {\bf 6})_{[2]}$
are \emph{only} sterile under $\SU(5)$ (thus sterile to SM forces), but they do carry
 gauge charges as fundamental, anti-fundamental, or spinor presentations of $\SU(4)=\Spin(6)$.

We can check that whether the $\Spin \times  \SU(5) \times \Spin(6)$ chiral fermion theory 
of additional matter $\Big((   {\bf 5} \oplus \overline {\bf 10} ),\; ( {\bf 4} \oplus  \overline{\bf 4}) \Big)$
are free from the anomaly classified by \cite{WanWangv2}
\bea
&& \Omega_5^{\Spin \times \SU(5)\times \Spin(6)} =0, \quad \\ 
&&\TP_5({\Spin \times \SU(5) \times \Spin(6)})  = \Z^2. \label{eq:TP5-SU5SU4-SU5Spin6}
\eea
The $\Z^2$ are two $\Z$ classes of 
4d perturbative local anomalies captured by one-loop triangle  Feynman diagrams and generated by 5d cobordism invariants (see footnote \ref{footnote:notations} for notations):
\bea \label{eq:anomaly-SU5SU4-SU5Spin6}
 \quad \exp( \ii    \;  \int \Big( n_{su(5)}\;  \frac{1}{2}\text{CS}_5(V_{\SU(5)}) + n_{so(6)} \; \frac{1}{2}\text{CS}_{5,e}(V_{\SO(6)}) \Big)  ),
\eea
 where $(n_{su(5)}, n_{5}) \in \Z^2$.
They are also related to the 6d anomaly polynomial of the 6th bordism group $\Omega_6$, generated by
the 3rd Chern class of SU(5) gauge bundle \cite{WangWen2018cai1809.11171, 2019arXiv191011277D}
and the 6th Euler class of Spin(6) or SO(6) gauge bundle \cite{2019arXiv191011277D},\footnote{Follow \cite{2019arXiv191011277D}, we use $\text{CS}_{2n-1}(V)$ to denote the Chern-Simons $2n-1$-form for the Chern class (if $V$ is a complex vector bundle) or the Pontryagin class (if $V$ is a real vector bundle) where $p_i(V)=(-1)^ic_{2i}(V\otimes\C)$. The relation between the Chern-Simons form and the Chern class is
$
c_n(V)=\dd \text{CS}_{2n-1}(V)
$
where the $\dd$ is the exterior differential and the $c_n(V)$ is regarded as a closed differential form in de Rham cohomology.
There is another kind of Chern-Simons form for Euler class {$e_{2n}(V)$}, we denote its Chern-Simons form 
by $\text{CS}_{2n-1,e}(V)$, it satisfies
$
e_{2n}(V)=\dd \text{CS}_{2n-1,e}(V).
$
} see more details in \cite{WanWangv2}.
\begin{itemize}
\item
It is straightforward to check the representation multiplet in \Eq{eq:SU5SU4-SU5Spin6-multiplet} , 
equivalently as
$( {\bf 5} \oplus \overline  {\bf 10} , {\bf 4})$ and 
$( {\bf 5} \oplus \overline  {\bf 10} , \overline{\bf 4})$,
is free from
the 4d perturbative local anomaly of 5d $\frac{1}{2}\text{CS}_5(V_{\SU(5)})$, see \Sec{sec:SU(5)3} for the exemplary calculation.
In fact, generally a ${\bf 5} \oplus \overline  {\bf 10}$ has no 4d perturbative local anomaly of 5d $\frac{1}{2}\text{CS}_5(V_{\SU(5)})$.

\item
It is straightforward to check the representation multiplet in \Eq{eq:SU5SU4-SU5Spin6-multiplet}, 
equivalently as
$( {\bf 5}     , {\bf 4} \oplus \overline{\bf 4})$ and 
$(  \overline  {\bf 10} , {\bf 4} \oplus \overline{\bf 4})$,
is also free from
the 4d perturbative local anomaly of 5d $\frac{1}{2}\text{CS}_{5,e}(V_{\SO(6)})$. For example, we can show that 
$( {\bf 5}     , {\bf 4} \oplus \overline{\bf 4})$ and 
$(  \overline  {\bf 10} , {\bf 4} \oplus \overline{\bf 4})$ of the ${\bf  2}_L$ left-handed Weyl fermions
can be regarded as
$( {\bf 5}     , {\bf 4}  ) \oplus (  \overline  {\bf 10} , {\bf 4}  )$ of the ${\bf  2}_L$ left-handed Weyl spinors 
and 
$( {\bf 5}     , {\bf 4}  ) \oplus (  \overline  {\bf 10} , {\bf 4}  )$ of the ${\bf  2}_R$ right-handed Weyl spinors of the Spin(3,1) Lorentz spinors.
Since the 
$$( {\bf 5}     , {\bf 4}  ) \oplus (  \overline  {\bf 10} , {\bf 4}  ) \text{ of } {\bf  2}_L \oplus {\bf  2}_R \text{ Weyl spinors }$$ 
have the same quantum number but the opposite chirality,
they do not have the 4d perturbative chiral anomaly of 5d $\frac{1}{2}\text{CS}_{5,e}(V_{\SO(6)})$.

\item We can also ask: How many, among the 16 copies of the 16th Weyl fermions in \Eq{eq:SU5SU4-SU5Spin6-neutrino-multiplet},
$\Big(  {\bf 1}   ,\; ( {\bf 1} \oplus {\bf 1} \oplus {\bf 4} \oplus  \overline{\bf 4}\oplus  {\bf 6}) \Big)$,
can be gapped out 
while still preserving the ${\Spin \times \SU(5) \times \Spin(6)}$ symmetry?\\

Clearly, there are two copies of $({\bf 1}, {\bf 1})$ neutral under all gauge forces. 
(The only possible anomalies for the gauge neutral matter $({\bf 1}, {\bf 1})$ are the gravitational anomalies, if any.)
It is easy to see that $({\bf 1}, {\bf 1})$ is free from
all the anomalies in \Eq{eq:TP5-SU5SU4-SU5Spin6}
and \Eq{eq:anomaly-SU5SU4-SU5Spin6}.
Therefore we can gap $({\bf 1}, {\bf 1})$ without breaking ${\Spin \times \SU(5) \times \Spin(6)}$ symmetry, for example, by adding it a Majorana mass.
However, as shown in \Ref{JW2006.16996},
 if we wish to preserve an extra $\Z_{4,X} \supset \Z_2^F$,
we cannot gap $({\bf 1}, {\bf 1})$ by adding a single Majorana mass which breaks $\Z_{4,X}$, 
thus there must be an additional anomaly.\\

We also confirm that
$({\bf 1}, {\bf 4} \oplus  \overline{\bf 4})=({\bf 1}, {\bf 4})  \oplus ({\bf 1}, \overline{\bf 4})$ 
of the ${\bf  2}_L$ left-handed Weyl spinors
can be written as $({\bf 1}, {\bf 4})$ 
of the ${\bf  2}_L \oplus {\bf  2}_R$,  left-handed and right-handed Weyl spinors,
thus they are free from the perturbative chiral anomaly of the ${\Spin \times \SU(5) \times \Spin(6)}$ symmetry in \Eq{eq:anomaly-SU5SU4-SU5Spin6}.
In short, the $({\bf 1}, {\bf 4} \oplus  \overline{\bf 4})$  can also be gapped 
out while still preserving the ${\Spin \times \SU(5) \times \Spin(6)}$ symmetry.
 
If we do not preserve an extra symmetry (such as the $\Z_{4,X}$) but only preserve the ${\Spin \times \SU(5) \times \Spin(6)}$, then the
$$
\Big(  {\bf 1}   ,\; ( {\bf 1} \oplus {\bf 1} \oplus {\bf 4} \oplus  \overline{\bf 4} ) \Big)
$$
can be fully gapped because they are free from all the anomalies in \Eq{eq:TP5-SU5SU4-SU5Spin6}
and \Eq{eq:anomaly-SU5SU4-SU5Spin6}.\\[-3mm]

The $((   {\bf 5} \oplus \overline {\bf 10} ), {\bf 6})$ or $({\bf 1}, {\bf 6})$ has the ${\bf 6}$ in the vector Rep of Spin(6) and SO(6) whose perturbative local anomaly is captured by
$\exp( \ii  \;  \int  \; \frac{1}{2}\text{CS}_{5,e}(V_{\SO(6)}))$ 
with a coefficient
$A({\bf R})=N-4=0$ for the matter field in the anti-symmetric Rep ${\bf R}$ of $\SU(N)$.
Since $A({\bf R})=0$ at $N=4$ is anomaly free, the ${\bf 6}$ \emph{can} be gapped while preserving the ${\Spin \times \Spin(6)}$ symmetry.
 
 \item On the other hand,  the  $((   {\bf 5} \oplus \overline {\bf 10} ), {\bf 4})$ or $({\bf 1}, {\bf 4})$ has the ${\bf 4}$ in the irreducible spinor Rep of Spin(6) and SO(6) whose perturbative local anomaly captured by
$\exp( \ii  \;  \int  \; \frac{1}{2}\text{CS}_{5,e}(V_{\SO(6)}))$ 
with a coefficient
$A({\bf R})=1$ for the matter in a fundamental Rep ${\bf R}$ of $\SU(N)$.
Since $A({\bf R})=1$ is anomalous, the ${\bf 4}$ alone (also $\overline{\bf 4}$ alone) \emph{cannot} be gapped while preserving the ${\Spin \times \Spin(6)}$ symmetry.

\end{itemize}
In summary, above we have shown that the extra matter multiplet in \Eq{eq:SU5SU4-SU5Spin6-multiplet} is free from all anomalies classified in
\Eq{eq:TP5-SU5SU4-SU5Spin6} and \Eq{eq:anomaly-SU5SU4-SU5Spin6}.   
 Thus we can gap ${(   {\bf 5}, {\bf 4}) \oplus (   {\bf 5} , \overline {\bf 4}) \oplus ( \overline  {\bf 10} , {\bf 4}) \oplus ( \overline  {\bf 10} , \overline{\bf 4})}
 =\Big((   {\bf 5} \oplus \overline{\bf 10} ),\; (  {\bf 4} \oplus  \overline {\bf 4}) \Big) 
$ by nonperturbative interactions, 
but we keep the \Eq{eq:SU5SU4-SU5Spin6-gapless-multiplet} 
gapless 
$$
\Big(( \overline  {\bf 5} \oplus {\bf 10} ),\; ( {\bf 1} \oplus {\bf 1} \oplus  {\bf 6}) \Big) 
$$
intact, while preserving the ${\Spin \times \SU(5) \times \Spin(6)}$ symmetry.
%
{If there is an energy gap scale for this gapping $\Big((   {\bf 5} \oplus \overline{\bf 10} ),\; (  {\bf 4} \oplus  \overline {\bf 4}) \Big)$ scenario, we can name it as:
\bea
\Delta_{\mathrm{KW}.su(5)\times so(6)},
\eea
which is around the breaking scale $M_{su(5)\times so(6)}$. This seems to be fine because we are left with the nearly
gapless $\Big(( \overline  {\bf 5} \oplus {\bf 10} ),\; ( {\bf 1} \oplus {\bf 1} \oplus  {\bf 6}) \Big)$ more than three generations of quarks and leptons in SM, which 
we can gap out some of them further if we break down Spin(6) to a smaller subgroup Spin(5) in the next \Sec{sec:SO(18)SU(9)SU(5)SO(5)GUTgap}.}

 .

\subsection{$so(18)$ GUT to $su(9)$ GUT and $su(5) \times so(5)$ GUT, $\Delta_{\mathrm{KW}.su(5) \times so(5)}$, and Three Generations}
\label{sec:SO(18)SU(9)SU(5)SO(5)GUTgap}

\item {\bf Break $so(18)$ GUT from $\Spin(18)$ to $\SU(5) \times \Spin(5)$ and their GUT}:\\

From another viewpoint, we consider
${\Spin \times \SU(9)} \to \Spin \times  \SU(5) \times \Spin(5)= \Spin \times  \SU(5) \times \U\Sp(4)= \Spin \times  \SU(5) \times \Sp(2)$,
we break the vector representation ${\bf 6}$ of Spin(6) to a vector and a trivial representation ${\bf 1} \oplus {\bf 5}$ in Spin(5).
We also reduce  the spinor representation ${\bf 4}$ and $ \overline {\bf 4}$ of Spin(6) to the same spinor representation ${\bf 4}$ in Spin(5).
As proposed by \cite{BenTov2015graZee1505.04312}, the 240 Weyl fermions from 
\Eq{eq:SU5SU4-SU5Spin6-gapless-multiplet} and \Eq{eq:SU5SU4-SU5Spin6-multiplet}
out of the ${\bf 256}^+$  have the representation of $\SU(5) \times \Spin(5)$
as 
\bea
&&\text{Representation of }  \SU(5) \times \Spin(5): \nn \\[2mm]
&&( \overline  {\bf 5} \oplus {\bf 10}, {\bf 1} \oplus {\bf 1} \oplus {\bf 1}) \oplus
( \overline  {\bf 5} \oplus {\bf 10},  {\bf 5})
\oplus
(   {\bf 5} \oplus \overline{\bf 10},  {\bf 4} \oplus {\bf 4})\cr
&&=\Big(( \overline  {\bf 5} \oplus {\bf 10} ),\; ( {\bf 1} \oplus {\bf 1} \oplus {\bf 1} \oplus  {\bf 5}  ) \Big)
\oplus
\Big((   {\bf 5} \oplus \overline {\bf 10} ),\;   ({\bf 4} \oplus {\bf 4}) \Big)
\label{eq:SU5Spin5-multiplet}
\eea
The $( \overline  {\bf 5} \oplus {\bf 10}, {\bf 1} \oplus {\bf 1} \oplus {\bf 1})$ forms precisely the fermion matter of
3 families of $su(5)$ GUT from the spacetime-internal symmetry structure $({\Spin \times \SU(5)})_{\text{3-Family}}$.

Traditional approaches use \emph{hypercolor} or \emph{technicolor} type of ideas \cite{Weinberg1975gmDynamical,Susskind1978msDynamical,Farhi1980xsSusskindDynamical1981,Hill2002apDynamical0203079} to conceal the additional matter; however, the dynamics of hypercolor and dynamical symmetry breaking is not fully understood.
%
\Ref{BenTov2015graZee1505.04312} proposed that we can gap the additional matter 
$( \overline  {\bf 5} \oplus {\bf 10},  {\bf 5})
\oplus
( \overline  {\bf 5} \oplus {\bf 10},  {\bf 4} \oplus {\bf 4})$
via 
Kitaev-Wen (KW) mechanism.
Here we can check explicitly by
a cobordism theory.
To establish this claim, we can check that the $\Spin \times  \SU(5) \times \Spin(5)$ chiral fermion theory 
of additional matter
are free from the anomaly \cite{WanWangv2}
\bea
&& \Omega_5^{\Spin \times \SU(5)\times \Spin(5)} = \Z_2, \quad \\ 
&&\TP_5({\Spin \times \SU(5) \times \Spin(5)})  = \Z \times  \Z_2. \label{eq:TP5-SU5Spin5}
\eea
In fact, the $\Z$ class is a 4d perturbative local anomaly, captured by a perturbative one-loop triangle Feynman diagram calculation
and by a 5d cobordism invariant $\frac{1}{2}\text{CS}_5(V_{\SU(5)})$.
The $\Z_2$ class is a 4d nonperturbative global anomaly, captured by a mod 2 class similar to the 4d Witten SU(2) $=\Spin(3) \subset \Spin(5)$ anomaly,
which requires an odd number of isospin-1/2 fermion 
${\bf 2}$ of SU(2) = Spin(3) to realize the anomaly.

\begin{itemize}
\item
We can easily see both $( \overline  {\bf 5} \oplus {\bf 10},  {\bf 5})$
and 
$(   {\bf 5} \oplus \overline{\bf 10},  {\bf 4} \oplus {\bf 4})$ is free from the 4d local anomaly of $\frac{1}{2}\text{CS}_5(V_{\SU(5)})$,
since we can check that the $\overline  {\bf 5} \oplus {\bf 10}$ and ${\bf 5} \oplus \overline{\bf 10}$ multiplets have the $\frac{1}{2}\text{CS}_5(V_{\SU(5)})$ anomaly cancelled,
see \Sec{sec:SU(5)3} for instance.

\item 
Next we check that the
$( \overline  {\bf 5} \oplus {\bf 10},  {\bf 5})$
and 
$(   {\bf 5} \oplus \overline{\bf 10},  {\bf 4} \oplus {\bf 4})$ are free from the $\Z_2$ class global anomaly of ${\Spin \times \SU(5) \times \Spin(5)}$ in \Eq{eq:TP5-SU5Spin5}.
We can do a branching rule to decompose the ${\bf 5}$ of Spin(5) as $({\bf 1}\oplus{\bf 1} \oplus{\bf 3})$ of ${\Spin(3) =  \SU(2)}$,
and we decompose the ${\bf 4}$ of Spin(5) as $({\bf 2} \oplus{\bf 2})$ of ${\Spin(3)}=  \SU(2)$.\footnote{See the footnote
\ref{ft:Spin5-122-113}, there could be other ways of decompositions, such as
the ${\bf 5}$ of Spin(5) as $({\bf 1}\oplus{\bf 2} \oplus{\bf 2})$ of ${\Spin(3)}$
and
the ${\bf 4}$ of Spin(5) as $({\bf 1} \oplus {\bf 1} \oplus{\bf 2})$ of ${\Spin(3)}$.
Thus the the ${\bf 4}\oplus  {\bf 4}$ of Spin(5) as $4{ (\bf 1})\oplus 2 ({\bf 2})$ of ${\Spin(3)}$.
Then we can still confirm that the
${\bf 5}$ and  ${\bf 4}\oplus  {\bf 4}$ of Spin(5) both have  
an even number of ${\bf 2}$ and no ${\bf 4}$ of SU(2) (thus their mod 2 classes are zeros), 
thus they are free from the old and the new SU(2) anomalies. 
The ${\bf 5}$ and  ${\bf 4}\oplus  {\bf 4}$ are thus free from the $\Z_2$ class global anomaly in \Eq{eq:TP5-SU5Spin5}.
\label{ft:Spin5-122-113-112}}
Then we can confirm that the
${\bf 5}$ and  ${\bf 4}\oplus  {\bf 4}$ of Spin(5) both have  
an even number of ${\bf 2}$ and no ${\bf 4}$ of SU(2) (so their mod 2 classes are zeros), 
thus they are free from the old \cite{Witten1982fp} and the new SU(2) anomalies \cite{WangWenWitten2018qoy1810.00844}. 
The ${\bf 5}$ and  ${\bf 4}\oplus  {\bf 4}$ of Spin(5) are thus free from the $\Z_2$ class global anomaly of ${\Spin \times \Spin(5)}$ in \Eq{eq:TP5-SU5Spin5}.
Therefore, we achieve the proof of the initial statement.

 \end{itemize}

In summary, above we have  established the KW analogous mechanism via establishing the all anomaly free conditions hold for
both $( \overline  {\bf 5} \oplus {\bf 10},  {\bf 5})$
and 
$(   {\bf 5} \oplus \overline{\bf 10},  {\bf 4} \oplus {\bf 4})$, free from all anomalies classified in
\Eq{eq:TP5-SU5Spin5}.  
Thus we can gap  $( \overline  {\bf 5} \oplus {\bf 10},  {\bf 5})$ and $(   {\bf 5} \oplus \overline{\bf 10},  {\bf 4} \oplus {\bf 4})$ by nonperturbative interactions. 

{Moreover, we can ask among the remaining 
16 Weyl fermions in ${\bf 256}^+$ analogous to 16 copies of a right-handed ``sterile'' neutrino from the \Eq{eq:SU5SU4-SU5Spin6-neutrino-multiplet},
now in a new $ \Spin(6) \to  \Spin(5)$ representation:
\bea
&&\text{Representation of }  \SU(5) \times \Spin(5): \nn \\[2mm]
&&
=\Big(  {\bf 1}   ,\; ( {\bf 1} \oplus {\bf 1} \oplus {\bf 1} \oplus {\bf 4} \oplus  {\bf 4}\oplus  {\bf 5}) \Big),
\label{eq:SU5Spin5-neutrino-multiplet}
\eea
whether part of the multiplet can be gapped without breaking $\Spin \times  \SU(5) \times \Spin(5)$?}
Recall that we had answered among $\Big(  {\bf 1}   ,\; ( {\bf 1} \oplus {\bf 1} \oplus {\bf 4} \oplus  \overline{\bf 4}\oplus  {\bf 6}) \Big)$
of ${\bf 2}_L$ Weyl fermion in the $\Spin \times  \SU(5) \times \Spin(6)$ in \Eq{eq:SU5SU4-SU5Spin6-neutrino-multiplet},
the $\Big(  {\bf 1}   ,\; ( {\bf 1} \oplus {\bf 1} \oplus {\bf 4} \oplus  \overline{\bf 4} ) \Big)$ can be gapped. 
%

After breaking $({\bf 1}, {\bf 6})$ of $\SU(5) \times \Spin(6)$ down to $({\bf 1}, {\bf 1}\oplus {\bf 5})$ of $\SU(5) \times \Spin(5)$, 
which we find that ${\bf 1}$ of $\SU(5)$ does not carry the perturbative $\Z$ class anomaly in \ref{eq:TP5-SU5Spin5}
the ${\bf 1}\oplus {\bf 5}$ of $\Spin(5)$ does not carry the nonperturbative $\Z_2$ class anomaly in \ref{eq:TP5-SU5Spin5} based on the derivation in 
footnote \ref{ft:Spin5-122-113} and \ref{ft:Spin5-122-113-112}.
In fact, based on the similar argument, we find that 
each of the $({\bf 1}, {\bf 1})$, $({\bf 1}, {\bf 5})$, and $({\bf 1}, {\bf 4} \oplus {\bf 4})$
 in \Eq{eq:SU5Spin5-neutrino-multiplet} can be gapped out while preserving $\Spin \times\SU(5) \times \Spin(5)$. 

{Moreover, we prefer to keep the remaining
\bea \label{eq:510-111}
\Big(( \overline  {\bf 5} \oplus {\bf 10} ),\; ( {\bf 1} \oplus {\bf 1} \oplus  {\bf 1}) \Big) 
\eea
gapless intact, while preserving the ${\Spin \times \SU(5) \times \Spin(5)}$ symmetry.
Gapping the extra matter
$( \overline  {\bf 5} \oplus {\bf 10},  {\bf 5})
\oplus
( \overline  {\bf 5} \oplus {\bf 10},  {\bf 4} \oplus {\bf 4})$ gives 
an energy scale 
\bea
\Delta_{\mathrm{KW}.su(5) \times so(5)}
\eea
that we show in \Fig{fig:energy-hierarchy-lie-algebra-1}.}

\end{enumerate}

\subsection{With an additional discrete $X=5({\bf B}- {\bf L})-4Y$ symmetry}
\label{sec:discreteX=5B-L-4Y}

\begin{figure*}[!h] 
\hspace{-22mm}
\includegraphics[width=8.4in]{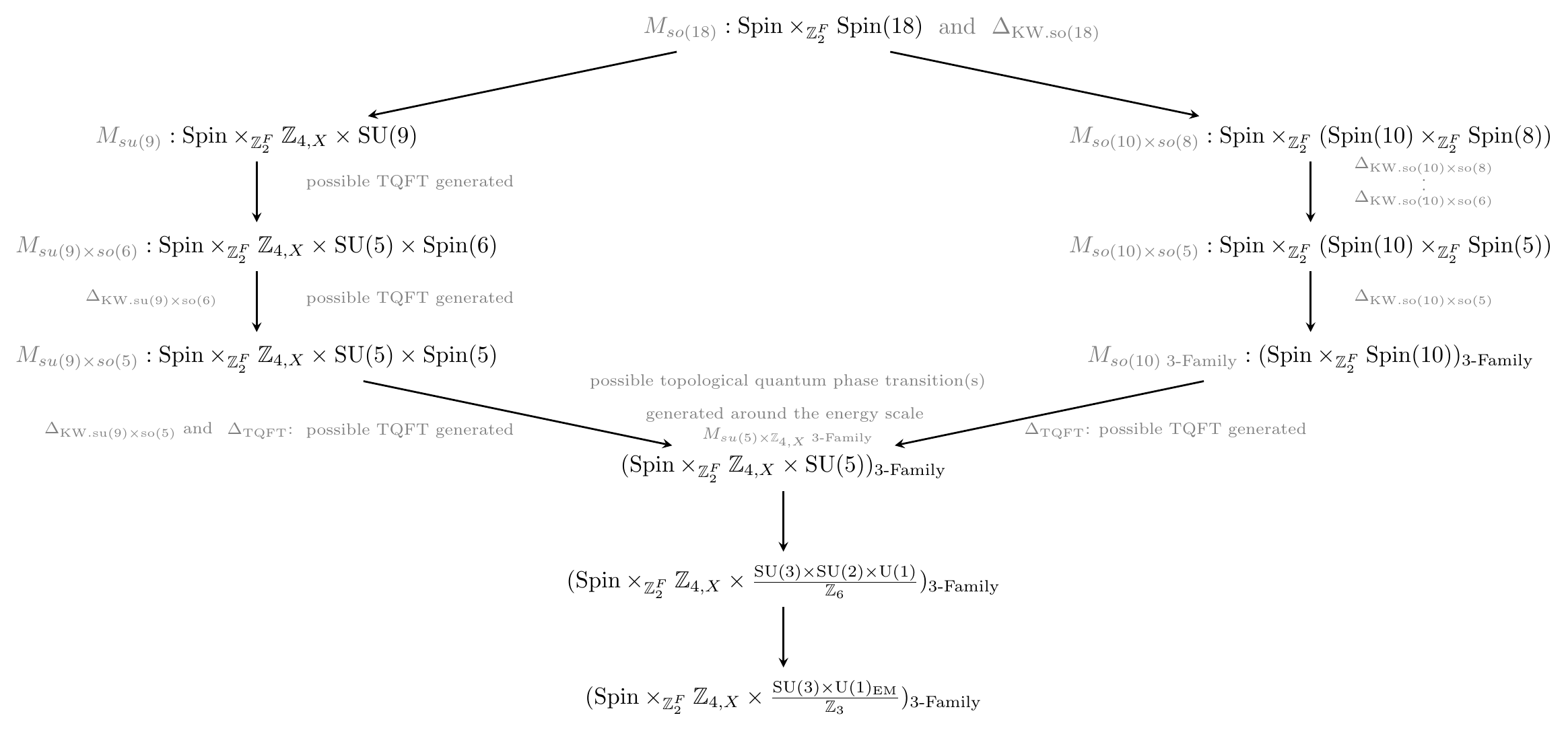}
\caption{The full spacetime-internal symmetry
$G={\frac{{G_{\text{spacetime} }} \ltimes  {{G}_{\text{internal}} }}{{N_{\text{shared}}}}}$ (the precise global symmetry before gauging the ${{G}_{\text{internal}} }$)
for the hierarchy starting from the $so(18)$ GUT with ${{\Spin \times_{\Z_2} \Spin(18)}}$, which can be placed on \emph{non-spin manifolds}.
The setup is similar to \Fig{table:sym-web-1}, 
but now we include the additional discrete symmetry sector
$\Z_{4,{X}} = Z(\Spin(10))  =Z(\Spin(18))$ sitting at the center $Z({{G}_{\text{internal}} })$ normal subgroup of
${{G}_{\text{internal}} }=\Spin(10)$ and $\Spin(18)$.
We follow the notations/explanations of \Fig{table:sym-web-1}'s caption.
 We compute the cobordism group 
 $\TP_d(G)$
 of these spacetime-internal symmetry group $G$ in \Ref{WanWangv2}. 
 {The arrow $G_1 \to G_2$ (with the condition $G_1 \supseteq G_2$) shows that a possible breaking process.
 We explore the two possible breaking patterns on the left-hand side (l.h.s) and right-hand side (r.h.s), with their possible energy hierarchy shown in  
 \Fig{fig:energy-hierarchy-lie-algebra-1} 
 and  \Fig{fig:energy-hierarchy-lie-algebra-2}. 
 Some of the arrows have a subtitle ``possible TQFT generated,'' which means that a noninvertible  TQFT may be generated to match 
 the 4d anomaly, especially from 
 the cobordism group $\TP_5({\Spin \times_{\Z_2^F} \Z_{4,X}}) =\Z_{16}$. 
 The l.h.s breaking pattern suggests (at least) three possible breaking steps to generate a possible TQFT. In particular, 
 thanks to mathematical and phenomenological constraints,  
the l.h.s step $\Spin \times_{\Z_2^F} \Z_{4,X} \times \SU(5) \times \Spin(5)  \to (\Spin \times_{\Z_2^F} \Z_{4,X} \times \SU(5))_{\text{3-Family}} $
and the r.h.s step  $(\Spin \times_{\Z_2^F} \Spin(10) )_{\text{3-Family}}   \to (\Spin \times_{\Z_2^F} \Z_{4,X} \times \SU(5))_{\text{3-Family}}$,
these two steps seem to be the most promising energy scale denoted $\Delta_{\text{TQFT}}$ to generate a TQFT with a gap size $\Delta_{\text{TQFT}}$. 
A possible interpretation of topological quantum phase transition(s) around this energy scale $M_{su(5) \times \Z_{4,X}\;\text{3-Family}}$ is given in \Table{table:scenarios}.
We also enlist other sequences of possible energy scales analogous to Kitaev-Wen (KW) mechanism, 
gapping the fully anomaly-free extra matter. 
We denote these KW-type energy scales as  $\Delta_{\text{KW}}$. See
 \Fig{fig:energy-hierarchy-lie-algebra-1} 
 and  \Fig{fig:energy-hierarchy-lie-algebra-2}. }}
 \label{table:sym-web-so18-Z4}
\end{figure*}

There is an embedding from the $so(10)$ GUT to Georgi-Glashow $su(5)$ GUT with a discrete $\Z_{4,X}$ of $X=5({\bf B}- {\bf L})-4Y$ symmetry (such that $\Z_{4,{X}} = Z(\Spin(10))$), as follows \cite{GarciaEtxebarriaMontero2018ajm1808.00009, WW2019fxh1910.14668, JW2006.16996},
\bea 
 \label{eq:SMembed3}
{\frac{\Spin(d) \times
\Spin(10)}{{\Z_2^F}} 
\supset 
\Spin(d) \times_{\Z_2} \Z_4 \times \SU(5) 
\supset 
\Spin(d) \times_{\Z_2} \Z_4 \times \frac{\SU(3) \times   \SU(2) \times \U(1)}{\Z_6} }.
\eea
Here we can generalize the above embedding from the $so(18)$ GUT to $su(9)$ GUT with a discrete $\Z_{4,X}$ of $X=5({\bf B}- {\bf L})-4Y$ symmetry
(such that $\Z_{4,{X}} = Z(\Spin(10))  =Z(\Spin(18))$) as follows, 
\begin{multline} 
 \label{eq:SMembedSpin18Z4}
\frac{\Spin(d) \times
\Spin(18)}{{\Z_2^F}} 
\supset 
\Spin(d) \times_{\Z_2} \Z_4 \times \SU(9) 
\supset \Spin(d) \times_{\Z_2} \Z_4 \times \SU(5) \times \Spin(6)  \supset\\
\Spin(d) \times_{\Z_2} \Z_4 \times \SU(5) \times \Spin(5) 
\supset
\Spin(d) \times_{\Z_2} \Z_4 \times \SU(5)
\supset
\Spin(d) \times_{\Z_2} \Z_4 \times \frac{\SU(3) \times   \SU(2) \times \U(1)}{\Z_6} .
\end{multline} 
By \Eq{eq:SMembedSpin18Z4}, we can modify \Fig{table:sym-web-1}'s spacetime-internal symmetry group embedding web to include the
discrete $\Z_{4,{X}}$ sector. We obtain  \Fig{table:sym-web-so18-Z4}.

We compute the full list of cobordism group $\TP_d(G)$ of these spacetime-internal symmetry group $G$ of \Fig{table:sym-web-so18-Z4} in \Ref{WanWangv2}. 
A crucial fact is that now many of these $G$  of \Fig{table:sym-web-so18-Z4} suitable for $su(5)$ GUT
contain also ${\Spin \times_{\Z_2^F} \Z_{4,X}}$, 
we know that part of their bordism group $\Omega_5$ 
and their cobordism group $\TP_5$: 
\bea \label{eq:Z16anomaly-cobordism} 
\Omega_5^{\Spin \times_{\Z_2^F} \Z_{4,X}} =\Z_{16}, \quad\quad\quad  \TP_5({\Spin \times_{\Z_2^F} \Z_{4,X}}) =\Z_{16}.
\eea
We can check this \Eq{eq:Z16anomaly-cobordism} implies a constraint from the $\Z_{16}$ class 4d nonperturbative global anomaly \Eq{eq:Z16anomaly} 
\cite{GarciaEtxebarriaMontero2018ajm1808.00009, JW2006.16996}. 
This implies that several previously discussed KW-type mechanisms in \Sec{sec:SO(18)SU(9)SU(5)SU(4)GUTgap} - \Sec{sec:SO(18)SU(9)SU(5)SO(5)GUTgap} may not work unless we have some multiple of 16 Weyl spinors ${\bf 2}_{L}$ of Lorentz spacetime. For example,
if we aim to gap the extra matter
${\bf 5} \oplus \overline{\bf 10}$ of SU(5)
in \Eq{eq:SU5SU4-SU5Spin6-multiplet}-\Eq{eq:SU5Spin5-multiplet}, or the 
${\bf 1}$ of SU(5) in \Eq{eq:SU5SU4-SU5Spin6-neutrino-multiplet}-\Eq{eq:SU5Spin5-neutrino-multiplet} under the extra $\Z_{4,X}$ symmetry while still matching the $\Z_{16}$ anomaly,
we need to choose either one of the following ways:
 \begin{enumerate}
\item Combine ${\bf 5} \oplus \overline{\bf 10}$ with ${\bf 1}$ of SU(5) to form a 16 Weyl Lorentz spinors ${\bf 2}_{L}$, 
in order to let the combined ${\bf 5} \oplus \overline{\bf 10} \oplus {\bf 1}$ of SU(5) be free from the $\Z_{16}$ anomaly. Then we can apply the KW mechanism on the anomaly free sector of 16 Weyl spinors. 

\item If we want to gap only ${\bf 5} \oplus \overline{\bf 10}$ of SU(5)
in \Eq{eq:SU5SU4-SU5Spin6-multiplet}-\Eq{eq:SU5Spin5-multiplet} alone, then we need to go beyond the KW mechanism. 
We can seek for the anomalous symmetric  3+1d TQFT construction to match the anomaly $\nu=15 = -1 \in \Z_{16}$, such as using the symmetry extension approach \cite{Wang2017locWWW1705.06728}.
This 3+1d TQFT is a generalization of 2+1d anomalous symmetric surface topological order (see a review \cite{Senthil1405.4015}) to the 3+1d case.

\item If we want to gap only ${\bf 1}$ of SU(5) in \Eq{eq:SU5SU4-SU5Spin6-neutrino-multiplet}-\Eq{eq:SU5Spin5-neutrino-multiplet} alone,
 then we need to go beyond the KW mechanism. 
We can seek for the anomalous symmetric 3+1d TQFT construction to match the anomaly 
 $\nu=+1 \in \Z_{16}$, such as using the symmetry extension approach \cite{Wang2017locWWW1705.06728}.
This 3+1d TQFT is a generalization of 2+1d anomalous symmetric surface topological order (see a review \cite{Senthil1405.4015}) to the 3+1d case.
\end{enumerate}

Therefore, other than KW-type energy scales  $\Delta_{\text{KW}}$, we may have
another energy scale $\Delta_{\text{TQFT}}$, for the anomalous symmetric 3+1d TQFT (the energy gap of 3+1d  topological order, there are fractionalized excitations such as anyonic strings above the gap, e.g. see \cite{WangLevin1403.7437, Jiang1404.1062, Wang1404.7854, 1602.05951WWY,Putrov2016qdo1612.09298PWY,Wang2019diz1901.11537} and References therein).
We show several candidate  $\Delta_{\text{KW}}$ and  $\Delta_{\text{TQFT}}$ energy scales in \Fig{table:sym-web-so18-Z4}, 
in companion with \Fig{fig:energy-hierarchy-lie-algebra-1} 
 and  \Fig{fig:energy-hierarchy-lie-algebra-2}.

\subsection{Kinematics vs Dynamics}

We should remind ourselves that the anomaly can be determined from the \emph{kinematics} of QFT --- namely, given the action, partition function, or path integral of QFT,
we could already determine whether the anomaly occurs. For example, for perturbative local anomalies, see Fig.~1  of \Ref{JW2006.16996}:\\[-8mm]
 \begin{itemize}[leftmargin=2.mm]
\item 
When
$G=({\frac{{G_{\text{spacetime} }} \ltimes  {{G}_{\text{internal}} }}{{N_{\text{shared}}}}})$
is treated as a global symmetry, then we determine the $(d-1)$d 't Hooft anomalies of such a global symmetry $G$ by turning on all possible background fields
via the cobordism group $\TP_d(G)$ in \Eq{eq:TPG}, e.g., Fig.~1 ({\it 2}) of \Ref{JW2006.16996}.
\item When ${{G}_{\text{internal}} }$ is dynamically gauged, part of the anomalies of the cobordism group $\TP_d(G)$ in \Eq{eq:TPG}
become 
dynamical gauge anomalies, e.g., Fig.~1 ({\it 1}) of \Ref{JW2006.16996},
while part of the anomalies
become to be interpreted as the ABJ type anomalies, e.g., Fig.~1 ({\it 3}) of \Ref{JW2006.16996},
\end{itemize}
In this subsection, we organize previous statements on the 
constraints from anomalies and cobordisms in \Sec{sec:so(18)GUTnofermion doubling} - \Sec{sec:discreteX=5B-L-4Y} into a list. 
A priori based on our anomaly and cobordism analysis alone, we \emph{cannot} fully determine the gauge dynamics, 
but we \emph{can} suggest possible dynamics.
We shall comment on how the anomalies
obtained from the \emph{kinematics} of QFT
can constrain the \emph{dynamics} of QFT at different energy scales afterward in \Sec{sec:Conclusion}:
\begin{enumerate}[leftmargin=4.mm, label=\textcolor{blue}{(\arabic*)}., ref={(\arabic*)}]
\item \label{list:Spin18}
When $G={\Spin \times_{\Z_2^F} {\Spin(18)}}$,
the following 
Weyl fermion matter field ${\bf 2}_L$ in the ${\Spin(18)}$ representation is fully $G$-anomaly-free thus
 can be gapped by KW mechanism while preserving $G$ (without breaking $G$) by generic nonperturbative interactions:
 \begin{itemize}
\item the ${\bf 256}^-$ in \Eq{eq:256-}  can be gapped so without fermion doubling suggested in  \cite{WangWen2018cai1809.11171}.
\item the ${\bf 256}^+$ can be gapped, but we keep ${\bf 256}^+$ nearly gapless to match the lower energy hierarchy and SM phenomenology.
 
\end{itemize}

\item When $G={\Spin \times_{\Z_2^F} ({\Spin(10) \times_{\Z_2^F} \Spin(8)})}$,
the following 
Weyl fermion matter field ${\bf 2}_L$ in the $({\Spin(10) \times_{\Z_2^F} \Spin(8)})$ representation
 is fully $G$-anomaly-free thus
 can be gapped by KW mechanism while preserving $G$ (without breaking $G$) by generic nonperturbative interactions:
  \begin{itemize}
\item the $({\bf 16}^+, {\bf 8}^-)$ and  $({\bf 16}^-, {\bf 8}^+)$   in \Eq{eq:so10so8R}, but they can be already gapped out as ${\bf 256}^-$ in 
${\Spin \times_{\Z_2^F} {\Spin(18)}}$ in \ref{list:Spin18} at a higher energy.
\item  the $({\bf 16}^-, {\bf 8}^-)$ in \Eq{eq:so10so8L}.
\item  the $({\bf 16}^+, {\bf 8}^+)$ in \Eq{eq:so10so8L} can be gapped, but we keep it 
nearly gapless to match the lower energy hierarchy and SM phenomenology.

\end{itemize}

\item When $G={\Spin \times_{\Z_2^F} ({\Spin(10) \times_{\Z_2^F} \Spin(6)})}$,
the following 
Weyl fermion matter field ${\bf 2}_L$ in the $({\Spin(10) \times_{\Z_2^F} \Spin(6)})$ representation
is fully $G$-anomaly-free thus
 can be gapped by KW mechanism while preserving $G$ (without breaking $G$) by generic nonperturbative interactions:
\begin{itemize}
\item $({\bf 16}^+, {\bf 6})$ in \Eq{eq:so10so6-16-116}.
\item $({\bf 16}^+, {\bf 1} \oplus {\bf 1})$ in \Eq{eq:so10so6-16-116}  can be gapped, but we keep it 
nearly gapless to match the lower energy hierarchy and SM phenomenology. The disadvantage is that there are only two generations of SM particles 
in $({\bf 16}^+, {\bf 1} \oplus {\bf 1})$.

\end{itemize}

\item When $G={\Spin \times_{\Z_2^F} ({\Spin(10) \times_{\Z_2^F} \Spin(5)})}$, the following 
Weyl fermion matter field ${\bf 2}_L$ in the $({\Spin(10) \times_{\Z_2^F} \Spin(5)})$ representation
is fully $G$-anomaly-free thus
 can be gapped by KW mechanism while preserving $G$ (without breaking $G$) by generic nonperturbative interactions:
\begin{itemize}
\item the 
  $({\bf 16}^+, {\bf 5})$ in \Eq{eq:so10so5-16-1115}.
\item the 
  $({\bf 16}^+, {\bf 1}\oplus{\bf 1}\oplus{\bf 1})$  in \Eq{eq:so10so5-16-1115}, but we keep it 
nearly gapless to match the lower energy hierarchy and SM phenomenology. Its advantage is that there are three generations of SM particles. 

\end{itemize}

\item \label{SpinSU(5)Spin(6)}
When $G=\Spin \times  \SU(5) \times \Spin(6)$,
the following Weyl fermion matter field ${\bf 2}_L$ in the $\SU(5) \times \Spin(6)$ representation
individually is fully $G$-anomaly-free thus
can be gapped by KW mechanism while preserving $G$ (without breaking $G$) by generic nonperturbative interactions:
\begin{itemize}
\item the $((   {\bf 5} \oplus \overline {\bf 10} ),\; ( {\bf 4} \oplus  \overline{\bf 4}) )$ in \Eq{eq:SU5SU4-SU5Spin6-multiplet}.
\item the $(  {\bf 1}   ,\; ( {\bf 1} \oplus {\bf 1} \oplus {\bf 4} \oplus  \overline{\bf 4} ) )$ in \Eq{eq:SU5SU4-SU5Spin6-neutrino-multiplet}.
\item the $(( \overline  {\bf 5} \oplus {\bf 10} ),\; ( {\bf 1} \oplus {\bf 1}  ) )$ in  \Eq{eq:SU5SU4-SU5Spin6-gapless-multiplet}, but we can keep it nearly gapless to match 
SM phenomenology.
\item the $(( \overline  {\bf 5} \oplus {\bf 10} ),\;  {\bf 6} )$ in  \Eq{eq:SU5SU4-SU5Spin6-gapless-multiplet}.
\item the $(  {\bf 1}   ,\;  {\bf 6}  )$ in \Eq{eq:SU5SU4-SU5Spin6-neutrino-multiplet}.
\end{itemize}
The following Weyl fermion matter field ${\bf 2}_L$ in the $\SU(5) \times \Spin(6)$ representation by itself individually is \emph{not} $G$-anomaly-free thus
\emph{cannot} be gapped by KW mechanism alone while preserving $G$ due to some non-vanishing anomaly:
\begin{itemize}
\item the $((   {\bf 5} \oplus \overline {\bf 10} ),\;  {\bf 4}   )$ alone or
$((   {\bf 5} \oplus \overline {\bf 10} ),\;    \overline{\bf 4} )$ alone in \Eq{eq:SU5SU4-SU5Spin6-neutrino-multiplet}.
\item the $({\bf 1},\;  {\bf 4}   )$ alone or
$({\bf 1},\;    \overline{\bf 4} )$ alone in \Eq{eq:SU5SU4-SU5Spin6-multiplet}.
\end{itemize}

\item  \label{SpinSU(5)Spin(5)}
When $G=\Spin \times  \SU(5) \times \Spin(5)$,
the following 
Weyl fermion matter field ${\bf 2}_L$ in the $\SU(5) \times \Spin(5)$ representation
 is fully $G$-anomaly-free thus
 can be gapped by KW mechanism while preserving $G$ (without breaking $G$) by generic nonperturbative interactions:
\begin{itemize}
\item the $( \overline  {\bf 5} \oplus {\bf 10},  {\bf 5})$ in \Eq{eq:SU5Spin5-multiplet}.
\item the $(   {\bf 5} \oplus \overline{\bf 10},  {\bf 4} \oplus {\bf 4})$  in \Eq{eq:SU5Spin5-multiplet}.
\item the $( \overline  {\bf 5} \oplus {\bf 10},  {\bf 1} \oplus {\bf 1} \oplus {\bf 1})$ in \Eq{eq:SU5Spin5-multiplet}, but we prefer to keep it nearly gapless to match 
SM phenomenology.
\item the $( {\bf 1}, {\bf 1})$ and any number (e.g.,$1,2,3$) of copies of it in $( {\bf 1}, {\bf 1}\oplus {\bf 1} \oplus{\bf 1})$ in  \Eq{eq:SU5Spin5-neutrino-multiplet}.
This $( {\bf 1}, {\bf 1})$ degree of freedom relates to the right-hand sterile neutrino. The conventional phenomenology suggests to use 
(1) Dirac mass or (2) Majorana mass and seesaw mechanism to gap this $( {\bf 1}, {\bf 1})$. We will also consider (3) Topological Mass from TQFT to gap this $( {\bf 1}, {\bf 1})$.
\item the $( {\bf 1}, {\bf 5})$ in  \Eq{eq:SU5Spin5-neutrino-multiplet} proposed in \cite{JW2006.16996}.
\item the 
$( {\bf 1}, {\bf 4}\oplus {\bf 4})$ in  \Eq{eq:SU5Spin5-neutrino-multiplet}.

\end{itemize}

\item When $G= \Spin \times_{\Z_2^F} \Z_{4,X} \times \SU(5) \times \Spin(6)$ with an additional discrete $\Z_{4,X}$ in contrast to Scenario \ref{SpinSU(5)Spin(6)}, 
due to a $\Z_{16}$ class 4d nonperturbative global anomaly from \Eq{eq:Z16anomaly-cobordism}, 
we are no longer allowed to gap $((   {\bf 5} \oplus \overline {\bf 10} ),\; ( {\bf 4} \oplus  \overline{\bf 4}) )$ or gap
$({\bf 1},\; ( {\bf 4} \oplus  \overline{\bf 4}) )$
individually alone in Scenario \ref{SpinSU(5)Spin(6)}, because they have only 15n Weyl fermions and 1n Weyl fermions, instead of 16n Weyl fermions.
However, we are allowed to gap their combination 
$((   {\bf 5} \oplus \overline {\bf 10} ) \oplus {\bf 1},\; ( {\bf 4} \oplus  \overline{\bf 4}) )$ with 16n Weyl fermions.
We can no longer take the gapping conditions in Scenario \ref{SpinSU(5)Spin(6)}, but we can modify them ---
The following 16n Weyl fermion matter field ${\bf 2}_L$ in the $\Spin \times_{\Z_2^F} \Z_{4,X} \times \SU(5) \times \Spin(6)$ 
representation\footnote{All the Weyl fermions carry an odd $\Z_{4,X}$ charge, see Table 1 in  \Ref{JW2006.16996}.}
is fully $G$-anomaly-free thus
can be gapped by KW mechanism while preserving $G$ (without breaking $G$) by generic nonperturbative interactions:
\begin{itemize}
\item the $((   {\bf 5} \oplus \overline {\bf 10} ) \oplus {\bf 1},\; ( {\bf 4} \oplus  \overline{\bf 4}) )$.
\item the $(( \overline  {\bf 5} \oplus {\bf 10} )\oplus {\bf 1},\; {\bf 1}  )$ and their multiple copies, but we can keep them nearly gapless to match 
SM phenomenology.
\item the $(( \overline  {\bf 5} \oplus {\bf 10} ) \oplus {\bf 1},\;  {\bf 6} )$.
\end{itemize}
The following 16n
Weyl fermion matter field ${\bf 2}_L$ in the  $\Spin \times_{\Z_2^F} \Z_{4,X} \times \SU(5) \times \Spin(6)$ representation
by itself individually is anomaly free from the $\Z_{16}$ anomaly of \Eq{eq:Z16anomaly-cobordism},
but \emph{not} fully $G$-anomaly-free, thus
\emph{cannot} be gapped by KW mechanism alone while preserving $G$ due to some non-vanishing anomaly:
\begin{itemize}
\item the $((   {\bf 5} \oplus \overline {\bf 10} ) \oplus {\bf 1},\;  {\bf 4}   )$ alone or
$((   {\bf 5} \oplus \overline {\bf 10} ) \oplus {\bf 1},\;    \overline{\bf 4} )$ alone.
\end{itemize}

\item When $G= \Spin \times_{\Z_2^F} \Z_{4,X} \times \SU(5) \times \Spin(5)$  with an additional discrete $\Z_{4,X}$ in contrast to \ref{SpinSU(5)Spin(5)},
due to a $\Z_{16}$ class 4d nonperturbative global anomaly from \Eq{eq:Z16anomaly-cobordism}, 
we can no longer take the gapping conditions in Scenario \ref{SpinSU(5)Spin(5)}, but we can modify them ---
The following 16n Weyl fermion matter field ${\bf 2}_L$ in the $\Spin \times_{\Z_2^F} \Z_{4,X} \times \SU(5) \times \Spin(5)$ 
representation with an odd $\Z_{4,X}$ charge is fully $G$-anomaly-free thus
can be gapped by KW mechanism while preserving $G$ (without breaking $G$) by generic nonperturbative interactions:
\begin{itemize}
\item the $( (\overline  {\bf 5} \oplus {\bf 10}) \oplus {\bf 1},  {\bf 5})$.
\item the $(  ( {\bf 5} \oplus \overline{\bf 10}) \oplus {\bf 1},  {\bf 4} \oplus {\bf 4})$.
\item the $( (\overline  {\bf 5} \oplus {\bf 10}) \oplus {\bf 1},  {\bf 1} )$ and their multiple copies, but we may prefer to keep part of it nearly gapless to match 
SM phenomenology.
\end{itemize}

\end{enumerate}

The above we have summarized what degrees of freedom are fully $G$-anomaly free for various given $G$ at different energy scales
(in the hierarchy of \Fig{fig:energy-hierarchy-lie-algebra-1} and  \Fig{fig:energy-hierarchy-lie-algebra-2}).
Based on the modern understanding \cite{WangWen2018cai1809.11171, NSeiberg-Strings-2019-talk}, 
the $G$-anomaly free degrees of freedom can be fully gapped while still preserving the $G$-symmetry, for example
by adding nonperturbative $G$-symmetric interactions to gapless modes.   
In fact, the anomaly derived from the \emph{\bf kinematics of QFT}
only suggest many possible fates of  \emph{\bf dynamics of QFT} at long distances.

The desirable task is: Could we reveal more information and eliminate some of possibilities to constrain more on the \emph{dynamics of QFT} at different energy scales?
We address this in the \Sec{sec:Conclusion}.

\section{Conclusion: 
}
\label{sec:Conclusion}

\subsection{Energy hierarchy, possible dynamics, and topological quantum phase transitions}

We had mentioned there are some primary goals and questions in this work:

\begin{enumerate}[leftmargin=4.mm, label=\textcolor{blue}{$\bullet$\arabic*)}., ref={$\bullet$\arabic*)}]

\item \label{goal1}
Study the anomalies systematically from the classification by the cobordism group --- 
we take into account the perturbative local and nonperturbative global anomalies. 
Check the full theory of various GUT models can be consistent
under (i) dynamical gauge anomaly-free conditions and (ii) 't Hooft anomaly-matching conditions.

\item \label{goal2} 
Given a spacetime-internal symmetry group $G$ and 
(a subset or the full set of) matter fields in some representations of ${\bf R}$, we can ask are there anomalies associated with this
set of matter fields? We especially consider the three separate cases for (i) the chiral matter associated with SM, (ii) extra matter,  and (iii) mirror matter 
in \Sec{sec:EnergyandMassHierarchy}

\item \label{goal3} 
Are there {\bf non-perturbative constraints} from anomalies and cobordism, given the {\bf low energy physics at SM}, guiding us toward 
discovering something {\bf heavy at higher energy}? (We especially ask this question under the Consideration \ref{goal2}
(i) chiral matter, (ii) extra matter, and (iii) mirror matter. If they have anomalies (or not), how could they manifest their dynamics at different energy scales?)

\end{enumerate}

Considerations \ref{goal1} and \ref{goal2} are mostly answer in the earlier sections (and also in Appendices).
Now we focus on Consideration \ref{goal3}.
Given our previous results in \Sec{sec:GaugeGroupHierarchy} and \Sec{sec:EnergyandMassHierarchy}, 
indeed we could attempt to address this Consideration \ref{goal3},
if we take these additional 
phenomenology and math/theoretical 
inputs into account: 
\begin{enumerate}[leftmargin=4.mm, label=\textcolor{blue}{\arabic*)}., ref={\arabic*)}]
\item \label{pheno1}
\emph{Phenomenology inputs of 15n Weyl fermions}: From Standard Model physics, we already 
know that there are nearly gapless degrees of freedom of 15n Weyl fermions (15n of Lorentz spinor ${\bf 2}_L$ with n $=3$ for 3 generations) 
whose masses are  smaller than the electroweak scale $v \sim 246$ G$e$V, while we are exploring around or above the $su(5)$ GUT and other GUT scales 
(conventionally for the gauge coupling unification $\sim 10^{16}$ G$e$V  in \Fig{fig:energy-hierarchy-lie-algebra-1} and  \Fig{fig:energy-hierarchy-lie-algebra-2}).
\item \label{pheno2}
\emph{Phenomenology inputs of 16th Weyl fermion and right-handed neutrinos}: We do not or have not yet observed the 16th Weyl fermions in any of 3 generations, which is commonly referred to be
the sterile right-handed neutrinos. As summarized in \Ref{JW2006.16996}, since the 16th Weyl fermions are not observed at the SM or T$e$V energy scales,
we shall give them a higher energy gap by:
\begin{enumerate}[leftmargin=4.mm, label=\textcolor{blue}{(\arabic*)}., ref={(\arabic*)}]
\item 
{\bf Dirac mass} (and the seesaw mechanism). 
\item 
{\bf  Majorana mass} (and the seesaw mechanism).
\item 
\label{topomass1}
{\bf  Topological Mass} from the excitation energy gap of a 4d noninvertible TQFT: In this way, the 16th Weyl fermion(s) would be missing from the vacua of our 
Universe --- the TQFT degrees of freedom cannot be described by any particle-like QFT or 
a perturbative (nearly free) quadratic QFT description conventionally used in particle physics.
The 16th Weyl fermion degrees of freedom would be smeared out to a long-range entangled 4d topological order (whose low energy is the 4d TQFT).
For example, we may apply the method of {\bf symmetry-extension} or {\bf higher-symmetry-extension}  in \cite{Wang2017locWWW1705.06728, Wan2018djlW2.1812.11955}. 
See Sec.~5 of \Ref{JW2006.16996} for details. 
\item 
\label{topomass2}
{\bf  Topological Mass} in an extra dimension from a 5d invertible TQFT: This approach is valid when there is an 4d invertible anomaly 
associated with the missing 16th Weyl fermion(s). In that case, we can do the anomaly-matching by introducing a gapped
5d invertible TQFT (or 5d SPTs in condensed matter) to cancel the missed anomaly. See Sec.~5 of \Ref{JW2006.16996} for details.

\end{enumerate}

\item \label{pheno3}
\emph{Phenomenology inputs of mirror fermions and extra matter}:
We do not observe any mirror fermions and extra matter so we better introduce ways to gap them.
If the extra matter carries no $G$-anomaly, we can gap them while preserving $G$ via: 
\begin{enumerate}[leftmargin=4.mm, label=\textcolor{blue}{(\arabic*)}., ref={(\arabic*)}]
\setcounter{enumii}{4}
\item
{\bf Kitaev-Wen (KW) mechanism} or an {\bf  anomaly-free symmetric mass/energy gap}. The KW mechanism is also used in the chiral fermion or chiral gauge theory 
problem \cite{Wen2013ppa1305.1045, Wang2013ytaJW1307.7480, You2014oaaYouBenTovXu1402.4151, YX14124784, Kikukawa2017gvk1710.11101, 
Wang2018ugfJW1807.05998, WangWen2018cai1809.11171, BenTov2015graZee1505.04312}).

\end{enumerate}
If the extra matter carries some 't Hooft anomaly in $G$, we may attempt to gap them via the aforementioned Topological Mass from \ref{topomass1} and \ref{topomass2} while 
still preserving $G$. 
%
%
%
%
\item \label{input4-so10so5}
From the $so(18)$ GUT, breaking the Spin(18)
via the r.h.s route in \Fig{table:sym-web-so18-Z4},
let us do a comparison of 
\Sec{sec:SO(10)SO(8)SO(6)GUTgap}
and
\Sec{sec:SO(10)SO(8)SO(5)GUTgap}.
We may ask whether
breaking to ${\Spin(10) \times_{\Z_2^F} \Spin(6)} $ or ${\Spin(10) \times_{\Z_2^F} \Spin(5)} $ is favored 
dynamically and at which range of energy scales?
(Here and below we assume and apply the old wisdom  \cite{WilczekZee1981iz1982Spinors,Fujimoto1981SO18Unification, BenTov2015graZee1505.04312} 
that dynamical symmetry breaking may make these Scenarios \ref{SO(10)SO(8)SO(6)GUTgap} and  \ref{SO(10)SO(8)SO(5)GUTgap} happened.)
Sine ${\Spin(10) \times_{\Z_2^F} \Spin(6)}$  can have either 2 generations of 16 Weyl fermions or extra matter in \Eq{eq:so10so6-16-116}, 
while ${\Spin(10) \times_{\Z_2^F} \Spin(5)}$  can have exactly the 3 generations of 16 Weyl fermions $({\bf 16}^+, {\bf 1}\oplus{\bf 1}\oplus{\bf 1})$ 
in \Eq{eq:so10so5-16-111}, 
we expect that eventually the ${\Spin(10) \times_{\Z_2^F} \Spin(5)}$ is a more viable option at a wider energy sale,
before we encounter $({\Spin \times_{\Z_2^F} {\Spin(10)}})_{\text{3-Family}}$.
Below Spin(18), it is possible to firstly encounter ${\Spin(10) \times_{\Z_2^F} \Spin(6)}$ at a higher energy, but it shall be broken down to ${\Spin(10) \times_{\Z_2^F} \Spin(5)}$
before eventually we encounter the energy scale of 3 generations of 16 Weyl fermions at $({\Spin \times_{\Z_2^F} {\Spin(10)}})_{\text{3-Family}}$.
This gives a partial reasoning for the energy hierarchy presented on the r.h.s in \Fig{table:sym-web-so18-Z4}.

\item  \label{input5-su5so5}
From the $so(18)$ GUT, breaking the Spin(18)
via the l.h.s route in \Fig{table:sym-web-so18-Z4},
let us do a comparison of 
\Sec{sec:SO(18)SU(9)SU(5)SU(4)GUTgap},
\Sec{sec:SO(18)SU(9)SU(5)SO(5)GUTgap},
and
\Sec{sec:discreteX=5B-L-4Y}.
We may ask whether
breaking to $\SU(5) \times \Spin(6)$ or $\SU(5) \times \Spin(5)$ is favored 
dynamically and at which range of energy scales?
Sine $\SU(5) \times \Spin(6)$  can have either 2 generations of 15 (or $+1$) Weyl fermions or extra matter in \Eq{eq:SU5SU4-SU5Spin6-gapless-multiplet}, 
while ${\SU(5) \times \Spin(5)}$  can have exactly the 3 generations of 15 (or $+1$) Weyl fermions $(( \overline  {\bf 5} \oplus {\bf 10} ),\; ( {\bf 1} \oplus {\bf 1} \oplus  {\bf 1}))$ 
in \Eq{eq:510-111}, 
we expect that eventually the ${\SU(5) \times \Spin(5)}$ is a more viable option for a wider energy sale,
before we encounter $({\SU(5)})_{\text{3-Family}}$.

Moreover, in \Sec{sec:discreteX=5B-L-4Y}, by taking into account the 
extra discrete $\Z_{4,X}$ of $X=5({\bf B}- {\bf L})-4Y$ symmetry with the enriched spacetime structure
${\Spin  \times_{\Z_2^F} \Z_{4,X}}$ in \Eq{eq:SMembedSpin18Z4}, we have to also match the extra $\Z_{16}$ anomaly 
in \Eq{eq:Z16anomaly-cobordism}.
Together with 
\ref{pheno2}'s \emph{Phenomenology inputs of 16th Weyl fermion and neutrinos},
we suggest that there is a huge mass gap associated with the unobserved 16th Weyl fermion
above 
the scale of 
$({\Spin  \times_{\Z_2^F} \Z_{4,X } \times \frac{\SU(3)\times \SU(2)\times \U(1)}{\Z_6}})_{\text{3-Family}}$.

So the $(( \overline  {\bf 5} \oplus {\bf 10} ),\; ( {\bf 1} \oplus {\bf 1} \oplus  {\bf 1}))$ 
in \Eq{eq:510-111} alone cannot be enough to match the $\Z_{16}$ anomaly 
in \Eq{eq:Z16anomaly-cobordism}.
We suggest schematically, following Sec.~6 of \Ref{JW2006.16996}, the anomaly can be matched by hidden sectors:
\bea \label{eq:anomaly-match}
3 \cdot (15 \text{ Weyl fermions}) +n_{\nu} \cdot (16\text{th Weyl fermions}) + \nu_{\rm{4d}}\cdot(\text{4d TQFT})  + \nu_{\rm{5d}}\cdot(\text{5d iTQFT})
\eea
with the anomaly matching condition for the $\Z_{16}$:
\bea \label{eq:anomaly-match-nu}
\nu = \nu_{\rm{5d}} - \nu_{\rm{4d}} - n_{\nu}=-N_{\text{generation}} = -3 \mod 16.
\eea
\begin{itemize}
\item
The $n_{\nu}$ means the number of the right-handed neutrinos (=16\text{th Weyl fermions}).\footnote{Accidentally, there is a collision of the notations: 
the $\nu$ may refer to as the neutrino such as $\nu_e, \nu_\mu, \nu_\tau$, 
or as the topological index $\nu$ in the class of $\nu \in \Z_{16}$.}
\item The $\nu_{\rm{4d}} \in \Z_{16}$ implies the anomaly index of the {4d TQFT}, if the {4d TQFT} is realized in the theory at a certain energy scale.
\item The $\nu_{\rm{5d}} \in \Z_{16}$ implies the anomaly index of the {5d iTQFT}, if the {5d iTQFT}  is realized in the theory at a certain energy scale.
\end{itemize}

In \Table{table:scenarios}, we present \emph{only} a \emph{possible} set of data of $(n_\nu,      \nu_{\rm{4d}},    \nu_{\rm{5d}})$  obeying 
 the anomaly matching condition in \Eq{eq:anomaly-match} plausibly 
 with different values of $(n_\nu,      \nu_{\rm{4d}},    \nu_{\rm{5d}})$ at different energy scales.
\begin{table}[!h]
$\hspace{-9mm}
\begin{array}{|c | c | c c c c|}
\hline
\text{Theory (l.h.s)}     & \text{Theory (r.h.s)}          &       n_\nu &      \nu_{\rm{4d}} &    \nu_{\rm{5d}} &      \nu \\
\hline
\multicolumn{2}{|c|}{{\Spin \times_{\Z_2^F} \Spin(18)}}           &  3   & 0 & 0   &   -3     \\
\hline
{\Spin  \times_{\Z_2^F} \Z_{4,X} \times \SU(9)} &   {\Spin \times_{\Z_2^F} ({\Spin(10) \times_{\Z_2^F} \Spin(8)})}     & n_\nu''''\text{ vs }3  & 3-n_\nu''''+ \nu_{\rm{5d}}'' \text{ vs }0 &  \nu_{\rm{5d}}''    &    -3  \\
\hline
{\Spin  \times_{\Z_2^F} \Z_{4,X} \times \SU(5)\times \Spin(6)} & {\Spin \times_{\Z_2^F} ({\Spin(10) \times_{\Z_2^F} \Spin(6)})}    & n_\nu'''\text{ vs }3  & 3-n_\nu'''+  \nu_{\rm{5d}}'' \text{ vs }0&  \nu_{\rm{5d}}''  &    -3   \\
\hline
\multirow{ 2}{*}{$\Spin  \times_{\Z_2^F} \Z_{4,X} \times \SU(5)\times \Spin(5)$} & {\Spin \times_{\Z_2^F} ({\Spin(10) \times_{\Z_2^F} \Spin(5)})}      & 
\multirow{ 2}{*}{$n_\nu''\text{ vs }3$}  & \multirow{ 2}{*}{$3-n_\nu''+ \nu_{\rm{5d}}'' \text{ vs }0$} &  \nu_{\rm{5d}}''  &    -3\\
\cline{2-2}\cline{5-6}
 & {({\Spin \times_{\Z_2^F} {\Spin(10)}})_{\text{3-Family}}}     &   &  &  \nu_{\rm{5d}}''  &    -3 \\
 \hline
\hline
 \multicolumn{2}{|c|}{({\Spin  \times_{\Z_2^F} \Z_{4,X} \times \SU(5)})_{\text{3-Family}}}    &  n_\nu'  & 3- n_\nu' +  \nu_{\rm{5d}}' & \nu_{\rm{5d}}'  &    -3\\
 \hline
 \multicolumn{2}{|c|}{({\Spin  \times_{\Z_2^F} \Z_{4,X } \times \frac{\SU(3)\times \SU(2)\times \U(1)}{\Z_6}})_{\text{3-Family}} }   &  n_\nu'  & 3- n_\nu' +  \nu_{\rm{5d}}' & \nu_{\rm{5d}}'  &    -3\\
\hline
 \multicolumn{2}{|c|}{({\Spin  \times_{\Z_2^F} \Z_{4,X } \times  \frac{\SU(3)\times \U(1)_{\text{EM}}}{\Z_3} })_{\text{3-Family}}}    &  n_\nu'  & 3- n_\nu' +  \nu_{\rm{5d}}' & \nu_{\rm{5d}}' &    -3 \\
 \hline
\end{array}\quad
$
\caption{We show only a \emph{possible} set of data of  $(n_\nu,      \nu_{\rm{4d}},    \nu_{\rm{5d}})$  obeying 
 the anomaly matching condition in \Eq{eq:anomaly-match} and \Eq{eq:anomaly-match-nu} so that $\nu = \nu_{\rm{5d}} - \nu_{\rm{4d}} - n_{\nu}=-N_{\text{generation}} = -3 \mod 16$,
 at different energy scales, see  \Fig{fig:energy-hierarchy-lie-algebra-1}, 
   \Fig{fig:energy-hierarchy-lie-algebra-2}, and \Fig{table:sym-web-so18-Z4}. We present possible different results of $(n_\nu,      \nu_{\rm{4d}},    \nu_{\rm{5d}})$
   for the l.h.s and r.h.s route between the $so(18)$ GUT and SM shown in \Fig{table:sym-web-so18-Z4}.
   Whenever we show distinct possibilities of data for l.h.s versus r.h.s, we write in the entry as the l.h.s data vs the r.h.s data.
   The apostrophe $','',''',''''$ on $(n_\nu,      \nu_{\rm{4d}},    \nu_{\rm{5d}})$ implies possible different sets of data.
   A possible interpretation can be that  $(n_\nu'=0,      \nu_{\rm{4d}}'=3,    \nu_{\rm{5d}}'=0)$ below the 4d TQFT gap scale $\Delta_{\rm TQFT}$ which occurs 
   naturally around the energy scale of ${({\Spin  \times_{\Z_2^F} \Z_{4,X} \times \SU(5)})_{\text{3-Family}}}$,
    {$\Spin  \times_{\Z_2^F} \Z_{4,X} \times \SU(5)\times \Spin(5)$}, and ${({\Spin \times_{\Z_2^F} {\Spin(10)}})_{\text{3-Family}}}$.
 {Topological quantum phase transition(s)} may happen around these energy scale 
 (above $M_{su(5) \times \Z_{4,X}\;\text{3-Family}}$ in \Fig{fig:energy-hierarchy-lie-algebra-1} and \Fig{fig:energy-hierarchy-lie-algebra-2})
 drawn with the double horizontal lines (hlines) between the rows.
   If we eventually climb to the $so(18)$ GUT scale with the spacetime-internal structure ${{\Spin \times_{\Z_2^F} \Spin(18)}}$,
   then it is naturally to have some multiple of ${\bf 16}$ of Spin(10), so we have all the right-handed neutrino $n_\nu=3$ joining the $3 \cdot {\bf 16}$,
   so $(n_\nu=3,      \nu_{\rm{4d}}=0,    \nu_{\rm{5d}}=0)$. Tuning the energy scale 
   from the low energy SM or $su(5)$ GUT to a higher energy $so(18)$ GUT
   may result in a \emph{topological quantum phase transition}: The $ \nu_{\rm{4d}}=3$ on one end with a long-range entangled 4d TQFT (intrinsic topological order), and 
   the $n_\nu=3$ on another end with three generations of right-handed neutrinos in some multiple of ${\bf 16}$.
   }
 \label{table:scenarios}
\end{table}

\item \label{input6-TopologicalMass}
{\bf  Topological Mass and TQFT energy gap scale} $\Delta_{\rm TQFT}$ in \Eq{eq:DeltaTQFT}:
We argue that it is more natural to generate the 4d TQFT gap scale $\Delta_{\rm TQFT}$ between these 
energy scales: 
{${({\Spin  \times_{\Z_2^F} \Z_{4,X } \times \frac{\SU(3)\times \SU(2)\times \U(1)}{\Z_6}})_{\text{3-Family}} }$,
${({\Spin  \times_{\Z_2^F} \Z_{4,X} \times \SU(5)})_{\text{3-Family}}}$, and {$\Spin  \times_{\Z_2^F} \Z_{4,X} \times \SU(5)\times \Spin(5)$}.}
The $\Delta_{\rm TQFT}$ shall also be below the scale of ${({\Spin \times_{\Z_2^F} {\Spin(10)}})_{\text{3-Family}}}$.
In short, we may tentatively propose that
\be \label{eq:TQFT-hierarchy}
M_{su(3)\times su(2)\times u(1) \;\text{3-Family}} < M_{su(5) \times \Z_{4,X}\;\text{3-Family}} \lesssim  \Delta_{\rm TQFT} \lesssim M_{so(10) \;\text{3-Family}} 
\text{ or }
M_{su(5) \times so(5)}. \quad
\ee
Around the $\Delta_{\rm TQFT}$ scale may be where the {Grand Unification $+$ Topological Force and Matter}, proposed as Ultra Unification \cite{JW2006.16996}
manifest itself. Let us comment briefly why
the hierarchy \Eq{eq:TQFT-hierarchy} makes sense:
\begin{itemize}
\item The $\Delta_{\rm TQFT}$ is above $M_{su(3)\times su(2)\times u(1) \;\text{3-Family}}$ and $M_{su(5) \times \Z_{4,X}\;\text{3-Family}}$ (for the 
${({\Spin  \times_{\Z_2^F} \Z_{4,X } \times \frac{\SU(3)\times \SU(2)\times \U(1)}{\Z_6}})_{\text{3-Family}}}$ and ${({\Spin  \times_{\Z_2^F} \Z_{4,X} \times \SU(5)})_{\text{3-Family}}}$ spacetime-internal structure), because there are only 15 Weyl fermions and the $( {\bf 5} \oplus \overline{\bf 10})$ of SU(5) around those energy scales.
\item The $\Delta_{\rm TQFT}$ is likely below $M_{so(10) \;\text{3-Family}}$ because it is natural to have 
the 4d TQFT transforming to right-handed neutrino(s) to become part of a multiple of ${\bf 16}$ of Spin(10) in 
${({\Spin \times_{\Z_2^F} {\Spin(10)}})_{\text{3-Family}}}$ around those energy scales.

\item The $\Delta_{\rm TQFT}$ is likely below $M_{su(5) \times so(5)}$ and $M_{su(5) \times so(6)}$,
likewise below $M_{so(10) \times so(5)}$ and $M_{so(10) \times so(6)}$. Why?\\[2mm]
$\diamond 1)$ One reason is that the three generation multiplet
$(( \overline  {\bf 5} \oplus {\bf 10} ) \oplus {\bf 1},\; ( {\bf 1} \oplus {\bf 1} \oplus  {\bf 1}))$
appears naturally in $M_{su(5) \times so(5)}$  but not in $M_{su(5) \times so(6)}$;
the
$( {\bf 16} ,\; ( {\bf 1} \oplus {\bf 1} \oplus  {\bf 1}))$
appears naturally in $M_{so(10) \times so(5)}$  but not in $M_{so(10) \times so(6)}$.\\

$\diamond 2)$ Another reason is that there is an 
inclusion\footnote{Caveat: Part of this embedding is \emph{different} from \Fig{table:sym-web-so18-Z4}, 
so we have $(\Spin  \times_{\Z_2^F} \Z_{4,X} \times \SU(5))\times_{\Z_2^F}(\Spin(5+ \varepsilon) \times_{\Z_2^F}  \Spin(3- \varepsilon)) \supset
(\Spin  \times_{\Z_2^F} \Z_{4,X} \times \SU(5))  \times_{\Z_2^F} \Spin(5+  \varepsilon)   $ 
instead of 
$(\Spin  \times_{\Z_2^F} \Z_{4,X} \times \SU(5))  \times \Spin(5+  \varepsilon)$  of \Fig{table:sym-web-so18-Z4}.}
\begin{multline}\hspace{-30mm}
{{\Spin \times_{\Z_2^F} \Spin(18)}} \supset
{\Spin \times_{\Z_2^F} ({\Spin(10) \times_{\Z_2^F} (\Spin(5+ \varepsilon) \times_{\Z_2^F}  \Spin(3- \varepsilon))})}
\supset 
{\Spin \times_{\Z_2^F} ({\Spin(10) \times_{\Z_2^F} \Spin(5+ \varepsilon)})} \supset \dots\\
\hspace{-46mm}\cup\quad\quad\quad\quad\quad\quad\quad\\
\quad\;
(\Spin  \times_{\Z_2^F} \Z_{4,X} \times \SU(5))\times_{\Z_2^F}(\Spin(5+ \varepsilon) \times_{\Z_2^F}  \Spin(3- \varepsilon)) \supset
(\Spin  \times_{\Z_2^F} \Z_{4,X} \times \SU(5))  \times_{\Z_2^F} \Spin(5+  \varepsilon) \supset \dots,
\end{multline}
here $\supset$ or $\cup$ means the former includes the later as a subgroup/subset.
This inclusion implies the analogous embedding arrow in \Fig{table:sym-web-so18-Z4}.
The $\varepsilon$ can be chosen to be $\varepsilon=0$ for $(\Spin(5) \times_{\Z_2^F}  \Spin(3))$ or 
$\varepsilon=1$ for
$(\Spin(6) \times_{\Z_2^F}  \Spin(2))$ respectively for our purpose. 
When $\varepsilon=0$, not only we have the $\Spin(5)$ that suits for the multiplet 
$( {\bf 16} ,\; ( {\bf 1} \oplus {\bf 1} \oplus  {\bf 1}))$ or $(( \overline  {\bf 5} \oplus {\bf 10} ) \oplus {\bf 1},\; ( {\bf 1} \oplus {\bf 1} \oplus  {\bf 1}))$,
but also the internal subgroup $\Spin(3)$ is highly relevant for the gauge structure of the required candidate 4d TQFT \Ref{JW2006.16996}.
In particular, a certain dimensional-reduced analogous 3d TQFT requires an  
$\Spin(3)=\SU(2)$ or more precisely the $\SO(3)=\Spin(3)/\Z_2$ gauge group as 3d Chern-Simons TQFTs (CS$_3$) \cite{Fidkowski1305.5851, WangLevin1610.04624} denoted as:
\bea \label{eq:3dCS}
\Spin(3)_6 \text{ CS}_3=\SU(2)_6 \text{ CS}_3,\quad  \text{ or } \quad \SO(3)_3 \text{ CS}_3. 
\eea
$\diamond 3)$ The last reason is that $\Spin(8)  \supset (\Spin(5) \times_{\Z_2^F}  \Spin(3))$, where the triality plays an important rule in Spin(8).
The triality of representation in \Sec{sec:SO(10)SO(8)SO(5)GUTgap} likely hints that there is a quantum phase transition with emergent and enlarge symmetry
so that the triality can be generated.
These three reasons motivate us to suggest the $\Delta_{\rm TQFT}$ is around the energy scale $M_{su(5) \times so(5)}$ 
at the $\varepsilon=0$.\footnote{{\bf {Three family (three generation) puzzle}}: 
\Ref{BenTov2015graZee1505.04312} suggests the structure $\Spin(8)  \supset (\Spin(5) \times_{\Z_2^F}  \Spin(3))$ from $\Spin(18)$ may be one key to resolve the family puzzle,
once we apply the {\bf Kitaev-Wen} type mechanism for the {\bf anomaly-free symmetric mass generation}. What \Ref{JW2006.16996} proposes was possibly
another key: {\bf Topological mass} mechanism from the {\bf anomalous symmetric gapped topological order} absorbs part of the gauge structure $\Spin(3)$ in $\Spin(8)  \supset (\Spin(5) \times_{\Z_2^F}  \Spin(3))$, since $\Spin(3)=\SU(2)$ or $\SO(3)=\Spin(3)/\Z_2$ in 3d CS theories \Eq{eq:3dCS}.}
\end{itemize}

\end{enumerate}

\subsection{Energy Scale of Ultra Unification: Grand Unification $+$ Topological Force and Matter} 
\label{subsec:ConclusionUU}
With these phenomenology inputs \ref{pheno1}, \ref{pheno2}, and \ref{pheno3},
and theoretical or mathematical inputs \ref{input4-so10so5},
\ref{input5-su5so5},
\ref{input6-TopologicalMass},
we can provide some tentative but more restricted answers for Consideration \ref{goal3}:
Are there {non-perturbative constraints} from anomalies and cobordism, given the {low energy physics at SM}, guiding us toward 
discovering something {heavy at higher energy}? 
Together with \Ref{JW2006.16996}, we suggest that 
the anomaly can be matched at different energy scales in different manners: 
\begin{enumerate}[leftmargin=4.mm, label=\textcolor{blue}{\underline{\arabic*}]}., ref=\underline{\arabic*]}]
\item In SM, electroweak and Higgs energy scales: 
Around {$M_{su(3)\times u(1)_{\text{EM}} \;\text{3-Family}}$} and below 
{$M_{su(3)\times su(2)\times u(1) \;\text{3-Family}}$} in \Fig{fig:energy-hierarchy-lie-algebra-1} and \Fig{fig:energy-hierarchy-lie-algebra-2}, the $\Z_{4,X}$ symmetry  
is broken by Yukawa-Higgs Dirac mass term.
The $\Z_{16}$ anomaly in \Eq{eq:Z16anomaly-cobordism} is manifestly matched (in fact killed) once the $\Z_{4,X}$ is broken.

\item In $su(5)$ GUT energy scale: Above {$M_{su(5)\;\text{3-Family}}$} 
and around {$M_{su(5) \times \Z_{4,X}\;\text{3-Family}}$}
 in \Fig{fig:energy-hierarchy-lie-algebra-1} and \Fig{fig:energy-hierarchy-lie-algebra-2},
the  $\Z_{4,X}$ symmetry can be restored and regarded a global symmetry. 
Conventionally, the $\Z_{16}$ anomaly \Eq{eq:Z16anomaly-cobordism} can be matched by the 16th Weyl fermion with heavy Dirac or Majorana masses by a seesaw mechanism, 
but those conventional {\bf symmetry-breaking} masses again breaks the $\Z_{4,X}$.

\Ref{JW2006.16996} suggests an alternative to assume the $\Z_{4,X}$ 
is preserved and the $\Z_{16}$ anomaly \Eq{eq:Z16anomaly-cobordism} can still be matched by 4d TQFT or 5d iTQFT (with 't Hooft anomaly)
replacing the gapless or gapped 16th Weyl fermion.
Topological mass here is a {\bf  symmetry-preserving} mass.
\Ref{JW2006.16996} also suggests a linear combination of the three scenarios: Dirac mass + Majorana mass + Topological mass, to match the
\Eq{eq:anomaly-match} and \Eq{eq:anomaly-match-nu}. 

\item In $so(10)$ GUT energy scale: 
The $\Z_{4,X}=Z(\Spin(10))$ symmetry as the center of Spin(10) is dynamically gauged, since the $so(10)$ GUT has dynamical Spin(10) gauge feilds.
The $\Z_{4,X}$ gauge field ${\CA_{{\Z_4}} \in \H^1(M, \Z_{4,X})}$ is locally a one-form mod 4 gauge field (or a $\Z_{4,X}$-valued 1-cocycle).
\begin{itemize}
\item
Since $\Z_{4,X}\supset \Z_2^F$, the fermion parity $(-1)^F$ symmetry is also gauged at a higher $so(10)$ GUT scale, 
the fermionic system becomes \emph{bosonized} by gauging Spin(10) in ${\Spin \times_{\Z_2^F} {\Spin(10)}}$.
\item Another way to say this is that dynamical spin structure is generated when breaking $so(10)$ GUT to $su(5)$ GUT at the lower energy scale \cite{WangWenWitten2018qoy1810.00844}.
\end{itemize}
The $\Z_{4,X}$ gauge field can couple and communicate between  `the 4d SM or GUT sector' and `the 4d TQFT sector or 5d iTQFT sector.'
See the quantum communication by Topological Force of the $\Z_{4,X}$ gauge field in \Ref{JW2006.16996}'s Sec.~6.2.
Since all the quarks and leptons in SM and all the ${\bf 16}$ of Spin(10) carries an odd $\Z_{4,X}$ charge $q_{X}=1 \mod 4$ (see Table 1 of \cite{JW2006.16996}),
the SM/GUT sectors, say with an action $S_{\text{4d-SM/GUT}}$, 
in fact couple to $\Z_{4,X}$ gauge field $\CA_{{\Z_4}}$ in this way: The covariant derivative should be promoted from the SM/GUT coupling to:
\bea
(\nabla_{\mu} - \ii g_{\text{SM/GUT}} A_{\mu} )\psi \Longrightarrow
(\nabla_{\mu} - \ii g_{\text{SM/GUT}} A_{\mu} - \ii q_{X} \CA_{{\Z_4}} )\psi
\eea
with ${\CA_{{\Z_4}} \in \H^1(M, \Z_{4,X})}$ and ${\CA_{{\Z_2}} =(\CA_{{\Z_4}} \mod 2)  \in \H^1(M, \Z_{4,X}/\Z_2^F)}$,
where the restriction may be formulated by a Lagrange multiplier constraint of BF theory term \cite{Putrov2016qdo1612.09298PWY}.
The schematic 
partition function defined via summing all inequivalent gauge configurations in the path integral thus includes a contribution, see Sec.~6.1 of  \cite{JW2006.16996},
\begin{multline}  \label{eq:UU-GUT-1}
{\bf Z}_{\overset{\text{5d-iTQFT/}}{\text{4d-QFT}}}[\CA_{{\Z_4}}]
=
\exp(\frac{2\pi \ii}{16} \cdot\nu_{\rm{5d}} \cdot \eta(\text{PD}(\CA_{{\Z_2}} )) \rvert_{M^5})  \cdot
\int [{\cal D} {\psi}] [{\cal D}\bar{\psi}][{\cal D} A][{\cal D} \phi_H][{\cal D}  \mathscr{A}] [{\cal D}  \mathscr B]\cdots \\
\exp( \ii \left. S_{\text{4d-SM/GUT}}^{(n_\nu)} [\psi, \bar{\psi}, A, \phi_H, \dots,  \CA_{\Z_4}] \right  \rvert_{M^4}
+ \ii \left. S_{\text{4d-TQFT}}^{(\nu_{\rm{4d}})} [\mathscr{A}, \mathscr{B}, \dots , \CA_{{\Z_4}}] \right\rvert_{M^4}
) 
\bigg\rvert_{
\nu = \nu_{\rm{5d}} - \nu_{\rm{4d}} - n_{\nu}=-N_{\text{generation}}}.
\quad\quad
\end{multline} 
The $S_{\text{4d-SM/GUT}}$ is the 4d SM or GUT action.
The $\psi, \bar{\psi}, A, \phi_H$ are SM and GUT quantum fields,
where $\psi, \bar{\psi}$ are the 15 or 16 Weyl spinor fermion fields,
the $A$ are gauge bosons (with 12 components in SM, 24 in $su(5)$ GUT, 45 in $so(10)$ GUT, etc.) 
given by gauge group Lie algebra generators,
and $\phi_H$ is the Higgs (electroweak and GUT Higgs). 
The $S_{\text{4d-TQFT}}^{(\nu_{\rm{4d}})}$ is a 4d noninvertible TQFT outlined in \cite{JW2006.16996}.
The $ \mathscr{A}$ and $ \mathscr{B}$ (and possibly others fields)
are TQFT gauge fields (locally differential 1-form and 2-form anti-symmetric tensor gauge connections).
The theory of \Eq{eq:UU-GUT-1} includes the physics and mathematical constructions of
\begin{enumerate}[leftmargin=4.mm, label=\textcolor{blue}{[\roman*]}., ref={[\roman*]}]
\item \label{th1}
3+1d {\bf Maxwell} (U(1)) and  {\bf Yang-Mills} (SU($N$) and Spin($N$)) gauge theory with some gauge group $G_{\text{internal}}$
and gauge fields $A$:
The gauge field is a gauge connection on a $G_{\text{internal}}$-bundle.

\item  \label{th2}
3+1d fermion field theory of  {\bf Dirac} spinors (the complex ${\bf 4}_{\mathbb{C}}$),  {\bf Weyl} spinors $\psi$ or $\bar{\psi}$ 
(the complex ${\bf 2}_{\mathbb{C}}$ as left-handed ${\bf 2}_{L}$ or right-handed ${\bf 2}_{R}$), 
or  {\bf Majorana} spinors (the real ${\bf 4}_{\mathbb{R}}$) in the representation of  Lorentzian spacetime Spin(3,1),
and in various representation ${\bf R}$ of the gauge group $G_{\text{internal}}$.
Mathematically, the spinors are the sections of the spinor bundle 
with odd-degree fibers in supergeometry or spin spacetimemanifold geometry.

\item  \label{th3}
{\bf Higgs} boson $\phi_H$ scalar field theory. The $\phi_H$ is a scalar ${\bf 1}$ in Lorentzian spacetime Spin(3,1),
and again in some representation ${\bf R}$ of the gauge group $G_{\text{internal}}$.
The action contains possible Higgs potential term $U(\phi_H)$ such as quadratic or quartic terms.
The action can also contain some  {\bf Yukawa-Higgs Dirac} terms or  {\bf Yukawa-Higgs Majorana} terms.

\item  \label{th4}
The  {\bf $\theta$-term}, well-known as $F \wedge F$ or $F\tilde{F}$ in the particle physics community,
is in fact related to the  {\bf second Chern class} $c_2(V_{G})$ and the square of the  {\bf first Chern class} $c_1(V_{G})$
of the associated vector bundle of the gauge group $G$:
\begin{multline}
\theta\; c_2(V_{G})=  -\frac{\theta}{8\pi^2} \Tr(\widehat{F}\wedge \widehat{F}) +\frac{\theta}{8\pi^2} (\Tr \widehat{F}) \wedge(\Tr \widehat{F})
=  -\frac{\theta}{8\pi^2} \Tr(\widehat{F}\wedge \widehat{F}) 
  +\frac{\theta}{2} c_1(V_{G})^2\\
  \Rightarrow
  \frac{\theta}{8\pi^2} \Tr(\widehat{F}\wedge \widehat{F}) =\frac{\theta}{2} c_1(V_{G})^2 - \theta\; c_2(V_{G}).
\end{multline} 
In particular, here we consider $G$ as the U($N$) or SU($N$) gauge group,
so we can define the Chern characteristic classes associated with complex vector bundles.
This  $\theta$-term is a topological term, but it is summed over as a weighted factor to define a 
Yang-Mills gauge theory partition function \cite{AharonyASY2013hdaSeiberg1305.0318} \cite{Gaiotto2017yupZoharTTT,Wan2019oyr1904.00994}. 
This  $\theta$-term is not a quantum phase of matter by itself,
so it is very different from the 4d TQFT with intrinsic topological order and 5d iTQFT with SPTs (as certain quantum phases of matter).

\item  \label{th5}
4d TQFT is mathematically a 4d non-invertible TQFT whose partition function ${\bf Z}$ on some closed manifold $M$
has an absolute value
$|{\bf Z}(M)|\neq 1$. In the case $M= M^3 \times S^1$,
the ${\bf Z}(M^3 \times S^1)=\GSD = \dim \CH_{M^3}$ is known as the number of ground states (GSD: ground state degeneracy) or the dimension of 
TQFT Hilbert space $\CH$ on the spatial  $M^3$. 
In general, $\GSD\neq 1$ on a spatial  $M^3$ is related to the counting of distinct topological superselection sectors of fractionalized excitations 
(from particles of 1-line operators or strings of 2-surface operators).
The 4d TQFT is the low energy field theory description of 
some intrinsic topological order in the sense of quantum matter.
The gauge fields for 4d TQFT here are cocycles in differential cohomology.
This 4d TQFT is a new addition from \cite{JW2006.16996} to SM and particle physics.

\item  \label{th6}
5d iTQFT is mathematically  a 5d invertible TQFT whose partition function ${\bf Z}$ on any closed manifold $M=M^5$ has an absolute value
$|{\bf Z}(M)|=1$. So that the number ${\bf Z}(M)^*= {\bf Z}(M)^\dagger= {\bf Z}(M)^{-1}$ defines an inverted phase of the original iTQFT ${\bf Z}(M)$.
The combined phase ${\bf Z}(M)^* {\bf Z}(M)=1$ describes a trivial phase with no SPT nor topological order.
This 5d iTQFT is a new addition from \cite{JW2006.16996} to SM and particle physics.
Is is an analogous interacting $\Z_{16}$ class of topological superconductor in condensed matter physics 
\cite{Kitaev2015, 2010RMP_HasanKane, 2011_RMP_Qi_Zhang, CWang1401.1142, Metlitski20141406.3032} \cite{1711.11587GPW}
but now in one higher dimension in 4+1d.

\end{enumerate}
We should emphasize repeatedly that 
this topological $\theta$-term is totally \emph{different} from the new topological sector (4d TQFT or 5d iTQFT) introduced in \cite{JW2006.16996}.
The previous Grand Unification contains a framework to include  \ref{th1},  \ref{th2},  \ref{th3},  \ref{th4},
but the Ultra Unification is proposed to include Grand Unification plus additional new topological sectors of TQFTs in \ref{th5} and  \ref{th6}.

Some more comments:
\begin{itemize}
\item
If only the $\Z_{4,X}$ gauge field are dynamical and summed over in the partition function, then we deal with a QFT problem with
SM/GUT and 4d TQFT or 5d iTQFT sector as in \cite{JW2006.16996}.
\item 
If not only the $\Z_{4,X}$ gauge field but also the $\eta$ invariant together with the underlying spacetime topology/geometry are dynamical and summed over in the partition function
(i.e., the $\eta\big(\text{PD}(\CA_{{\Z_2}})  \big)$ in \Eq{eq:Z16anomaly} are summed over), 
then we will have to deal with a QFT coupling to a dynamical gravity problem: a more challenging topological or quantum gravity issue.  
\end{itemize}

\end{enumerate}

\Ref{JW2006.16996} proposal ends at the  $su(5)$ GUT and below the $so(10)$ GUT scale. 
In the present work, we continue to explore higher energy spectra to the hypothetical 
$so(18)$ GUT scale (compare with \Fig{fig:energy-hierarchy-lie-algebra-1}, \Fig{fig:energy-hierarchy-lie-algebra-2}, \Fig{table:sym-web-so18-Z4}, and \Table{table:scenarios}):
\begin{enumerate}[leftmargin=4.mm, label=\textcolor{blue}{\underline{\arabic*}]}., ref=\underline{\arabic*]}]
\setcounter{enumi}{3}
\item Above the {$M_{su(5)\;\text{3-Family}}$} and below {$M_{su(5) \times \Z_{4,X}\;\text{3-Family}}$}, if there are Dirac or Majorana masses given to the sterile neutrinos,
then their masses could be around these scales. So the $\Z_{4,X}$ is broken below {$M_{su(5) \times \Z_{4,X}\;\text{3-Family}}$} due to the explicit Dirac/Majorana masses.
  
\item Above the  {$M_{su(5) \times \Z_{4,X}\;\text{3-Family}}$}  and below {$M_{so(10)\;\text{3-Family}}$} (on the r.h.s of \Fig{table:sym-web-so18-Z4}),
or below $M_{su(5)\times so(5)}$ (on the l.h.s of \Fig{table:sym-web-so18-Z4}),
there could be a 4d TQFT gap scale $\Delta_{\mathrm{TQFT}}$ given by \Eq{eq:DeltaTQFT} for the 4d TQFT described in \ref{th5}

In addition, the KW mechanism can take place, at a scale $\Delta_{\rm{KW}.{su(9) \times so(5)}}$, to gap out extra matter.

\item Above the {$M_{so(10)\;\text{3-Family}}$} and ${M_{so(10) \times so(5)}}$ (on the r.h.s of \Fig{table:sym-web-so18-Z4})
or above $M_{su(5)\times so(5)}$ (on the l.h.s of \Fig{table:sym-web-so18-Z4}),
below the ${M_{so(10) \times so(6)}}$ (on the r.h.s of \Fig{table:sym-web-so18-Z4}) or $M_{su(5)\times so(6)}$ (on the l.h.s of \Fig{table:sym-web-so18-Z4}),
there could be a topological quantum phase transition (ideally tuning at the zero temperature $T=0$, 
increasing the energy scale at $T=0$
but by probing the shorter distance). 
The  topological quantum phase transition occurs due to that part of the 4d TQFT degrees of freedom may become a nearly free-particle description of 16th Weyl fermion (right-handed neutrino).

In addition, the KW mechanism can take place, at scales {$\Delta_{\rm{KW}.{su(9) \times so(6)}}$} and {$\Delta_{\rm{KW}.{so(10) \times so(6)}}$}, etc. in sequence, 
to gap out extra matter, steps by steps.

\item Above the ${M_{so(10) \times so(6)}}$ or $M_{su(5)\times so(6)}$ (respectively on the r.h.s and l.h.s of \Fig{table:sym-web-so18-Z4}),
below the ${M_{so(10) \times so(8)}}$ or $M_{su(9)}$ (respectively on the r.h.s and l.h.s of \Fig{table:sym-web-so18-Z4}),
there could be additional topological quantum phase transitions 
due to that other remained part of the 4d TQFT degrees of freedom may eventually become nearly free-particle description of 16th Weyl fermion(s) coupling to GUT gauge fields.

In addition, the KW mechanism can take place, at the scale {$\Delta_{\rm{KW}.{so(10) \times so(8)}}$}, etc. in sequence, 
to gap out extra matter, steps by steps.

\item At  ${M_{so(18)}}$, if all matter fields are eventually in ${\bf 256}^+$,
then it may be possible that 
\Eq{eq:anomaly-match} and
\Eq{eq:anomaly-match-nu} are satisfied by
$(n_\nu=3,      \nu_{\rm{4d}}=0,    \nu_{\rm{5d}}=0)$.

In addition, the KW mechanism can take place, at the scale {$\Delta_{\rm{KW}.{so(18)}}$} above $M_{so(18)}$, 
to gap out the ${\bf 256}^-$ mirror matter, steps by steps.

\item If $\nu_{\rm{5d}}\neq 0$ for any scale above ${M_{so(10)}}$, then, since
$\Z_{4,X}=Z(\Spin(10))=Z(\Spin(18))$ is dynamically gauged above the scale ${M_{so(10)}}$,
then there is a topological force mediated between 4d SM/GUT to 5d gauged theory. (Note: Gauging the $\Z_{4,X}$ of 5d iTQFT \Eq{eq:Z16anomaly} 
becomes a 5d noninvertible TQFT 
[plus gravity if the $\eta\big(\text{PD}(\CA_{{\Z_2}})  \big)$ in \Eq{eq:Z16anomaly} are also dynamical and summed over].)
This agrees with the proposal in \cite{JW2006.16996}.

\item {\bf Dark Matter as Topological Matter from Extended Objects}?
Above the  {$M_{su(5) \times \Z_{4,X}\;\text{3-Family}}$}  and below {$M_{so(10)\;\text{3-Family}}$} 
or below $M_{su(5)\times so(5)}$ (respectively on the r.h.s or  l.h.s of \Fig{table:sym-web-so18-Z4}),
the possible 4d TQFT gap scale $\Delta_{\mathrm{TQFT}}$ 
 in \Eq{eq:DeltaTQFT} is precisely the \emph{energy gap}
  of heavy fractionalized extended object excitations 
  from 4d intrinsic topological order. (See the previous remark \ref{th5}.)
  It is possible these heavy fractionalized extended objects (from particles of 1-line operators or strings of 2-surface operators)
  can account for the heavy Dark Matter. If so, the Dark Matter is not formulated in terms of the conventional 
  point-particle QFT physics, but the Dark Matter may be formulated in terms of extended objects of the (4d or 5d) TQFT physics.

\end{enumerate}

In summary, in this work, we had checked explicitly that the anomaly can be matched by novel scenarios, 
not only in the energy scales below $su(5)$ GUT, 
but also between $su(5)$ GUT and $so(10)$ GUT, and to $so(18)$ GUT,
for various scenarios in the proposal \cite{JW2006.16996}.
{In the Appendices, we list down some additional explicit computations of anomaly matching.}



\appendix
\section{Dynamical Gauge Anomaly Cancellation}
\label{sec:DynamicalGaugeAnomalyCancellation}

In Appendix \ref{sec:DynamicalGaugeAnomalyCancellation},
we include the calculations of dynamical gauge anomaly cancellations
for the $su(5)$ GUT, the two version (on spin or non-spin manifolds) of the $so(10)$ GUTs and $so(18)$ GUT.
See \Table{table:SU5SO10SO18} for the anomalies classified by cobordism, including
\begin{itemize}
\item\emph{perturbative local anomalies}, classified by $\Z$ classes (known as free classes), and  
\item \emph{nonperturbative global anomalies},  classified by $\Z_n$ classes (known as torsion classes). 
\end{itemize}

Let us check explicitly that 
the 
dynamical gauge anomaly cancellation holds for $su(5)$ GUT and two version of $so(10)$ and $so(18)$ GUTs.
In fact, there is only a local $\Z$ class anomaly captured by Feynman-Dyson graph for $su(5)$ GUT,
and a global  $\Z_2$ class anomaly for $so(10)$ and $so(18)$ GUT placed on non-Spin manifolds.
Let us check below.

\begin{table}[H]
\centering
\hspace*{-7mm}
\begin{tabular}{c c c }
\hline
\multicolumn{3}{c}{Cobordism group $\TP_d(G)$ for Grand Unifications}\\
\hline
\hline
$d$d  & classes & cobordism invariants\\
\hline
\hline\\[-2mm]
 \multicolumn{3}{c}{$G=\Spin\times {\SU(5)}$ for $su(5)$ GUT}\\[2mm]
\hline
5d & $\Z$ & 
$\frac{1}{2}\text{CS}_5^{\SU(5)}$
 \\
\hline
\hline
\multicolumn{3}{c}{$G=\Spin_{}\times {\Spin(N)}$ for $N\geq 7$,}\\[2mm]
\multicolumn{3}{c}{e.g. ${\Spin(N)}={\Spin(10)}$ or ${\Spin(18)}$  for $so(10)$ or $so(18)$ GUT}\\[2mm]
\hline
5d & $0$ &  None
 \\
\hline
\multicolumn{3}{c}{$G=\Spin_{}\times_{\Z_2} {\Spin(N)}$ for $N\geq 7$,}\\[2mm]
\multicolumn{3}{c}{e.g. ${\Spin(N)}={\Spin(10)}$ or ${\Spin(18)}$   for $so(10)$ or $so(18)$ GUT}\\[2mm]
\hline
5d & $\Z_2$ & 
$w_2(TM)w_3(TM) = w_2(V_{\SO(N)}) w_3(V_{\SO(N)})$
 \\
\hline
\hline
\end{tabular}
\caption{The 4d anomalies can be written as 5d cobordism invariants
of
$\Omega^{d=5}_{G} \equiv
\TP_{d=5}(G)$, 
which are 5d iTQFTs. 
These 5d cobordism invariants/iTQFTs are derived in \cite{WW2019fxh1910.14668}.
We summarized the group classifications of 4d anomalies and their 5d cobordism invariants for the $su(5)$ GUT and the two versions of $so(10)$ GUT (placed on
Spin vs non-Spin manifolds).
See our notational conventions  in \cite{JW2006.16996} and in Sec.~1 and Sec.~1.2.4 of \Ref{WW2019fxh1910.14668}.}
 \label{table:SU5SO10SO18}
\end{table}

\subsection{SU(5)$^3$ for $su(5)$ GUT: 4d local anomaly from 5d $\frac{1}{2}{\text{CS}_5^{\SU(5)}}$ and 6d $\frac{1}{2}{c_3(\SU(5))}$}

\label{sec:SU(5)3}

For {$G=\Spin\times {\SU(5)}$ of $su(5)$ GUT}, 
we read from \Ref{WW2019fxh1910.14668} and Table \ref{table:SU5SO10SO18} for a $\Z$ class of 5d cobordism invariants of the following: 
5d $\frac{1}{2}{\text{CS}_5^{\SU(5)}}$ and 6d $\frac{1}{2}{c_3(\SU(5))}$.
These 5d cobordism invariants correspond to the 4d perturbative local anomalies captured by the one-loop Feynman graph:
\bea \label{eq:anomaly-SU3SU3SU3}
\includegraphics[width=2.2in]{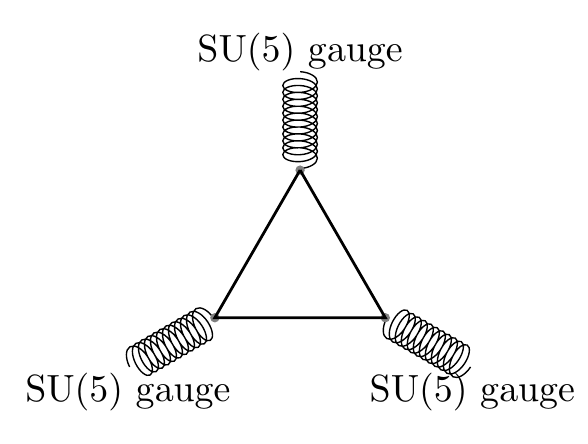}.
\eea
This is the 4d anomaly {SU(5)$^3$} 
from 
5d $\frac{1}{2}{\text{CS}_5^{\SU(5)}}$, 
which also descends from 6d $\frac{1}{2}{c_3(\SU(5))}$
 of bordism group $\Omega_6$ in \Ref{WW2019fxh1910.14668}.
We can check the anomaly \Eq{eq:anomaly-SU3SU3SU3} vanishes,
by taking all of the SU(5) generators. It is sufficient to take the diagonal SU(5) generator $Y$ as
\bea
\hat Y=
\frac{1}{2} \hat{Y}'
=\frac{1}{6} \hat {\tilde{Y}}
=
\begin{pmatrix}
-  1/3 & 0& 0 & 0 & 0\\
0 & -  1/3 & 0 & 0&0 \\
0& 0& -  1/3&0 &0\\
0& 0&0 &1/2& 0\\
0& 0&0& 0&1/2
\end{pmatrix} =
\frac{1}{6}
\begin{pmatrix}
-  2 & 0& 0 & 0 & 0\\
0 & - 2 & 0 & 0&0 \\
0& 0& -  2&0 &0\\
0& 0&0 &3& 0\\
0& 0&0& 0&3
\end{pmatrix}. 
\eea
We can check the anomaly of $su(5)$ GUT with matter $\overline{\bf 5}+  {\bf 10}$ indeed cancels:
\bea
\Tr \hat Y^3 \bigg\vert_{\overline{\bf 5}}+  
\Tr \hat Y^3 \bigg\vert_{\bf 10}
=(\frac{1}{6})^3\bigg(\Big(3 (2)^3 +2 (-3)^3\Big) + \Big(3 (-2-2)^3 +6 (-2+3)^3 + (6)^3\Big)\bigg)=0.
\eea
Other Lie algebra generators for the $\overline{\bf 5}+  {\bf 10}$ also cancel.

There is a similar calculation of $G=\Spin\times {\SU(9)}$ for the $su(9)$ GUT because
the cobordism group $\TP_5(G)=\Z$.
It is a 4d local anomaly from 5d $\frac{1}{2}{\text{CS}_5^{\SU(9)}}$ and 6d $\frac{1}{2}{c_3(\SU(9))}$.
We can easily check that the $su(9)$ GUT descended from the $so(18)$ GUT in \Fig{table:sym-web-so18-Z4} is free from this 4d local anomaly.

\subsection{Witten SU(2) anomaly vs New SU(2) anomaly}
\label{subsec:SU(2)anomaly}

We summarize the 't Hooft anomalies of 4d $\SU(2) = \Spin(3)$ symmetry theory in \Eq{eq:SU2anomaly} and \Table{table:SU2}.  
When SU(2) is gauged, these anomalies become dynamical gauge anomalies.
There are two kinds of SU(2) anomalies, both are  nonperturbative global anomalies.
We will use the Witten SU(2) anomaly \cite{Witten1982fp} and the new SU(2) anomaly \cite{WangWenWitten2018qoy1810.00844}
in 4d to characterize the anomalies in the $so(N)$ GUT for $N \geq 7$, such as $N=10,18$.
The $\checkmark$ mark  in \Eq{eq:SU2anomaly}  means the anomaly exists for that matter representation ${\bf R}$.
\bea \label{eq:SU2anomaly}
\begin{array}{c |c c c c c c c c c  |c  c c c}
\hline
\text{SU(2) isospin} &0  & \frac{1}{2} &1  & \frac{3}{2} &2  & \frac{5}{2}&3 & \frac{7}{2} &\text{mod }4 & 2 r +\frac{1}{2} & 4 r +\frac{3}{2} & \text{mod }4 \\
\hline
\text{SU(2) Rep {\bf R} (dim)} & {\bf 1}  & {\bf 2} &{\bf 3}  & {\bf 4}& {\bf 5}  & {\bf 6} &{\bf 7} & {\bf 8} &\text{mod }8 & {\bf 4}  r +{\bf 2} & {\bf 8}  r +{\bf 4}  & \text{mod }8\\
\hline
\text{Witten SU(2) anomaly \cite{Witten1982fp}} & &\checkmark &  & &  &\checkmark& & & & \checkmark & &\\
\hline
\text{New SU(2) anomaly \cite{WangWenWitten2018qoy1810.00844}}    &  &  &  & \checkmark & &&& & &  & \checkmark & &\\
\hline
\end{array}
\eea
For a 4d SU(2) symmetry theory,
\Eqn{eq:SU2anomaly} shows that: 
\begin{itemize}
\item
when the fermions (the spacetime spinors)
are in the {SU(2) isospin} $2 r +\frac{1}{2}$ (namely the SU(2) representation dimension {\bf R} is ${\bf 4}r+{\bf 2}$ for some integer $r$),
we have the Witten SU(2) anomaly \cite{Witten1982fp} as 't Hooft anomaly detectable on both
$\Spin_{}\times {\SU(2)}$ and
$\Spin_{}\times_{\Z_2} {\SU(2)}$ spacetime-internal structures. 
When SU(2) is gauged, the dynamical SU(2) gauge theory becomes inconsistent even on spin manifolds.

\item when the fermions (the spacetime spinors)
are in the {SU(2) isospin} $4 r +\frac{3}{2}$ (namely the SU(2) representation dimension {\bf R} is $ {\bf 8}  r +{\bf 4}$),
we have the new SU(2) anomaly \cite{WangWenWitten2018qoy1810.00844} as 't Hooft anomaly detectable only 
on $\Spin_{}\times_{\Z_2} {\SU(2)}$ spacetime-internal structures.
When SU(2) is gauged, the dynamical SU(2) gauge theory can still be  consistent on Spin or Spin$^c$ manifolds;
 the dynamical SU(2) gauge theory  becomes inconsistent \emph{only} on certain non-spin manifolds.

\end{itemize}

\begin{table}[H]
\centering
\hspace*{-7mm}
\begin{tabular}{c c c }
\hline
\multicolumn{3}{c}{Cobordism group $\TP_d(G)$ for SU(2) anomalies}\\
\hline
\hline
$d$d  & classes & cobordism invariants\\
\hline
\hline
\multicolumn{3}{c}{$G=\Spin_{}\times {\Spin(3)} =\Spin_{}\times \SU(2)$}\\[2mm]
\hline
5d & $\Z_2$ &   $c_2(V_{\SU(2)})\tilde\eta$
 \\
\hline
\multicolumn{3}{c}{$G=\Spin_{}\times_{\Z_2} {\Spin(3)}=\Spin_{}\times_{\Z_2} {\SU(2)}$}\\[2mm]
\hline
5d & $(\Z_2)^2$ & 
$\begin{matrix}
(N_0'^{(5)} \mod 2),\\
w_2(TM)w_3(TM) = w_2(V_{\SO(3)}) w_3(V_{\SO(3)})
\end{matrix}$
 \\
\hline
\hline
\end{tabular}
\caption{The 4d anomalies can be written as 5d cobordism invariants
of
$\Omega^{d=5}_{G} \equiv
\TP_{d=5}(G)$, 
which are 5d iTQFTs. 
These 5d cobordism invariants/iTQFTs are derived in \cite{WanWang2018bns1812.11967}.
We summarized the group classifications of 4d anomalies and their 5d cobordism invariants for two versions of SU(2) symmetric theory (placed on
Spin vs non-Spin manifolds). One of the $\Z_2$ class global anomaly is the familiar Witten SU(2) anomaly \cite{Witten1982fp}, 
captured by  $c_2(V_{\SU(2)})\tilde\eta$ or {$N_0'^{(5)} \mod 2$}. 
The $N_0'^{(5)}$ is the number of the zero modes of the Dirac operator in 5d. 
The $N_0'^{(5)}$ mod 2 is a spin-topological invariant known as the mod 2 index defined in \cite{Witten1982fp, WangWenWitten2018qoy1810.00844}.
(We find that the cobordism invariant of 
{$N_0'^{(5)} \mod 2$} read from Adams chart has the similar form  related to $\tilde w_3 \text{Arf}$, where Arf is an Arf invariant \cite{Arf1941}
and $\tilde w_3$ is a twisted version of the third Stiefel-Whitney class $w_3$.)
Another $\Z_2$ class global anomaly is the new SU(2) anomaly \cite{WangWenWitten2018qoy1810.00844}.
The $\tilde\eta$ is a mod 2 index of 1d Dirac operator.
See our notational conventions  in \cite{JW2006.16996} and in Sec.~1 and Sec.~1.2.4 of \Ref{WW2019fxh1910.14668}.}
 \label{table:SU2}
\end{table}

\subsection{A new SU(2) = Spin(3) $\subset$ Spin(10) $\subset$ Spin(18) anomaly for $so(10)$ and $so(18)$ GUT  on non-Spin manifolds}
\label{sec:newSU2anom}

There is $\Z_2$ classification of possible anomaly for SO(10) and $so(18)$ GUT shown in \Table{table:SU5SO10SO18}, 
 \bea
&&\Omega_5^{{\Spin \times_{\Z_2} \Spin(10)}} = 
\Omega_5^{{\Spin \times_{\Z_2} \Spin(18)}} = \Z_2, \quad \\ 
&&\TP_5({{\Spin \times_{\Z_2} \Spin(10)}})= 
\TP_5({{\Spin \times_{\Z_2} \Spin(18)}})= \Z_2.
\eea
This implies that there is only a 
5-dimensional topological invariant 
written in terms of a bulk partition function
on a 5-manifold $M^5$,
\begin{align}
\label{topinv}
{\bf Z} = \exp({\ii \pi \int_{M^5} w_2(TM) \cup w_3(TM)})= \exp({\ii \pi \int_{M^5}  w_2(V_{\SO(3)}) \cup w_3(V_{\SO(3)})}),
\end{align}
where $w_n(TM)$ is the $n$th-Stiefel-Whitney
class for the tangent bundle of 5d spacetime manifold
$M^5$, and the $\cup$ is the cup product (which we may omit writing  $\cup$).
We note that on $M^5$, we have a $\frac{\Spin(D=5) \times
\Spin(N)}{{\Z_2^F}}$ connection --- a mixed gravitational and gauge connection,
rather than a pure gravitational $\Spin(D=5)$ connection. The mixed gravitational and gauge structure in 
$\Spin_{}\times_{\Z_2} {\Spin(3)}$ gives a constraint
{$w_2(TM)=w_2(V_{\SO(N)})$ and $w_3(TM)=w_3(V_{\SO(N)})$, where
$w_n(V_{\SO(N)})$ is the $n$th-Stiefel-Whitney class for an ${\SO(N)}$
gauge bundle}.  
Thus, $M^5$ can be a non-spin manifold due to $w_2(TM) \neq 0$ (note that a spin manifold iff $w_2(TM)=0$).

{We can detect the 5d cobordism invariant by
its 4d boundary state. In our case, the 5d state has a boundary described by
4d $\Spin(N)$ chiral Weyl fermion theory with Weyl fermion as the Lorentz spinor ${\bf 2}_L$ of the spacetime structure Spin(3,1). 
Then we can detect the 5d cobordism invariant via the $\Spin(N)$ representation of the chiral Weyl
fermions on the boundary.  Here we use a fact that the 5d cobordism
invariant can be detected by restricting to a subgroup 
$\SU(2)=\Spin(3) \subseteq
\Spin(N)$ \cite{WangWen2018cai1809.11171, WangWenWitten2018qoy1810.00844}:  Let $n_j$ be the number of isospin-$j$
representations of $\SU(2)= \Spin(3) \subseteq \Spin(N)$ 
(so the dimension of representation is ${\bf R}=2 j+1$)
for
4d boundary chiral Weyl fermions, then
the 5d cobordism invariant  is
absent if the following two numbers are zero mod 2:
\begin{align}
\label{njnj}
\sum_{r=0}^\infty n_{2r+\frac12} = 0 \mod 2,
\ \ \ \
\sum_{r=0}^\infty n_{4r+\frac32} = 0 \mod 2. 
\end{align}
}
To check
how the representation of $\Spin(N)$ reduces to the representations of
$\Spin(3)$, let us study the representation of $\Spin(N)$ (the spinor
representation of $\Spin(N)$), assuming $N \in $ even. We first introduce
$\ga$-matrices $\ga_a$, $a=1,\cdots,N$:
\begin{align}
 \ga_{2k-1}&=
\underbrace{\si^0\otimes \cdots \otimes \si^0}_{\frac{N}{2}-k  \ \si^0{\text{'s}}}
\otimes \si^1\otimes
\underbrace{\si^3\otimes \cdots \otimes \si^3}_{k-1 \ \si^3{\text{'s}}},
\nonumber\\
 \ga_{2k}&=
\underbrace{\si^0\otimes \cdots \otimes \si^0}_{\frac{N}{2}-k \ \si^0{\text{'s}}}
\otimes \si^2\otimes
\underbrace{\si^3\otimes \cdots \otimes \si^3}_{k-1  \ \si^3{\text{'s}}},
\end{align}
$k=1,\cdots,\frac{N}{2}$, which satisfy $ \{\ga_a,\ga_b\}=2\del_{ab}$ and
$\ga_a^\dag=\ga_a$.  Here $\si^0$ is the rank-2 identity matrix, and $\si^l$ with
$l=1,2,3$ are the rank-2 Pauli matrices.  The $\frac{N(N-1)}{2}$  hermitian matrices $
\ga_{ab}=\frac{\ii}{2}[\ga_a,\ga_b]= \ii \ga_a\ga_b$  for $a<b$, generate a
$2^{N/2}$-dimensional representation of $\Spin(N)$.  The above
$2^{N/2}$-dimensional representation is reducible.  To obtain an irreducible
representation, we introduce
\begin{align}
\ga_\text{FIVE}&=(-\ii )^{N/2} \ga_1 \cdots \ga_{N}
=
\underbrace{\si^3\otimes \cdots \otimes \si^3}_{\frac{N}2 \ \si^3{\text{'s}}}.
\end{align}
We have $(\ga_\text{FIVE})^2=1$, its trace $\Tr (\ga_\text{FIVE})=0$, and
$\{\ga_\text{FIVE},\ga_a\}=[\ga_\text{FIVE},\ga_{ab}]=0$.  This allows us to
obtain two $2^{N/2-1}$-dimensional irreducible representations: one representation survive under the projection
$\frac{1+\ga_\text{FIVE}}{2}$ (known as the original chiral matter in physics),
the other representation survive under the projection
$\frac{1-\ga_\text{FIVE}}{2}$ (known as the mirror matter in physics).

Now, let us consider an $\SU(2)=\Spin(3)$ subgroup of $\Spin(N)$, generated by
$\ga_{12} = I\otimes \si^0\otimes \si^3$, $\ga_{23} = I\otimes \si^1\otimes
\si^1$,  and $\ga_{31} = I\otimes \si^1\otimes \si^2$, where $I$ is an identity matrix from $\si^0$.  We see that the
$2^{N/2-1}$-dimensional irreducible representation of $\Spin(N)$ becomes $2^{N/2-2}$
isospin-1/2 representations (${\bf R}={\bf 2}$) of $\SU(2)$.
This means
 $$
\text{the $2^{N/2-1}$-dimensional irreducible spinor representation of $\Spin(N)$ $\sim 2^{(N/2)-2} \cdot ({\bf 2})$ of $\Spin(3)=\SU(2)$ }.
 $$
{In short, we see that for an even $N\geq 8$, the
4d boundary chiral Weyl fermions only reduces to an even number of
isospin-1/2 representations (${\bf R}={\bf 2}$) of $\SU(2)$,  and, according to (\ref{njnj}), the 5d
cobordism invariant $\e^{\ii \pi \int_{M^5} w_2(TM)w_3(TM)}$ is absent.
Thus the 4d $so(N\geq 8)$ GUTs including the $so(10)$ and $so(18)$ GUT
 are free from all dynamical gauge anomalies.
 These GUTs are free from perturbative local anomalies are well-known since 1970-80s,
 but these  GUTs are free from nonperturbative global anomalies are known only recently in \cite{WangWen2018cai1809.11171, WangWenWitten2018qoy1810.00844}.

\section{Anomaly Matching for GUT with Extra Symmetries} 
\label{Sec:discretesymmetries}

For the $su(5)$ GUT, we can introduce the $X=5({\bf B}- {\bf L})-4Y$ symmetry as an $\U(1)_{X}$ or 
$\Z_{4,X}$ symmetry. This gives an $\Spin^c$ or $\Spin\times_{\Z_2} \Z_4$ structure respectively.
See \Table{table:SU5-discrete} for the anomalies classified by cobordism.
For $so(10)$ and $so(18)$ GUT, the $\Z_{4,X}= Z(\Spin(10))= Z(\Spin(18))$ is part of the gauge group, so we already classify all possible anomalies 
of $so(N \geq 7)$ GUT including $X=5({\bf B}- {\bf L})-4Y$ symmetry in \Table{table:SU5SO10SO18}.

\begin{table}[H]
\centering
\hspace*{-18mm}
\begin{tabular}{c c c }
\hline
\multicolumn{3}{c}{
 $\begin{array}{c}
\text{ Cobordism group $\TP_d(G)$ for Grand Unifications with extra symmetries}  
\end{array}$
}\\
\hline
\hline
$d$d & classes & cobordism invariants\\
\hline
\hline\\[-2mm]
\multicolumn{3}{c}{$G=\Spin\times_{\Z_2} \Z_4 \times \SU(5)
=\Spin\times_{\Z_2^F} \Z_{4,X} \times \SU(5)$}\\[2mm]
\hline
5d & $\Z\times\Z_2\times\Z_{16}$ & 
$\frac{(\CA_{{\Z_2}})^2\text{CS}_3^{\SU(3)}+\text{CS}_5^{\SU(3)}}{2}$, \quad
$(\CA_{{\Z_2}}) c_2(\SU(5))$,\quad
$\eta(\text{PD}(\CA_{{\Z_2}}))$\\
\hline
\hline\\[-2mm]
\multicolumn{3}{c}{$G=\Spin^c \times  \SU(5)
=\Spin\times_{\Z_2^F} \U(1)_{X} \times \SU(5)$}\\[2mm]
\hline
5d & $\Z^4$ & 
captured by perturbative local anomalies.\\
\hline
\hline
\end{tabular}
\caption{Setup follows \Table{table:SU5SO10SO18}.
The 4d anomalies can be written as 5d cobordism invariants
of
$\Omega^{d=5}_{G} \equiv
\TP_{d=5}(G)$, 
which are 5d iTQFTs. 
These 5d cobordism invariants/iTQFTs are derived in \cite{WW2019fxh1910.14668, WanWangv2}.
We summarized the group classifications of 4d anomalies and their 5d cobordism invariants for the $su(5)$ GUT with U(1)$_X$ or 
$\Z_{4,X}$ symmetry. For $so(10)$ GUT, the $\Z_{4,X}$ is part of the gauge group, so we only need to look at \Table{table:SU5SO10SO18}'s result.
See our notational conventions  in \cite{JW2006.16996} and in Sec.~1 and Sec.~1.2.4 of \Ref{WW2019fxh1910.14668}.}
 \label{table:SU5-discrete}
\end{table}

It is well-known that $su(5)$ GUT with $X=5({\bf B}- {\bf L})-4Y$ symmetry is free from all perturbative local anomalies, perhaps since 1970s-80s.
(Namely, the $\Z$ class anomalies in \Table{table:SU5-discrete} would vanish in the $su(5)$ GUT.)
However, it is not clear whether $su(5)$ GUT with $X=5({\bf B}- {\bf L})-4Y$ symmetry is free
from all non-perturbative global anomalies.
Recent attempts to check global anomalies of $su(5)$ GUT with $X$ symmetry can be found in 
\Ref{GarciaEtxebarriaMontero2018ajm1808.00009, 2019arXiv191011277D} and  \cite{WW2019fxh1910.14668}.
We will check the 4d $\Z_2$ global anomaly from 5d $(\CA_{{\Z_2}}) c_2(\SU(5))$ 
in \Sec{sec:XSU52}, and check 4d  $\Z_{16}$ global anomaly from $\eta(\text{PD}(\CA_{{\Z_2}}))$
in \Sec{sec:etaPDAZ16}.


  \subsection{$X$-SU(5)$^2$: 4d local $\Z$ anomaly or 4d global $\Z_2$ anomaly from 5d $(\CA_{{\Z_2}}) c_2(\SU(5))$}
\label{sec:XSU52}
Recall the $\U(1)_{{ \mathbf{B}-  \mathbf{L}}}$ is not a proper symmetry of $su(5)$ GUT.
The ``baryon minus lepton number symmetry'' of $su(5)$ GUT is $\U(1)_X$.
 Plug in  to check 4d local anomaly of $X$-SU(5)$^2$:
\bea  \label{eq:anomaly-XSU5SU5}
\includegraphics[width=2.2in]{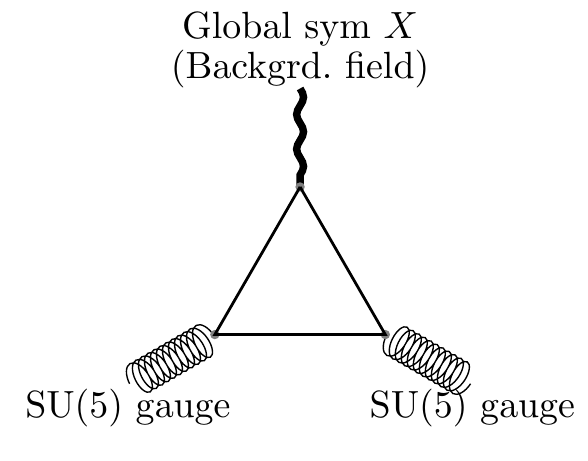}
\eea
\noindent
we find the anomaly factor contributed from the representation ${\bf{R}}$ of fermions in SU(5)
as the anti-fundamental ${\bf{R}} = \overline{\bf 5}$
and anti-symmetric ${\bf{R}} = {10}$, from the 15 Weyl fermions 
$\overline{\bf 5}\oplus {\bf  10}$ in one generation. Let us check the
$X$ current conservation or violation by ABJ type anomaly:
\bea \label{eq:XSU(5)2}
\dd \star (j_{X}) 
\propto
\sum_{\bf{R}} X_R \cdot  \Tr_{\bf{R}}[ F_{\SU(5)} \wedge F_{\SU(5)}]
\propto 
\sum_{\bf{R}} X_R \cdot c_2({\SU(5)}).
\eea
Here $c_2({\SU(5)})$ is the second Chern class of SU(5),
which is also related to the 4d instanton number of SU(5) gauge bundle.
For $\overline{\bf 5}\oplus {\bf  10}$ with $N_{\text{generation}}$, 
we get the $\U(1)_X$ charges for 
$$X_{\overline{\bf 5}}=-3, \quad X_{\bf  10}=1,$$ 
so\footnote{To evaluate the $c_2$ or the instanton number
in different representations,  ${\bf{R}}_1$ and  ${\bf{R}}_2$, 
we use the fact that
\bea
\Tr_{{\bf{R}}_1}[F \wedge F] / \Tr_{{\bf{R}}_2}[F \wedge F]  = (d({{\bf{R}}_1}) C_2({{\bf{R}}_1})) / (d({{\bf{R}}_2})C_2({{\bf{R}}_2}))
= (d(G) C({{\bf{R}}_1})) / (d(G)C({{\bf{R}}_2}))= C({{\bf{R}}_1}) / C({{\bf{R}}_2}),
\eea
here 
  $d({\bf{R}})$ and $C_2({\bf{R}})$  are respectively the dimension 
and  the quadratic
Casimir  of an irreducible representation ${\bf{R}}$.
Here $d(G)$ is the dimension of group and
$C(\bf{R})$ is the Dynkin index. We use a relation
$d(\bf{R}) C_2({\bf{R}})= d(G)C({\bf{R}})$ for a representation ${\bf{R}}$.
For the representation ${\bf{R}}$ of SU($N$) with $d(G)=N^2-1$, we have 
\bea
\begin{array}{ll l l}
\hline
{\bf{R}}      &      d(\bf{R})       &       C_2(\bf{R}) &          C(\bf{R}) \\
\hline
\text{Fundamental}     &   N         &   \frac{N^2-1}{2N}  &   \frac{1}{2}\\
\text{Antisymmetric}      &   N(N-1)/2   & \frac{(N+1)(N-2)}{N} &   \frac{N-2}{2}\\
\hline
\end{array}.
\eea
For SU(5) with $N=5$, we get
$\Tr_{{\bf{10}}}[F \wedge F]=(N-2)\Tr_{\bar{\bf{5}}}[F \wedge F]=3\Tr_{\bar{\bf{5}}}[F \wedge F]$.
\label{ft:SUNrep}
}
\bea
\dd \star (j_{X}) 
\propto
N_{\text{generation}} \Big(
X_{\bar{\bf 5}}\Tr_{\bar{\bf{5}}}[F \wedge F]
+ X_{\bf  10}\Tr_{{\bf{10}}}[F \wedge F]
\Big)
=N_{\text{generation}} \cdot 0 =0
\eea
vanishes. We confirm that the $\U(1)_X$ symmetry is ABJ anomaly free at least perturbatively in $su(5)$ GUT.

This anomaly matching is also true when we break down $\U(1)_X$ to $\Z_{4,X}$,
so that the mod 2 class 4d anomaly from 5d $(\CA_{{\Z_2}}) c_2(\SU(5))$ is still matched.

\subsection{$\eta(\text{PD}(\CA_{{\Z_2}}))$: 4d $\Z_{16}$ global anomaly}
 \label{sec:etaPDAZ16}
 
The 4d $\Z_{16}$ global anomaly from a 5d cobordism invariant $\eta(\text{PD}(\CA_{{\Z_2}}))$  in  \cite{JW2006.16996, WW2019fxh1910.14668} and Table \ref{table:SU5-discrete}
counts the number mod 16 of 4d left-handed Weyl spinors 
($\Psi_L \sim {\bf 2}_L \text{ of } \Spin(3,1)$ {or}  $\Psi_L \sim  {\bf 2}_L \text{ of } \Spin(4) = \SU(2)_L \times  \SU(2)_R$).
Given $N_{\text{generation}}$ (e.g., {3 generations}), 
for each generation, we have:
\bea
3 \cdot 2 + 3 \cdot 1 + 3 \cdot 1 + 1 \cdot 2   + 1 \cdot 1  = 15 = -1 \mod 16.
\eea 
For 1 generation, we need to saturates the anomaly with an index $\nu$: 
$$ \label{eq:nu-1}
\nu =-1 \mod 16.
$$
For 3 generations, we need
$$
3 \Bigg( 3 \cdot 2 + 3 \cdot 1 + 3 \cdot 1 + 1 \cdot 2   + 1 \cdot 1  \Bigg)= 45 = -3 \mod 16.
$$ 
Therefore we need to saturates the anomaly: 
$$
\nu =-3 \mod 16.
$$
For $N_{\text{generation}}$ generations, we need to saturates the anomaly: 
\bea \label{eq:Ngeneration}
\nu =-N_{\text{generation}} \mod 16.
\eea
This anomaly can be canceled by adding new degrees of freedom
\bea \label{eq:NgenerationNnuR}
\nu =N_{\text{generation}}\cdot (N_{\nu_R}=1) \mod 16.
\eea
This $\Z_{16}$ anomaly matching can be matched by adding a right-handed neutrino (the 16th Weyl spinor) per generation.
This also shows the robustness if we break down $\U(1)_{\bf{B}-\bf{L}}$ or $\U(1)_X$
down to $\Z_{4,\bf{B}-\bf{L}}$ or to $\Z_{4,X}$. Again this $\Z_4$ as the center  $Z(\Spin(10))=Z(\Spin(18))$ is important for the $so(10)$ or $so(18)$ GUT.

Are there other ways to match the anomaly other than introducing the right-handed neutrino (the 16th Weyl spinor) per generation? 
\Ref{JW2006.16996} introduces a new scenario by introducing 
a 4d TQFT or 5d TQFT in \Eq{eq:UU-GUT-1} to match the anomaly with a constraint \Eq{eq:anomaly-match-nu}.
In general, 
\Eq{eq:UU-GUT-1} schematically shows
the combinations of solutions by adding right-handed neutrino, or adding 4d non-invertible TQFT, or 5d invertible TQFT
to match the anomaly constraint \Eq{eq:anomaly-match-nu}.

\section{Bibliography}

\fontsize{11}{11pt} \selectfont

\bibliographystyle{Yang-Mills}
\bibliography{BSM-SU3SU2U1-cobordism-U.bib}

\newpage
\fontsize{12}{14pt} \selectfont

\section{Acknowledgements} 

JW is grateful to his previous collaborators for fruitful past researches  
as helpful precursors for the present work.
JW appreciates 
the conversations or collaborations with Miguel Montero, Kantaro Ohmori, Pavel Putrov, Ryan Thorngren \cite{PTWtoappear}, Zheyan Wan \cite{WanWangv2}, Yunqin Zheng,
Joe Davighi and Nakarin Lohitsiri,
and the mental support 
from Shing-Tung Yau.\footnote{
Instead of writing or drawing an image of the author's mental conditions, a piece of  
Ludwig van Beethoven's  music ``Piano Sonata No. 23 in F minor, Op. 57 Appassionata - the 2nd movement - Andante con moto'' may illuminate this well.
Listen:\\
\includegraphics[scale=.95]{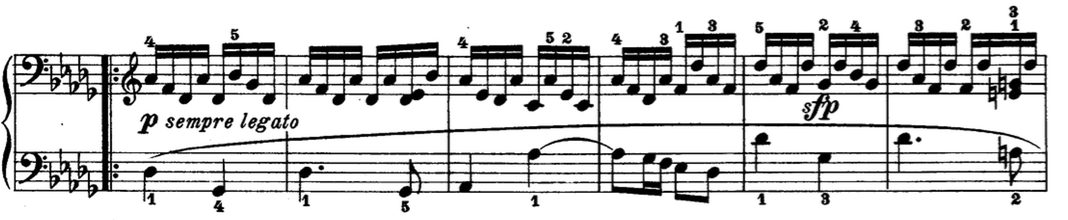}\\
\includegraphics[scale=.95]{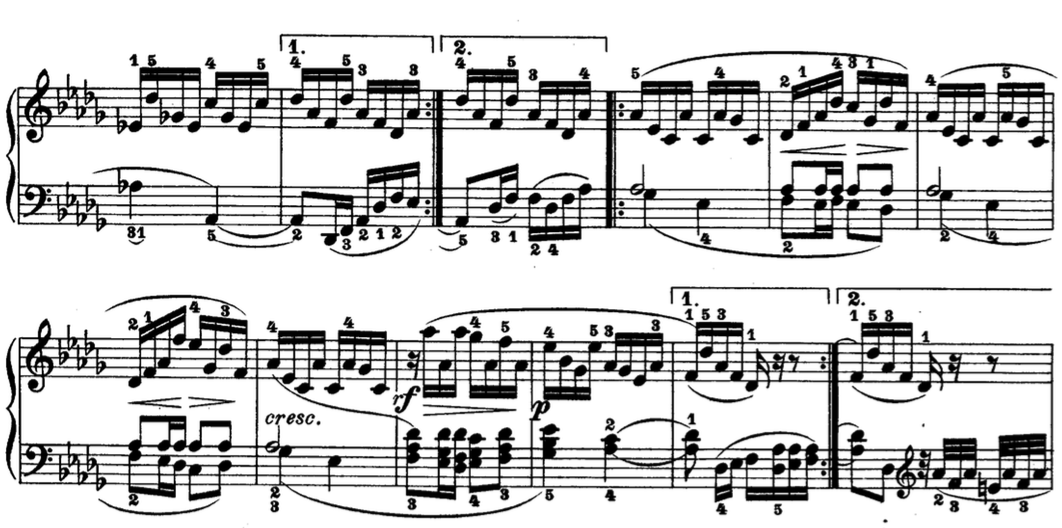}
}
JW thanks the participants of Quantum Matter in Mathematics and Physics program at
Harvard University CMSA for the enlightening atmosphere.
Part of this work is presented by JW in the workshop Lattice for Beyond the Standard Model physics 2019, on May 2-3, 2019 at Syracuse University 
and in the first week program of Higher Structures and Field Theory at Erwin Schr\"odinger Institute in Wien of August 4, 2020
\cite{ESIJWUltraUnification}.
%
JW was supported by
NSF Grant PHY-1606531. 
This work is also supported by 
NSF Grant DMS-1607871 ``Analysis, Geometry and Mathematical Physics'' 
and Center for Mathematical Sciences and Applications at Harvard University.

 \end{document}